\documentclass[12pt]{report}
\usepackage{graphicx}
\usepackage{epsfig}
\usepackage{mathtools}
\usepackage{amssymb}
\usepackage{slashed}
\usepackage[numbers,sort&compress]{natbib} 

\let\d\donothing

\newcommand{\bra}[1]{\langle #1|}
\newcommand{\ket}[1]{|#1\rangle}
\newcommand{\expt}[1]{\langle#1\rangle}
\newcommand{\braket}[2]{\langle #1|#2\rangle}

\newcommand{\tab}[0]{\,\,\,\,\,\,\,\,\,\,}

\newcommand{\d}[0]{\partial}
\usepackage{setspace}

\allowdisplaybreaks
\numberwithin{equation}{chapter}
\begin{document}

%

\thispagestyle{empty}

\newcommand{\thesisTitle}{}
\newcommand{\yourName}{Kevin Vanslette}
\newcommand{\yourSchool}{Physics}
\newcommand{\yourMonth}{May}
\newcommand{\yourYear}{2018}

\begin{center}

\textbf{\MakeUppercase{The inferential design of entropy and its application to quantum measurements}}\\
\doublespacing
\vspace{3\baselineskip}
by\\
\vspace{4\baselineskip}
\yourName\\
\vspace{3\baselineskip}
A Dissertation\\
Submitted to the University at Albany, State University of New York\\
in Partial Fulfillment\\
of the Requirements for the Degree\\
Doctor of Philosophy\\
\vspace{3\baselineskip}
College of Arts \& Sciences\\
Department of Physics\\
 \yourYear{}

\end{center}

\pagenumbering{roman}
\setcounter{page}{1}
\begin{center}

\textbf{}\\
\doublespacing
\vspace{3\baselineskip}

\vspace{4\baselineskip}
For Melissa, who told me never to give up on my dreams\\
\vspace{3\baselineskip}
\vspace{4\baselineskip}

\end{center}
\newpage
\doublespacing
\chapter*{Abstract}
\addcontentsline{toc}{chapter}{Abstract}
This thesis synthesizes probability and entropic inference with Quantum Mechanics and quantum measurement \cite{QRE,QBR,EDMeasurement,EDContextuality,MOPA,complexprob}. It is shown that the standard and quantum relative entropies are tools \emph{designed} for the purpose of updating probability distributions and density matrices, respectively \cite{QRE}. The derivation of the standard and quantum relative entropy are completed in tandem following the same inferential principles and design criteria. This provides the first design derivation of the quantum relative entropy while also reducing the number of required design criteria to two. 

The result of Caticha and Giffin \cite{GiffinBayes,Giffin} and Giffin's thesis \cite{GiffinThesis} is invaluable; it shows that the standard (relative) maximum entropy method is a ``universal method of inference" -- it is able to process information simultaneously that neither a Bayesian nor a standard maximum entropy method can process alone while being able to reproduce the results of both. An analogous conclusion is reached in this thesis, the quantum maximum entropy method derived in \cite{QRE} is the ``universal method of density matrix inference". This was accomplished by deriving, from the quantum maximum entropy method, a Quantum Bayes Rule \cite{QBR} and generalizations that cannot be obtained from a Quantum Bayes or a quantum (von Neumann) maximum entropy method \cite{Jaynes2} alone while being able to reproduce the results of both. The expanded results of \cite{QBR} more-or-less follow the main results of \cite{GiffinBayes,Giffin,GiffinThesis} in structure, but instead use density matrices rather than probability distributions. As the quantum maximum entropy method only uses the standard quantum mechanical formalism, the quantum maximum entropy method derived here may be appended to the standard quantum mechanical formalism and remove collapse as a required postulate, in agreement with \cite{Hellmann,Kostecki}.

The second part of this thesis revolves around the foundational theory of Quantum Mechanics called Entropic Dynamics \cite{EDNew}. Entropic Dynamics uses the standard maximum entropy method and information geometry to reformulate the Schr\"{o}dinger Equation and Quantum Mechanics as a theory of inference. We derive a density matrix formalism of Quantum Mechanics using the \emph{standard} maximum entropy method in Entropic Dynamics, and thus, the quantum maximum entropy method \cite{QRE} may be adopted by Entropic Dynamics wholesale. This implies that indeed the standard maximum entropy method is the ``universal method of inference" and that the quantum maximum entropy method holds the more specialized title of the ``universal method of density matrix inference" as it does not provide an intrinsic mechanism for the unitary evolution of pure states.

Entropic Dynamics is different than most interpretations of Quantum Mechanics because, rather than appending an interpretation to Quantum Mechanics, it states its interpretation, ``that particles have definite yet unknown positions and that entropic probability updating works", and only then does it derive Quantum Mechanics from these assumptions. This radical shift in interpretation allows one to solve the quantum measurement problem \cite{Johnson} (which was extended to included von Neumann and weak measurements in \cite{EDMeasurement}) and address quantum \emph{no-go} theorems \cite{EDContextuality}. Crucial to understanding why these apparent paradoxes pose no issue in Entropic Dynamics is understanding the foundation of inference that Entropic Dynamics is built upon. In particular, when it comes to measurement in Entropic Dynamics, we are able to divvy-up variables (observables, ect.) into two classes: they are the ontic \emph{beables} \cite{Bell1993}, which are the positions of particles, and the epistemic \emph{inferables} \cite{EDMeasurement} that in principle are inferred from detections of position and are therefore not predisposed to be part of the ontology. The fact that observables other than position are inferables in Entropic Dynamics, and that the positions of particles are the ontic \emph{beables}, allows the Entropic Dynamics formulation of Quantum Mechanics to \emph{not} be ruled out by the following pertinent \emph{no-go} theorems \cite{EDContextuality}: no $\psi$-epistemic \cite{Pusey}, Bell-Kochen-Specker \cite{BellKS,KS,Mermin}, and Bell's inequality \cite{BellEPR}. Entropic Dynamics is found to be viable theory of Quantum Mechanics.

\newpage

\chapter*{Acknowledgments}
\addcontentsline{toc}{chapter}{Acknowledgments}

\noindent \,\,\,\,\,\,\,\, First and foremost I must thank all of the members of my family, and of-course, Melissa Rhodes. Thank you for your never ending support, encouragement, and love along this extended academic journey -- I feel truly blessed having all of you in my life.   

To my advisor and friend, Professor Ariel Caticha, thank you for your guidance throughout my Ph.D. career. I have truly enjoyed the abundant number of philoso-physical conversations we shared over the years and the inspired research it has generated. Thank you for being a wonderful teacher, advisor, and friend.

To the many professors, academic colleagues, and friends in Albany and Worcester who have always made me feel welcome in your community's with your genuinely friendly, helpful, and fun attitudes -- you have all certainly made this journey more enjoyable and it has been a pleasure getting to know, and share experiences, with all of you. Thank you.

I would also like to thank the members of my Ph.D. defense committee, especially Ariel Caticha, for their interest, volunteered effort, time, and the insightful comments they have provided during the facilitation of this defense.
\newpage

%
%
\tableofcontents

\newpage
\setcounter{footnote}{0}

\chapter*{Introduction and Preface\label{Introduction}}
\addcontentsline{toc}{chapter}{Preface and Introduction}

The majority of the work in this thesis is inspired by Ariel Caticha's book, ``Entropic Inference and the Foundations of Physics" \cite{book}, which itself is inspired by the life work of E.T. Jaynes \cite{Jaynesbook}.  While facing push-back about Bayesian probability theory and his method of maximum entropy, Jaynes wrote \cite{JaynesProb}:\\\\
\emph{You cannot base a general mathematical theory on imprecisely defined concepts. You can make some progress that way; but sooner or later the theory is bound to dissolve into ambiguities which prevent you from extending it further. \,\,\,\,\,\,\,\,\,\,\,\,\,\,\,\,\,\,\,\,\,\,\,\,\,\,\,\,\,\,\,\,\,\,\,\,\,\,\,\,\,\,\,\,\,\,\,\,\,\,\,\,\,\,\,\,\,\,\,\,\,\,\,\,\,\,\,\,\,\,\,\,\,\,\,\,\,\,\,\,\,\,\,\,\,\,\,\,\,\,\,(A)}\\\\
By fine-tuning his notions of probability theory, Jaynes was able to reformulate Statistical Mechanics as an application of inference using the method of maximum entropy \cite{Jaynes1,Jaynes2}. Caticha takes this notion one step further and uses the maximum entropy method and information geometry to reformulate Quantum Mechanics (QM) as an application of inference called Entropic Dynamics \cite{EDNew,ED,book,ED2011,ED2010,ED2009}. Using Entropic Dynamics and precise notions of probability, inference, and entropy, we are able to resolve measurement problems and paradoxes in Quantum Mechanics through a newly prescribed lens \cite{Johnson,EDMeasurement,EDContextuality}. 

Caticha's book is enlightening in its ability to give meaning and purpose to mathematical objects that might otherwise be taken for granted. In particular, probability and entropy are derived, not by ansatz, but by \emph{design} as tools for inference \cite{Cox,Jaynesbook,book,CatichaEntropy,Shore1,Shore2}.\footnote{Cox derived probability theory as a measure of reasonable expectation \cite{Cox}. Later, E.T. Jaynes derived probability as a measure of plausibility \cite{Jaynesbook}. Shore and Johnson provided the first \emph{design} derivation of the relative entropy as a tool for inference. Many have since refined these design derivations of the relative entropy \cite{Csiszar1991,Skilling1, Skilling2, Skilling3,CatichaEntropy,GiffinBayes}.} The precise nature of these tools form the foundation for the rest of his work. This \emph{design} approach to inference and Physics not only reveals the functional form of the object in question, but also tells us how and when it should be used. For this reason, the design of probability will be reviewed in Chapter \ref{Probability} as it provides the foundation of this thesis. 

Perhaps the most compelling chapter in \cite{book} is Chapter 6: ``Entropy III: Updating Probabilities". In it, a functional of probabilities, which ends up being the relative entropy, is designed for the purpose of updating probability distributions, $\varphi\stackrel{*}{\rightarrow}\rho$, on the basis of new information $(*)$ \cite{CatichaEntropy,GiffinBayes,Shore1,Shore2,Csiszar1991,Skilling1, Skilling2, Skilling3}. I am happy that a large portion of my thesis is based on this chapter. In \cite{QRE}, I was able to refine and reduce the number of \emph{design criteria} needed to design the relative entropy from the first principles of inference. In addition, and using the \emph{same design criteria}, I was able to design a tool for the purpose of inferentially updating density matrices ($\hat{\varphi}\stackrel{*}{\rightarrow}\hat{\rho}$) -- it is the quantum relative entropy. This was accomplished though an (acknowledged) assistance by Caticha in \cite{QRE} -- he proposed the idea of designing a method of density matrix inference in the first place. These parallel derivations are reviewed in Chapter \ref{Entropy}.

Because the relative entropy is found to be the tool designed for updating probability distributions, we arrive at a refined description of Jaynes's maximum entropy method \cite{Jaynes1,Jaynes2}. The immediate result is, of course, consistency with Jaynes's notion that Statistical Mechanics may be  reformulated as a theory of inference; however, it was the work in Caticha and Giffin \cite{GiffinBayes,Giffin,GiffinThesis}, reviewed in \cite{book}, that proves the designed method of inference invaluable. In it, the maximum (relative) entropy method is revealed to be the universal method of inference. When joint probability distributions are constrained to satisfy expectation values over measured data, the maximum entropy method reproduces Bayes Rule and Jeffreys Rule as special cases. This shows that entropic methods are completely compatible with Bayesian methods of inference; however, the maximum entropy method may be updated with respect to arbitrary pieces of information (data, moments, both data and moments simultaneously, non-commuting constraints, etc.), and thus it forms a ``universal method of inference" \cite{GiffinBayes,Giffin,GiffinThesis}. 

Because both the standard and quantum relative entropies share the same design criteria and functional form \cite{QRE}, many of the well known applications (and restrictions) of the standard maximum entropy method manifest equally-well in the quantum maximum entropy method. Because density matrices are quantum mechanical, applications of the quantum maximum entropy method have direct ramifications in quantum mechanical experiments and inferences.  Inspired by the work of Caticha and Giffin \cite{GiffinBayes,Giffin,GiffinThesis}, I was able to derive the Quantum Bayes Rule \cite{Korotkov1,Korotkov2,Jordan} using the quantum maximum entropy method, and a few of its generalizations \cite{QBR}. This work therefore unifies topics in Quantum Information and quantum measurement through entropic inference in as much as Caticha and Giffin unified the maximum entropy method with Bayesian inference. In doing so, I reformulated L\"{u}ders' notion, that the operation of a measurement device must be to project the quantum state into a mixed state $\hat{\rho}\rightarrow \sum_x\hat{P}_x\hat{\rho}\hat{P}_x$ \cite{Luders}, from purely entropic (or quantum information-theoretic) arguments. L\"{u}ders' notion was found to be a manifestation of the ``unupdatability" of completely biased prior probability distributions in the density matrix formulation (i.e., if the state is known with certainty then it is in fact the state of the system).  I state this mathematical restriction on density matrix updating as a theorem, the ``Prior Density Matrix Theorem" (PDMT), that a prior density matrix may only be updated within the eigenspace that it spans. Thus the quantum maximum entropy method suggests that ``the collapse of the wavefunction" in the standard Quantum Mechanics formalism should really be rephrased as ``the decoherence of the wavefunction and then the collapse of the mixed state". The results and arguments from \cite{QBR} are reviewed in Chapter \ref{Entropy Applications}. Because the quantum maximum entropy method derived here only uses the standard quantum mechanical formalism, the quantum maximum entropy method may be appended to the standard quantum mechanical formalism.

A new conclusion from this thesis is that the quantum maximum entropy method is the ``universal method of density matrix inference". This was accomplished by deriving, from the quantum maximum entropy method, a Quantum Bayes Rule and generalizations that cannot be obtained from a Quantum Bayes or a quantum (von Neumann) maximum entropy method \cite{Jaynes2} alone while being able to reproduce the results of both. The results from \cite{QBR} are extended in Chapter \ref{Entropy Applications} and more-or-less follow the structure of \cite{GiffinBayes,Giffin,GiffinThesis} (here we use density matrices and the quantum relative entropy rather than probability distributions and the standard relative entropy).

A special case of the PDMT is that pure-state density matrices cannot be updated $\hat{\varphi}\stackrel{*}{\rightarrow}\hat{\varphi}$ via the quantum maximum entropy method. This implies that the quantum maximum entropy method does not provide an intrinsic dynamical evolution for pure quantum states, meaning that another mechanism is needed to generate the Schr\"{o}dinger Equation from the first principles of inference. The dynamical method for generating the Schr\"{o}dinger Equation as an application of inference may be done using the \emph{standard} maximum entropy method and information geometry -- it is Caticha's Entropic Dynamics formulation of Quantum Mechanics. An Entropic Dynamics-like formalism using the quantum relative entropy would ultimately miss the point -- it would not be foundational as it inevitably would require the use of Quantum Mechanics to derive Quantum Mechanics. Caticha's Entropic Dynamics is therefore reviewed in Chapter \ref{Entropic Dynamics}. In addition, Section \ref{MixedED} contributes a derivation of the beginnings of a density matrix formalism using the \emph{standard} maximum entropy method and information geometry within the Entropic Dynamics framework. Because Entropic Dynamics derives Quantum Mechanics from the standard maximum entropy method, as well as the density matrix formalism added here, the quantum maximum entropy method can be appended to Entropic Dynamics; and therefore, the \emph{standard} maximum entropy method retains its title as the ``universal method of inference". The quantum maximum entropy method thus holds the more specialized title of ``the universal method of density matrix inference". 

Having designed probability, entropy, quantum entropy, and derived Quantum Mechanics from the first principles of inference, my thesis is prepared to address, in Jaynes's words, the \emph{ambiguities which prevent you from extending (the theory) further} in Quantum Mechanics; these are the ``quantum measurement problem" and the various pertinent ``quantum \emph{no-go} theorems".

In \cite{Schlosshauer}, the ``quantum measurement problem" is broken into two related quantum measurement problems: the problem of \emph{definite outcomes} and the problem of \emph{preferred basis} (or \emph{degenerate basis}).  In QM, particles evolve unitarily from one pure state to another, and in some sense never ``settle down" to a definite final state, in all but the most trivial cases. Although the experimental results match the predictions of QM after a large number of experiments, they fail to match in formalism -- how do the \emph{definite outcomes} in experiment come about? ``Wavefunction collapse" is usually tacked onto the formalism \emph{ad hocly} as a second form of dynamical evolution to alleviate the problem of \emph{definite outcomes} \cite{vonNeumann}. The second quantum measurement problem of \emph{preferred basis} stems from the standard quantum formalism having an indifference toward a preferred ontology. The standard quantum formalism introduces Quantum Mechanics mathematically in terms of operators that act on vectors in a Hilbert space. There is no reason to believe one set of vectors should be more ``real" than any other set of vectors in the standard formalism. This lack of ontology becomes an issue when performing a von Neumann measurement \cite{vonNeumann}, which is arguably the most fundamental measurement technique in Quantum Mechanics. A von Neumann measurement uses quantum entanglement between the system of interest and the states of a detector to perform inference. If the ontology of the detector and system are not specified, when a detector state is measured, it can be unclear what (ontological) state was actually detected and thus what state of interest was actually inferred. Thus, without specification of a \emph{preferred basis}, quantum measurement procedures are insufficient in general.

 Pivotal to the arguments that allow Entropic Dynamics to solve the quantum measurement problem \cite{Johnson} (\cite{EDMeasurement} extends the treatment to von Neumann and weak measurements), and avoid being ruled out by quantum \emph{no-go} theorems \cite{EDContextuality}, is that Entropic Dynamics derives dynamical probability distributions whose evolution are constrained by the assertion that particles have definite and ontological, yet unknown, positions. This assertion is implemented via expectation value constraints that impose continuous motion, and (expected) drift, of particle positions using the standard maximum entropy method. The quantum measurement problem of preferred basis is solved by the fact that particles in Entropic Dynamics do have a preferred basis, it is the ontological position of the particles that Entropic Dynamics is built around \cite{Johnson,EDMeasurement}. The quantum measurement problem of \emph{definite values} or ``wavefunction collapse" is solved within the Entropic Dynamics framework by using Bayesian inference to update one's probability when confronted with information in the form of data \cite{GiffinBayes,Johnson,EDMeasurement}. This Bayesian treatment of collapse is completely consistent with the Entropic Dynamics framework because \emph{both Bayesian inference and Schr\"{o}dinger evolution are derived as special cases of entropic probability updating} -- the method of inference is generated by the available information at hand. Thus, Entropic Dynamics solves the quantum measurement problem in a way that afflicts other interpretations of Quantum Mechanics -- inference is the foundation of Entropic Dynamics. The results of \cite{Johnson,EDMeasurement} will be expanded upon in Chapter \ref{Measurement in Entropic Dynamics}. 

In \cite{EDMeasurement}, we differentiate between quantum ``observables" and what we call the \emph{inferables} of a theory, i.e., the quantities that may be inferred.\footnote{``Observables" other than position, and odd quantum mechanical quantities such as complex valued ``Weak Values" \cite{AAV}, are treated as purely epistemic \emph{inferables} (they need not have an ontological predisposition) \cite{EDMeasurement}. Rather than being directly measured, they are inferred from the preferred position basis.} Entropic Dynamics insists upon this change of language, \emph{observables} $\rightarrow$ \emph{inferables}, when it derives Quantum Mechanics as an application of inference; and although on the surface this change of language seems purely semantical, it has deep implications for the interpretation of Quantum Mechanics.
Using the concept of \emph{inferables}, as well as treating particles as having ontological (\emph{beable}) positions, we address the following quantum \emph{no-go} theorems in Entropic Dynamics \cite{EDContextuality}: the no $\psi$-epistemic theorem \cite{Pusey}, Bell's theorem \cite{BellEPR}, and the Bell-Kochen-Specker theorem \cite{BellKS,KS,Mermin}. The standard interpretation of the result of the Bell-Kochen-Specker Theorem is that operators in Quantum Mechanics are \emph{contextual}, that is, an observable's ``character", ``aspect", or ``value" depends on the remaining set of commuting observables in its measurement setting. Thus, quantum observables in general cannot be assigned definite, preexisting, yet potentially unknown, values before measurement. Because in Entropic Dynamics, variables other than position are epistemic \emph{inferables}, we do not need to assign them definite prexisting values before measurement. Because positions are ontological, but other \emph{inferables} are epistemic, and thus naturally \emph{contextual}, we classify Entropic Dynamics as a hybrid-contextual theory of Quantum Mechanics, which as it turns out, is not ruled out by the Bell-Kochen-Specker Theorem. As quantum \emph{no-go} theorems are proofs by contradiction, they only rule out theories or interpretations of Quantum Mechanics that strictly adhere to their (shown contradictory) set of assumptions. In short, Entropic Dynamics does not adhere to these set of assumptions in any of the previously mentioned \emph{no-go} theorems, and therefore it is not ruled out. This will be reviewed in Chapter \ref{No-go}.

\newpage
\setcounter{page}{0}
\pagenumbering{arabic}
\addcontentsline{toc}{part}{Part I: Universal Entropic Inference}
\part*{Part I: Universal Entropic Inference}

\chapter{Foundations of Probability Theory\label{Probability}}

This chapter reviews Caticha's \emph{design} derivation of probability theory \cite{book} and reiterates related concepts from \cite{Pragmatic}. A formal equivalence between, the ``degree of rational belief" one \emph{ought} to have that a proposition is true, and, the ``probability" a proposition is true,  is found. Cox derived probability theory as a measure of reasonable expectation by solving functional equations over Boolean equivalent propositions in 1946  \cite{Cox}. His derivation was appreciated by Jaynes \cite{Jaynesbook}, who instead \emph{designed} probability as an objective measure of plausibility. The approach taken by Caticha differs in a few ways from Cox and Jaynes; the goal is to design a tool for inductive inference such that objective statements may be made about one's rational, but informationally subjective, beliefs. The main differences will be noted within the derivation below.

\section{Designing a tool for inference}

When faced with incomplete information, complete and precise deduction may not always be possible. In these instances, we require a tool to perform inductive reasoning such that trustworthy inferences may be made. Usual deduction runs a straightforward course, ``If $a$ then $b$", and ``If $b$ then $c$", one may deduce: ``If $a$ then $c$". Deduction fails to address less certain situations, such as ``If $a$ suggests $b$ and $b$ suggests $c$" then, by what measure should we infer, ``$a$ suggests $c$"?  Although correlations between the propositions $\{a,b,c,..\}$ are known to exist, there is no rule for manipulating these ``suggestions" logically.  We are therefore motivated to design a tool capable of giving rational assessments in situations with limited information.  Not knowing if we should believe a proposition to be true or false, we seek to design rules for making inferences that maintain a high standard of honesty in our beliefs, \emph{for the purpose of doing science} \cite{Private}. We therefore seek a tool capable of quantifying \emph{the degree of rational belief} (DoRB) we \emph{ought} to hold that a proposition is true.\footnote{The question of whether degrees of rational \emph{belief} are scientific is discussed in a comment at the end of the derivation.}

In Cox's derivation of probability, he let $\{a,b,c,...\}$ denote ``propositions" rather than ``events". He states, ``speaking of events easily invokes the notion of sequence in time, and this may become a source of confusion. A proposition may, of course, assert the occurrence of an event, but it may just as well assert something else, for example, something about a physical constant." \cite{Cox}. Propositions are self-contained statements or assertions that may be true or false -- and therefore they may take on a wide variety of character; for example $a$ could represent `` the position of a particle is $x$", $b$ ``the gene sequence of a banana is $y$", to anything $c$ one might wish to infer ``$z$". 

In everyday situations it is apparent that some propositions are more believable than others. By design we would like our tool for inference to quantify the believability of a proposition. If $a$ is more believable than $b$ and $b$ more believable that $c$, then we wish degrees of rational belief (DoRBs) to represent this information, as well as $a$ being quantifiably more believable than $c$.  We are therefore inclined to represent DoRBs  by real numbers such that they may be ranked in a transitive manner. It is then by design that a DoRB should be a function that maps arbitrary propositions $\{a,b,c,...\}$ to values on the real line such that the believability of propositions are ranked. We will denote this function with square brackets $[...]$ such that $[z]\in\mathbb{R}$ represents the DoRB an arbitrary proposition $z$ is true. That is \cite{book},\\

\emph{Degrees of rational belief (or, as we shall later call them, probabilities)
are represented by real numbers. } 
\\

 When utilizing DoRBs, to be maximally informed \cite{Private}, we should make inferences only once all of the \emph{relevant} information has been taken into account. It is clear that the knowledge of one proposition may support or oppose another proposition. For instance, if we know $a:$``It is cloudy today", then $a$ logically supports the belief in the claim $b:$`` It will rain today". To remain honest about our beliefs, if we know $a$ to be true, we should logically assign a larger DoRB to $b$ rather than if we had not known $a$ to be true.  We introduce a vertical bar ``$|$" between propositions $b|a$ to denote a new proposition, which is read: ``$b$ is true given $a$ is true", ``$b$ given $a$", or ``$b$ conditional on $a$". As $b|a$ is a proposition itself, we may assign it a DoRB such that $[b|a]$ is read ``the degree of rational belief $b$ is true given that $a$ is true" or for short ``the DoRB of $b$ given $a$".  

Conditional propositions allow us to construct the upper and lower limit a DoRB may take. Because $a|a=v_t$ is true, the DoRB of $a|a$ must represent the ``most" believable situation for logical and rational consistency -- we denote the value of this DoRB as $[a|a]=[v_t]$. Conversely, the DoRB that not-$a$, denoted $\tilde{a}$, is true given $a$ is true is $\tilde{a}|a=v_f$ is false such that $[\tilde{a}|a]=[v_f]$ represents complete rational disbelief or the ``least" degree of rational belief. The amount of rational belief we assign to a certainly true proposition $a|a$ or a certainly false proposition $\tilde{a}|a$ is independent of the specific proposition $a$. This lets $[v_t]$ and $[v_f]$ be single, but distinct, numbers such that transitivity may be obeyed.

Compound propositions may be built by considering Boolean logic operations between propositions: the ``and" conjunction, denoted $\wedge$, constructs propositions of the form $a\wedge b$ (later denoted as $ab$ or with a comma $a,b$) that are true iff both $a$ and $b$ are true, as well as the Boolean disjunction ``or",  denoted $\vee$, that constructs propositions of the form $a\vee b$ that are true when $a$, $b$, or both are true while being false iff both are false. Because applying arbitrary conjunctions and disjunctions between propositions results in a new proposition itself, ``$z$", the set of all propositions is closed under Boolean operations. This realization is fundamental in moving forward because anything we learn about the functional form of $[a\vee b]$ and $[ a\wedge b]$ leads to us learning more about the desired functional form of $[z]$ for \emph{arbitrary} propositions $z$.

\section{Functional form of [$a$ or $b$] }
Consider the DoRB a proposition $z|d\equiv a\vee b|d$ is true. For consistency, it must be that $[z|d]=[a\vee b|d]$, which means that finding the functional form of $[a\vee b|d]$ will ultimately give insight into the functional form of $[z|d]$. Because of this, we expect the functional form of $[a\vee b|d]$ to somehow be related to the functions: $[a|d],[b|d],[a|bd],$ and $[b|ad]$. We let $F$ denote the functional dependence of a DoRB over a disjunction,
\begin{eqnarray}
[a\vee b|d]\equiv F([a|d],[b|d],[a|b,d],[b|a,d]),\label{F}
\end{eqnarray}
 with that of the constituting disjunct DoRBs. First consider the special case when $a$ and $b$ are mutually exclusive (i.e. given $b|a$ is false and vice versa), then,
\begin{eqnarray}
[a\vee b|d]=F([a|d],[b|d],[v_f],[v_f])\equiv F([a|d],[b|d]),
\end{eqnarray}
for brevity. The functional form of $F$ can be found by considering the associative property of three propositions,
\begin{eqnarray}
z\equiv a\vee (b\vee c)=(a\vee b)\vee c,
\end {eqnarray}
where again we will let $a,b,c$ be mutually exclusive propositions. The DoRB function must arrive at the same result for the logically equivalent forms of $z$. This imposes,
\begin{eqnarray}
[a\vee (b\vee c)|d]=[(a\vee b)\vee c|d].
\end{eqnarray}
$F$ may be applied two-fold in each instance, first,
\begin{eqnarray}
[a\vee b\vee c|d]=F([a\vee b|d],[c|d])=F([a|d],[b\vee c|d]),
\end{eqnarray}
and then again,
\begin{eqnarray}
[a\vee b\vee c|d]=F(F([a|d],[b|d]),[c|d])=F([a|d],F([a|d],[c|d])),
\end{eqnarray}
with equality due to the equivalency of the propositions and for self consistency. This is the well known Associativity Functional Equation \cite{Aczel}; it has
the general solution
\begin{eqnarray}
[a\vee b|d]=F([a|d],[b|d])=\phi^{-1}(\phi([a|d])+\phi([b|d])+\alpha)\label{Acc},
\end{eqnarray}
where $\phi$ is an arbitrary monotonic function and $\alpha$ is a constant \cite{book,Aczel}. It should be noted that in \cite{Cox,Jaynesbook}, the Associativity Functional Equation is used to specify the form of [$a$ and $b$] rather than [$a$ or $b$], the $\phi$ function of ours being related to theirs by its logarithm. This is consistent because the logarithm of a (positive) monotonic function is monotonic itself (theirs is a positive exponential). Because $\phi$ is a monotonic function, it necessarily preserves transitivity. Without loss of generality we may take $\phi$ of both sides
\begin{eqnarray}
\phi([a\vee b|d])=\phi([a|d])+\phi([b|d])+\alpha,
\end{eqnarray}
and regraduate (monotonically rescale) our DoRB function by letting,
\begin{eqnarray}
[a|d]\rightarrow \zeta(a|d)\equiv \phi([a|d])+\alpha,
\end{eqnarray}
such that the mutually exclusive disjunction rule for DoRBs takes the convenient form,
\begin{eqnarray}
\zeta(a\vee b|d)=\zeta(a|d)+\zeta(b|d).\label{or}
\end{eqnarray}

The special case of mutually exclusive propositions can be lifted by considering the logical identities: $a=ab\vee a\tilde{b}$, $b=ab\vee \tilde{a}b$, and $a\vee b=ab\vee \tilde{a}b\vee a\tilde{b}$. Because $ab,a\tilde{b},\tilde{a}b$ are propositions themselves we may use equation (\ref{or}) to find,
\begin{eqnarray}
\zeta(a\vee b|d)=\zeta(ab\vee \tilde{a}b\vee a\tilde{b}|d)=\zeta(ab|d)+\zeta(a\tilde{b}|d)+\zeta(\tilde{a}b|d)+[\zeta(ab|d)-\zeta(ab|d)]\nonumber\\
=\zeta(ab\vee a\tilde{b}|d)+\zeta(ab\vee\tilde{a}b|d) - \zeta(ab|d),
\end{eqnarray}
which means the general sum rule for DoRBs is,
\begin{eqnarray}
\zeta(a\vee b|d)=\zeta(a|d)+\zeta(b|d)- \zeta(ab|d).
\end{eqnarray}

\section{Functional form of [$a$ and $b$]}
Consider we wish to find the DoRB of a proposition $z\equiv a\wedge b\equiv ab$. Again this means that finding the functional form of $\zeta(a b|d)$ will give insight into the functional form of $\zeta(z|d)$. Due to this, we again expect the functional form of $\zeta(ab|d)$ to be related to the functions: $\zeta(a|d),\zeta(b|d),\zeta(a|bd),\zeta(b|ad)$, which we express as,
\begin{eqnarray}
\zeta(ab|d)=G(\zeta(a|d),\zeta(b|d),\zeta(a|bd),\zeta(b|ad)).
\end{eqnarray}
The function $G$ can be deconstructed from four arguments down to two arguments by following eliminative induction \cite{book,Smith}. The functional dependencies that capture conjunction ``and" as it pertains to the DoRB of $a$ and $b$ depend both on the DoRB of $a$ as well as our DoRB of $b$ given $a$ is true -- $b|a$,
\begin{eqnarray}
\zeta(ab|d)=G(\zeta(a|d),\zeta(b|ad))=G(\zeta(b|d),\zeta(a|bd))=\zeta(ba|d),\label{G}
\end{eqnarray}
while all other candidates fail to perform as desired. The functional form of $G$ can be found by considering the distributive property of propositions,
\begin{eqnarray}
z\equiv a(b\vee c)=ab\vee ac.
\end{eqnarray}
For our inference tool to not be self-refuting, $\zeta$ must arrive at the same result for the logically equivalent forms of $z$, which imposes the following relation,
\begin{eqnarray}
\zeta(a(b\vee c)|d)=\zeta((ab\vee ac)|d).
\end{eqnarray}
Letting $b$ and $c$ be mutually exclusive for simplicity, we have $\zeta(b\vee c|d)=\zeta(b|d)+\zeta(c|d)$ such that $\zeta((ab\vee ac)|d)=\zeta(ab|d)+\zeta(ac|d)$, so,
\begin{eqnarray}
\zeta(z)=G(\zeta(a|d),\zeta(b|ad)+\zeta(c|ad))=G(\zeta(a|d),\zeta(b|ad))+G(\zeta(a|d),\zeta(c|ad)).
\end{eqnarray}
Letting $u=\zeta(a|d)$, $v=\zeta(b|ad)$ and $w=\zeta(c|ad)$ recasts $G$ into a more obvious form,
\begin{eqnarray}
G(u,v+w)=G(u,v)+G(u,w).
\end{eqnarray}
This functional equation is linear in the second argument meaning it has the general solution \cite{Aczel},
\begin{eqnarray}
G(u,v)=A(u)v.\label{linear}
\end{eqnarray}
 Because $ab=ba$, if we let $u'=\zeta(a|bd)$ and $v'=\zeta(b|d)$ we have that $\zeta(ab)=\zeta(ba)$ implies $G(u,v)=G(v',u')$, so,
\begin{eqnarray}
A(u)v=A(v')u'.\label{Com}
\end{eqnarray}
In the special case that the propositions in question are independent, $u=u'$ and $v=v'$, one finds the functional form of $A(u)$ must be $A(u)\propto u$. The general solution for conjunction is therefore proportional to the simple product of its constituents, $G(u,v)\propto uv$. This means,
%
%
\begin{eqnarray}
\zeta(ab|d)=G(\zeta(a|d),\zeta(b|ad))=\chi\zeta(a|d)\zeta(b|ad)=\chi\zeta(b|d)\zeta(a|bd),
\end{eqnarray}
where $\chi$ is a constant, and the last equality is due to the commutativity of $a b$ (\ref{Com}). 

\paragraph{Cleaning up the constants:}
Lastly we find values for $\chi$, $\zeta(v_f)$, and $\zeta(v_t)$. Because a proposition or its negation is certainly true, $\zeta(v_t)=\zeta(a\vee \tilde{a})$, 
we have,
\begin{eqnarray}
\zeta(v_t)=\zeta(a\vee \tilde{a}|a)=\zeta(a|a)+\zeta(\tilde{a}|a)=\zeta(v_t)+\zeta(v_f).
\end{eqnarray}
This equation is satisfied for $\zeta(v_t)=\pm\infty$ or $\zeta(v_f)=0$ because by design $\zeta(v_t)\neq \zeta(v_f)$. Because the DoRBs are real numbers by design, $\zeta(v_t)$ is finite such that it is not a non-number, like infinity. This then allows DoRBs to be understood as a fraction of certainty, something that would be impossible if the DoRB of a true proposition was infinite. The only reasonable solution to the above equation is therefore $\zeta(v_f)=0$. 

We find $\chi$ by evaluating the DoRB of $ab|ab$ and using the product rule,
\begin{eqnarray}
\zeta(v_t)=\zeta(ab|ab)=\chi\zeta(v_t)^2,
\end{eqnarray}
and therefore $\chi=1/\zeta(v_t)$ because $\zeta(v_t)\neq \zeta(v_f)=0$. The product rule becomes,
\begin{eqnarray}
\zeta(ab)=\frac{\zeta(b|a)\zeta(a)}{\zeta(v_t)}.
\end{eqnarray}
We may further regraduate the DoRB  $\zeta(z)\rightarrow p(z)=\frac{\zeta(z)}{\zeta(v_t)}$ for convenience such that  $p(v_t)=p(z\vee\tilde{z})=1$ is normalized to unity, and therefore, $0\leq p(z)\leq 1$ for any proposition $z$. 
\section{Corollaries and Comments}
From now on we will use the standard comma notation between propositions to denote an ``and" conjunction. The sum rule,
\begin{eqnarray}
p(a\vee b|d)=p(a|d)+p(b|d)-p(a,b|d),\label{sumrule}
\end{eqnarray}
as well as the product rule,
\begin{eqnarray}
p(a,b|d)=p(a|d)p(b|a,d)=p(b|d)p(a|b,d),\label{productrule}
\end{eqnarray}
were found to be the desired strategies designed for manipulating \emph{degrees of rational belief}. These rules are formally equivalent to the sum and product probability rules, and thus, this derivation provides an interpretation for probability as ``a degree of rational belief" \emph{wherever and whenever probability theory is utilized}. This is perhaps the first suggestion that Statistical and Quantum Theory are theories of \emph{inference}, that is, they are theories that quantify, rank, and guide one's intuition of what one \emph{ought} to believe is true rather than providing an ultimate description of nature; however, this is not revealed to be the case until the maximum entropy method and the Entropic Dynamics formulation of Quantum Mechanics are derived. Future chapters in this thesis will provide the foundation for laws of Physics as applications of \emph{inference}. Note that at this point it is straightforward to introduce probability densities $p(x_i)\rightarrow \rho(x)$ for continuous propositions $x_i\rightarrow x$ that provide probabilities when taken with the product of their measure $\rho(x)dx=p(x)$. When there is no room for misinterpretation I may simply call $\rho(x)$ a probability or probability distribution rather than a probability density. It is interesting to note that when one is summing $\sum_{i}p(x_i)$ that it is a sum over propositions. In the continuous case $\int \rho(x)\, dx$ is, in the strictest interpretation, an integration over \emph{propositions}. I believe this interpretation is interesting in the later chapters pertaining to Entropic Dynamics.
\paragraph{Corollaries}
If we know $b$ to be true and potentially correlated to $a$, then as a matter of self honesty, we should include this information when considering the probability of $a$ (whenever available). The probability of $a$ conditional on $b$, $p(a|b)$ may be rewritten using the product rule (where logically $p(b)>0$), as
\begin{eqnarray}
p(a|b)=\frac{p(b|a)p(a)}{p(b)}.\label{BayesTheorem}
\end{eqnarray}
This is Bayes Theorem, and therefore it is designed to reflect informed and honest judgment concerning the DoRB $a$ is true given $b$ is true. In certain instances when the truth of a proposition is learned, one is obligated, for the purpose of making informed judgments and self honesty, to update an old probability $p(a)$ to a new probability, 
\begin{eqnarray}
p(a)\stackrel{*}{\rightarrow} p(a|b)\label{update},
\end{eqnarray}
from (\ref{BayesTheorem}). It is thus natural to define \emph{information} operationally $(*)$ as the \emph{rationale} that causes a probability distribution to change (inspired by \cite{book}), (\ref{update}) being one example. In another example, if it is unclear whether $b$ is true or $\tilde{b}$ is true, then one can use the fact that $b\vee\tilde{b}=(*)$ is certainly true, and make the following inference about $a$,
\begin{eqnarray}
p(a,b)\stackrel{*}{\rightarrow}p(a,b\vee\tilde{b}|d)=p(a,b|d)+p(a,\tilde{b}|d)=p(a|d)\sum_{b}p(b|a,d)=p(a|d).
\end{eqnarray}
This process is called \emph{marginalization}, in particular above, $b$ has been marginalized over and gives a best guess for $a$. If, later, the value of $b$ is learned, one may use (\ref{update}), or revert back to $p(a,b)$ if one inquires about the probability of $a$ and $b$.

The expectation value of an outcome is defined to be such that the variance about its value is minimal \cite{book}, and therefore the expected value is,
\begin{eqnarray}
\expt{x}=\sum_ix_ip(x_i).
\end{eqnarray}
It should be noted that in general $\expt{x}$ is not within $\{x_i\}$, and thus it cannot in general represent an element of reality even if $\{x_i\}$ do. In this sense, expectation values, like DoRBs, pertain to the characterization of knowledge about reality, rather than reality itself, and therefore they are potentially useful \emph{epistemic} quantities. 

A particularly interesting notion in probability theory is information geometry. There are several derivations of information geometry reviewed and given in \cite{book}, one of them in particular is as ``a measure of distinguishably between probability distributions". The measure of distinguishably is the invariant interval,
\begin{eqnarray}
d\ell^2=g_{ab}d\theta^a d\theta^b,
\end{eqnarray}
where $\theta^a$ is the $a$th parameter of $\theta=\{\theta^1, \theta^2,...\}$, which label the family of probability distributions $\{\rho(x|\theta)\}$ that populate a statistical manifold $\mathcal{M}$ having coordinates $\theta$. In particular, the information metric $g_{ab}$ assigns geometry to $\mathcal{M}$, and it is unique up to an overall scale factor $C$ \cite{Cencov,Campbell}, 
\begin{eqnarray}
g_{ab}=C\int dx\rho(x|\theta)\frac{\d \log \rho(x|\theta)}{\d \theta^a}\frac{\d \log \rho(x|\theta)}{\d \theta^b}.
\end{eqnarray}
Entropic Dynamics will take advantage of Information Geometry in the derivation of Quantum Mechanics in Chapter \ref{Entropic Dynamics}.

\paragraph{Comment: Are rational beliefs scientific?}
In short, ``Yes" \cite{Pragmatic}.  In science we have access to data, observations, and mathematical tools for accurately describing these data and observations. We are particularly complacent when a mathematical tool is able to predict new data and observations; however, if it fails to do so, we either modify it, abandon it, or claim it was being used in a region outside its operation. The set of mathematical tools that together form the body of scientific description are rigorous because they are subject to testing, peer-review, and may be adapted to reflect new evidence. This practice preserves honesty as well as allows one to make the most \emph{informed} predictions. It is therefore clear that \emph{rational practices} allow one to form \emph{rational beliefs}, and certainly, science aims to discredit the alternative. If the physics of today is changed tomorrow, our scientific process allows us to adjust accordingly such that we have the most informed description of nature as possible. This process may be exactly described by manipulating DoRBs, or as we find, probability.

\paragraph{Comment: Why not simply use Kolmogorov's Axioms?}

The axiomatic derivation of probability given by Kolmogorov in 1933 \cite{Kolmogorov}, is arguably incomplete. In it, probability is defined as a map from a set $E$ of events $\{e_i\}$ (which may be sets themselves) to the positive real line: $e_i\rightarrow P(e_i)\in \mathbb{R}$. The remaining axioms state that probability is normalized $P(E)=1$ and that under exclusive addition $P(e_1+e_2)=P(e_1)+P(e_2)$. This axiomatic characterization of probability is not favored on two accounts. 

The first is that it \emph{fails to provide an interpretation for probability}. Because at most the interpretation of probability in the Kolmogorov picture is a map from a set of events to the real line between $0$ and $1$, it is isomorphic to other instances represented in this interval, which may lead to misidentification. I dedicated an article \cite{complexprob} to argue against the notions of a ``generalized complex probability" or ``complex quasi-probability", because they fail to rank propositions by more or less probable -- the whole point of having probability in the first place. Some have poorly named these instances quasi-\emph{probabilities}, because when viewed axiomatically, they appear to simply relax Kolmgorov's third axiom that probabilities are real but fail to mention their inability to rank the DoRB, plausibility, or probability a proposition is true. I do give an explanation for how probabilities may be complex and still rank degrees of rational belief, but the complex representation of actual probabilities is ultimately not particularly useful other than for showing how quasi-probability distributions might fail to rank (a miniature \emph{no-go} theorem of sorts), so the full development is omitted here.

A second flaw is that the axiomatic characterization lacks an updating procedure or a reason to update.  This flaw may actually be regarded as an instance of the first flaw, since ultimately we would like to know what probability is \emph{and} how to use it. In the Kolmogorov paper \cite{Kolmogorov}, conditional probability (or Bayes Theorem) is defined and then used to ``derive" the product rule, neither of which make reference to the original axioms in an obvious way -- in some sense joint probability distributions are neglected all together in the axioms. In our derivation above, probability is a tool designed for the pragmatic purpose of making inferences; far more desirable for doing science than the axiomatic approach, which appears to design probability for the purpose of describing its set theoretic foundations.

\section{Conclusions and the desire for an updating tool}

Probability theory has been derived as the unique tool for manipulating degrees of rational belief, and thus, the ``degree of rational belief one ought to assign to a proposition being true" is formally equal to ``the probability" of that proposition. A probability distribution is therefore a statement of rational beliefs, or as it is sometimes called, a ``state of knowledge". Although we have derived rules for manipulating probability distributions using the sum and product rules, we have not given a definitive prescription for assigning numerical values to individual probability distributions \cite{Cox}.  This leads to the problem of priors \cite{book,Cox}, a problem Jaynes discusses at length \cite{JaynesPrior}. Ultimately it is found that the proper assignment of prior distributions is on the onus of the agent to use the tools of inference correctly -- they must be reasonable and use the information at hand, which may include, but is not limited to, the symmetries in the problem. Updating probability distributions has been motivated by practicality as well as self honesty, but a general, less \emph{ad hoc} method for updating probability distributions is desired. In the next Chapter we will design a rigorous tool for updating probability distributions as well as a tool for updating density matrices -- they are the standard and quantum relative entropies, respectively.


\chapter{The Shared Inferential Foundation of the Standard and Quantum Relative Entropies\label{Entropy}}

As is desired in the comments and conclusions of Chapter \ref{Probability}, we \emph{design} an inferential updating procedure for probability distributions and density matrices such that inductive inferences may be made. This chapter follows \cite{QRE}. 

 The inferential updating tools found in this derivation take the form of the standard and quantum relative entropy functionals, and thus we find that these functionals are \emph{designed} for the purpose of updating probability distributions and density matrices, respectively. Previously formulated design derivations that find the relative entropy to be a tool for inference originally required five \emph{design criteria} (DC) \cite{Shore1,Shore2,Csiszar1991}, this was reduced to four in \cite{Skilling1, Skilling2, Skilling3}, and then down to three in \cite{CatichaEntropy,GiffinBayes} (reviewed in \cite{book}). We reduced the number of required DC down to two while also providing the first \emph{design} derivation of the quantum relative entropy---\emph{using the same design criteria and inferential principles in both instances}.

 The designed quantum relative entropy takes the form of Umegaki's quantum relative entropy, and thus it has the ``proper asymptotic form of the relative entropy in quantum (mechanics)'' \cite{Petz1,Petz2,Petz3}. Recently, Wilming et al. \cite{Wilming} gave an axiomatic characterization of the quantum relative entropy that ``uniquely determines the quantum relative entropy''. Our derivation differs from their's, again in that we \emph{design} the quantum relative entropy for a purpose, but also that our DCs are imposed on what turns out to be the functional derivative of the quantum relative entropy rather than on the quantum relative entropy itself. The use of a quantum entropy for the purpose of inference has a large history: Jaynes~\cite{Jaynes1,Jaynesbook} invented the quantum maximum entropy method \cite{Jaynes2}, and this method was found to be useful by \cite{Balian1,Balian2,Balian3,Balian4,Balian5,Balian6,Partovi1,Partovi2} and many others. However, we find the quantum \emph{relative} entropy to be the suitable entropy for updating density matrices, rather than the von Neuman entropy \cite{vonNeumann}, as is suggested in \cite{Carlo}. Thus, \cite{QRE} gives the desired motivation for why the appropriate quantum relative entropy for updating density matrices, from prior to posterior, should be logarithmic in form while also providing a solution for updating non-uniform prior density matrices \cite{Carlo}.  The relevant results of these papers may be found using the quantum relative entropy with suitably chosen prior density matrices. 

One of the primary conclusions from \cite{Shore1,Shore2} is that because the relative entropies were reached by design, they may be interpreted as such, ``the relative entropies are tools for updating'', which means we no longer need to rely on an interpretation of entropy as a measure of disorder, an amount of missing information \cite{Shannon}, or amount of uncertainty \cite{Jaynes1}. This shifts the focus on entropy away from measures of ``information" or ``uncertainty", which may be misleading,\footnote{The following example from \cite{book} shows that interpreting entropy as an ``amount of missing information" (or uncertainty) can be misleading: Given someone who normally carries their keys in their pocket, the probable location of their keys is described by narrow probability distribution in their pocket. If this person makes a measurement, something that's normal utility is to gain information, but finds that their keys are not in their pocket, then the newly assigned probability distribution for the location of the keys is widespread. Thus, counterintuitively, by gaining the knowledge that their keys were outside of their pocket, they have lost information (as the entropy has increased). This leads \cite{book} to define information as that which causes probability distributions to change (independent of whether or not that change leads entropy to increase or decrease).} and rather toward a notion that entropies are objective tools for inference (as was utilized by Jaynes \cite{Jaynes1,Jaynes2,Jaynesbook}).   In this sense, the relative entropies were built for the purpose of saturating their own interpretation \cite{book,Skilling1}\footnote{Although, Skilling only accomplished this for ``intensities" rather than probabilities.}, and, therefore, the quantum relative entropy \emph{is the tool designed for updating density matrices}.


The remainder of the chapter is organized as follows: first, we will discuss some universally applicable principles of inference and motivate the design of an entropy functional able to rank probability distributions. This entropy functional will be designed such that it is consistent with inference by applying a few reasonable design criteria, which are guided by the principle of minimal updating. Using the same principles of inference and design criteria, we find the form of the quantum relative entropy suitable for inference. An example of updating \mbox{2 $\times$ 2} prior density matrices with respect to expectation values over spin matrices that do not commute with the prior, via the quantum maximum entropy method, is given in Appendix \ref{spinexampleQRE}. 

\section{The Design of Entropic Inference}
\label{s2}

Inference is the appropriate updating of probability distributions when new information is received. Bayes rule and Jeffreys rule are both equipped to handle information in the form of data; however, the updating of a probability distribution due to the knowledge of an expectation value was realized by Jaynes \cite{Jaynes1,Jaynesbook,Jaynes2} through the method of maximum entropy. The two methods of inference were thought to be separate until the work of \cite{GiffinBayes,Giffin,GiffinThesis}, which showed Bayes Rule and Jeffreys Rule are the derived forms of inference using the method of maximum entropy when the expectation value constraints are in the form of data. In the spirit of the derivation we will pretend the maximum entropy method is not known and show how it may be derived as an application of inference.

Given a probability distribution $\varphi(x)$ over a general set of propositions $x\in X$, if new information is learned, we are obligated to assign a new probability distribution $\rho(x)$ that somehow reflects this new information while also respecting our prior probability distribution $\varphi(x)$. The main question we must address is: ``Given some information, to what posterior probability distribution $\rho(x)$ should we update our prior probability distribution $\varphi(x)$?'', that is,
\begin{eqnarray}
\varphi(x)\stackrel{*}{\longrightarrow}\rho(x)?\nonumber
\end{eqnarray}
This specifies the problem of inductive inference.~Since ``information'' has many colloquial, and potentially counterintuitive, definitions (as discussed earlier), we remove potential confusion by defining {\bf information} operationally $(*)$ as the \emph{rationale} that causes a probability distribution to change (inspired by and adapted from \cite{book}). As probabilities are ``degrees of rational belief", a probability should only change value if there exists a rationale to do so. Thus, information can take many forms, for instance, learning the truth of a proposition or the value of an expectation value, \emph{but}, obtaining ``new" information also depends on one's current state of knowledge $\varphi$ \cite{book}.

The motivation for designing entropy is to build a function that allows for a systematic search of a preferred
posterior distribution in the presence of new information for the purpose of inductive inference \cite{book}. It is best described in \cite{book}:

\begin{quotation}
\noindent
 ``The central idea, first proposed in \cite{Skilling1},
is disarmingly simple: to select the posterior, first rank all candidate distributions
in increasing \emph{order of preference} and then pick the distribution that ranks
the highest. Irrespective of what it is that makes one distribution preferable over
another (we will get to that soon enough), it is clear that any ranking according
to preference must be transitive: if distribution $\rho_1$ is preferred over distribution
$\rho_2$, and $\rho_2$ is preferred over $\rho_3$, then $\rho_1$ is preferred over $\rho_3$. Such transitive
rankings are implemented by assigning to each $\rho(x)$ a real number $S[\rho]$, which
is called the entropy of $\rho$, in such a way that if $\rho_1$ is preferred over $\rho_2$, then
$S[\rho_1] > S[\rho_2]$. The selected distribution (one or possibly many, for there may
be several equally preferred distributions) is that which maximizes the entropy
functional."
\end{quotation}

Because we wish to update from prior distributions $\varphi$ to posterior distributions $\rho$ by ranking, the~entropy functional $S[\rho,\varphi]$ is a real function of both $\varphi$ and $\rho$.  In the absence of new information, there is no available \emph{rationale} to prefer any $\rho$ to the original $\varphi$, and thereby the relative entropy should be designed such that the selected posterior is equal to the prior $\varphi$ (in the absence of new information). The prior information encoded in $\varphi$ is valuable and we should not change it unless we are informed otherwise. Due to our definition of information, and our desire for objectivity, we state the predominate guiding principle for inductive inference:

\paragraph{The Principle of Minimal Updating (PMU):}
\emph{A probability distribution should only be updated to the extent required by the new information}. \\

This simple statement provides the foundation for inference \cite{book}. If the updating of probability distributions is to be done objectively, then possibilities should not be needlessly ruled out or suppressed. Being informationally stingy, that we should only update probability distributions when the information requires it, pushes inductive inference toward objectivity. Thus, using the PMU helps formulate a pragmatic (and objective) procedure for making inferences using (informationally) subjective probability distributions \cite{Pragmatic}. 

This method of inference is only as universal and general as its ability to apply \emph{equally well} to \emph{any} specific inference problem. The notion of ``specificity'' is the notion of independence; a specific case is only specific in that it is separable from other specific cases. The notion that systems may be ``sufficiently independent'' plays a central and deep-seated role in science and the idea that some things can be neglected and that not everything matters, is implemented by imposing criteria that tells us how to handle statistically independent systems and independent domains within a single system \cite{book}. Ironically, the universally \emph{shared} property by all specific inference problems is their ability to be \emph{independent} of one another---they share independence. Thus, a universal inference scheme based on the PMU permits:

\paragraph{Properties of Independence (PI):} 
\emph{Subdomain Independence: When information is received about one set of propositions, it should not affect or change the state of knowledge (probability distribution) of the other propositions (else information was also received about them too);} 
\begin{eqnarray}
\mbox{\emph{And,}}\nonumber
\end{eqnarray}
\noindent\emph{Subsystem Independence: When two systems are a priori believed to be
independent and we only receive information about one, then the state of knowledge of the other system remains unchanged.}\\

The PIs are special cases of the PMU that describe situations when someone should \emph{not} update pieces of their probability distribution. The PIs are ultimately implemented through \emph{design criteria} in this design derivation of the entropy $S[\rho,\varphi]$. The process of constraining the form of $S[\rho,\varphi]$ by imposing design criteria may be viewed as a process of \emph{eliminative induction} from which a single form for the entropy remains. Thus, the justification behind the surviving entropy is not that it leads to demonstrably correct inferences, but, rather, that all other candidate entropies demonstrably fail to perform as desired~\cite{book}. Rather than the \emph{design criteria} instructing one how to update, they instruct in what instances one should \emph{not} update. That is, rather than justifying one way to skin a cat over another, we tell you when \emph{not} to skin it, which is operationally unique---namely you don't do it---luckily enough for the cat.

\subsection*{The Design Criteria and the Standard Relative Entropy}
The following \emph{design criteria} (DC), guided by the PMU, are imposed and one finds that the standard relative entropy is a tool designed for inductive inference. The form of this presentation is inspired by \cite{book}.

\paragraph{DC1: Subdomain Independence}

 We keep DC1 from \cite{GiffinBayes,CatichaInfoEntropy,book} and follow it below. DC1 imposes the first instance of when one should not update---the Subdomain PI.  Suppose the information to be processed does \emph{not }refer to a
particular subdomain $\mathcal{D}$ of the space $\mathcal{X}$ of $x$'s. In
the absence of new information about $\mathcal{D}$, the PMU insists we do
not change our minds about probabilities that are conditional on $\mathcal{D}
$. Thus, we design the inference method so that $\varphi(x|\mathcal{D})$, the
prior probability of $x$ conditional on $x\in\mathcal{D}$, is not updated and therefore the selected conditional posterior is
\begin{equation}
P(x|\mathcal{D})=\varphi(x|\mathcal{D}).  \label{DC1a}
\end{equation}
(The notation will be as follows: we denote priors by $\varphi$, candidate
posteriors by lower case $\rho$, and the selected posterior by upper case $P$.)  Caticha emphasizes that the point is not that we make the unwarranted assumption that
keeping $\varphi(x|\mathcal{D})$ unchanged is guaranteed to lead to correct
inferences. It need not; induction is risky. The point is, rather, that, in
the absence of any evidence to the contrary, there is no reason to change our
minds and the prior information takes priority. \cite{book}

\paragraph{DC1 Implementation}

Consider the set of microstates $x_i\in \mathcal{X}$ belonging to either of two non-overlapping domains $\mathcal{D}$ or its complement $\mathcal{D}'$, such that $\mathcal{X}=\mathcal{D}\cup \mathcal{D}'$ and $\emptyset=\mathcal{D}\cap \mathcal{D}'$. For convenience, let $\rho(x_i)=\rho_i$.  Consider the following constraints:
\begin{eqnarray}
\rho(\mathcal{D})=\sum_{i\in \mathcal{D}}\rho_i\quad\mbox{and}\quad \rho(\mathcal{D}')=\sum_{i\in \mathcal{D}'}\rho_i,\label{DC1}
\end{eqnarray}
such that $\rho(\mathcal{D})+\rho(\mathcal{D}')=1$ (however this will not be imposed directly with a Lagrange multiplier as it would lead to inter domain dependencies), and the following ``local''  expectation value constraints over $\mathcal{D}$ and $\mathcal{D}'$,
\begin{eqnarray}
\expt{A}=\sum_{i\in \mathcal{D}}\rho_iA_i\quad\mbox{and}\quad \expt{A'}=\sum_{i\in \mathcal{D}'}\rho_iA'_i,\label{DC2}
\end{eqnarray}
where $A=A(x)$ is a scalar function of $x$ and $A_i\equiv A(x_i)$. As we are searching for the candidate distribution which maximizes $S$ while obeying (\ref{DC1}) and (\ref{DC2}), we maximize the entropy $S\equiv S[\rho,\varphi]$ with respect to these expectation value constraints using the Lagrange multiplier method,
\vspace{12pt}
\begin{eqnarray}
0=\delta\Big(S-\lambda[\rho(\mathcal{D})-\sum_{i\in \mathcal{D}}\rho_i]-\mu[\expt{A}-\sum_{i\in \mathcal{D}}\rho_iA_i]\nonumber\\
-\lambda'[\rho(\mathcal{D}')-\sum_{i\in \mathcal{D}'}\rho_i]-\mu'[\expt{A'}-\sum_{i\in \mathcal{D}'}\rho_iA_i]\Big),\nonumber
\end{eqnarray}
 and, thus, the entropy is maximized when the following differential relationships hold:
\begin{eqnarray}
\frac{\delta S}{\delta \rho_i}&=&\lambda+\mu A_i\quad\mbox{$\forall\, i\in \mathcal{D}$},\label{c3}\\
\frac{\delta S}{\delta \rho_i}&=&\lambda'+\mu' A'_i\quad\mbox{$\forall\, i\in \mathcal{D}'$}.\label{c4}
\end{eqnarray}
Equations (\ref{DC1})--(\ref{c4}), are $n+4$ equations we must solve to find the four Lagrange multipliers $\{\lambda,\lambda',\mu,\mu'\}$ and the $n$ probability values $\{\rho_i\}$ associated to the $n$ microstates $\{x_i\}$.

If the subdomain constraint DC1 is imposed in the most restrictive case, then it will hold in general. The most restrictive case requires splitting $\mathcal{X}$ into a set of domains $\{\mathcal{D}_i\}$ such that each $\mathcal{D}_i$ singularly includes one microstate $x_i$. This gives,
\begin{eqnarray}
\frac{\delta S}{\delta \rho_i}=\lambda_i+\mu_iA_i\quad\mbox{in each $\mathcal{D}_i$}.\label{dc1 7}
\end{eqnarray}
Because the entropy $S=S[\rho_1,\rho_2,...;\varphi_1,\varphi_2,...]$ is a functional over the probability of each microstate's posterior and prior distribution, its variational derivative is also a function of said probabilities in~general,
\begin{eqnarray}
\frac{\delta S}{\delta \rho_i}\equiv\phi_i(\rho_1,\rho_2,...;\varphi_1,\varphi_2,...)=\lambda_i+\mu_iA_i\quad\mbox{for each $(i,\mathcal{D}_i)$}.
\end{eqnarray}
DC1 is imposed by constraining the form of $\phi_i(\rho_1,\rho_2,...;\varphi_1,\varphi_2,...)=\phi_i(\rho_i;\varphi_1,\varphi_2,...)$ to ensure that changes in $A_i\rightarrow A_i+\delta A_i$ have no influence over the value of $\rho_j$ in domain $\mathcal{D}_j$, through $\phi_i$, for $i\neq j$. If there is no new information about propositions in $\mathcal{D}_j$, its distribution should remain equal to $\varphi_j$ by the PMU. We further restrict $\phi_i$ such that an arbitrary variation of $\varphi_j\rightarrow \varphi_j+\delta \varphi_j$ (a change in the prior state of knowledge of the microstate $j$) has no effect on $\rho_i$ for $i\neq j$ and therefore DC1 imposes $\phi_i=\phi_i(\rho_i,\varphi_i)$, as is guided by the PMU. At this point, it is easy to generalize the analysis to continuous microstates such that the indices become continuous $i\rightarrow x$, sums become integrals, and discrete probabilities become probability densities $\rho_i\rightarrow \rho(x)$.


\paragraph{Remark}

We are designing the entropy for the purpose of ranking posterior probability distributions (for the purpose of inference); however, the highest ranked distribution is found by setting the variational derivative of $S[\rho,\varphi]$ equal to the variations of the expectation value constraints by the Lagrange multiplier method,
\begin{eqnarray}
\frac{\delta S}{\delta \rho(x)}=\lambda+\sum_i\mu_i A_i(x).
\end{eqnarray}
Therefore, the real quantity of interest is $\frac{\delta S}{\delta \rho(x)}$ rather than the specific form of $S[\rho,\varphi]$. \emph{All} forms of $S[\rho,\varphi]$ that give the correct form of $\frac{\delta S}{\delta \rho(x)}$ are \emph{equally valid} for the purpose of inference. Thus, every design criteria may be made on the variational derivative of the entropy rather than the entropy itself, which we do. When maximizing the entropy, for convenience, we will let,
\begin{eqnarray}
\frac{\delta S}{\delta \rho(x)}\equiv \phi_x(\rho(x),\varphi(x)),
\end{eqnarray}
and further use the shorthand $\phi_x(\rho,\varphi)\equiv\phi_x(\rho(x),\varphi(x))$, in all cases. 

\paragraph{DC1':} \emph{In the absence of new information, our new state of knowledge $\rho(x)$ is equal to the old state of knowledge~$\varphi(x)$.}\\

This is a special case of DC1, which is implemented differently in \cite{book}. The PMU is in principle a statement about informational honestly---that is, one should not ``jump to conclusions'' in light of new information, and in the absence of new information, one should not change their state of knowledge. If no new information is given, the prior probability distribution $\varphi(x)$ does not change, that is, the posterior probability distribution $\rho(x)=\varphi(x)$ is equal to the prior probability. If we maximize the entropy without applying constraints,
\begin{eqnarray}
\frac{\delta S}{\delta \rho(x)}=0,
\end{eqnarray}
then DC1' imposes the following condition on $\phi_x$:
\begin{eqnarray}
\frac{\delta S}{\delta \rho(x)}=\phi_x(\rho,\varphi)=\phi_x(\varphi,\varphi)=0,\label{18}
\end{eqnarray}
for all $x$ in this case. This special case of the DC1 and the PMU turns out to be incredibly constraining as we will see over the course of DC2.

\paragraph{Comment}

If the variable $x$ is continuous, DC1 requires that when information refers to points infinitely close but just outside the
domain $\mathcal{D}$, that it will have no influence on probabilities conditional on $%
\mathcal{D}$. This may seem surprising as it may lead to updated probability
distributions that are discontinuous. Is this a problem? No. \cite{book}

In certain situations (e.g., physics) we might have explicit reasons
to believe that conditions of continuity or differentiability should be
imposed and this information might be given to us in a variety of ways. The
crucial point, however -- and this is a point that we keep and will keep
reiterating -- is that unless such information is explicitly given,
we should not assume it. If the new information leads to discontinuities, so
be it. \cite{book}

\paragraph{DC2: Subsystem Independence}

DC2 imposes the second instance of when one should not update -- the Subsystem PI. We emphasize that DC2 \emph{is not a consistency requirement}. The argument we deploy is \emph{not} that both the prior \emph{and} the new
information tells us the systems are independent, in which case consistency
requires that it should not matter whether the systems are treated jointly
or separately. Rather, DC2 refers to a situation where the new information does not
say whether the systems are independent or not, but information is given about each subsystem. The updating is
being \emph{designed} so that the independence reflected in the prior is
maintained in the posterior by default via the PMU and the second clause of the PIs. \cite{book} 

The point is not that when we have no evidence for
correlations we draw the firm conclusion that the systems must necessarily
be independent. They could indeed have turned out to be correlated and then
our inferences would be wrong. Again, induction involves risk. The point is rather
that if the joint prior reflects independence and the new evidence is
silent on the matter of correlations, then the prior independence takes precedence. As before, in this case subdomain independence, the probability distribution should not be
updated unless the information requires it. \cite{book} \enlargethispage{0.5cm} 

\paragraph{DC2 Implementation}

Consider a composite system, $x=(x_{1},x_{2})\in \mathcal{X}=\mathcal{X}%
_{1}\times \mathcal{X}_{2}$. Assume that all prior evidence led us to
believe the subsystems are independent. This belief is reflected in the prior
distribution: if the individual system priors are $\varphi_{1}(x_{1})$ and $%
\varphi_{2}(x_{2})$, then the prior for the whole system is their product $%
\varphi_{1}(x_{1})\varphi_{2}(x_{2})$. Further suppose that new information is acquired
such that $\varphi_{1}(x_{1})$ would by itself be updated to $P_{1}(x_{1})$ and
that $\varphi_{2}(x_{2})$ would itself be updated to $P_{2}(x_{2})$. By design, the implementation of DC2
constrains the entropy functional such that, in this case, the joint product prior $%
\varphi_{1}(x_{1})\varphi_{2}(x_{2}) $ updates to the selected product posterior $P_{1}(x_{1})P_{2}(x_{2})$ \cite{book}. 

The argument below is considerably simplified if we expand the space of
probabilities to include distributions that are not necessarily normalized. This does
not represent any limitation because a normalization constraint can always be applied. We consider a few special cases below: \\

\noindent \textbf{Case 1:} We receive the extremely constraining information
that the posterior distribution for system $1$ is completely specified to be 
$P_{1}(x_{1})$ while we receive no information at all about system $2$. We
treat the two systems jointly. Maximize the joint entropy $S[\rho(x_1,x_2),\varphi(x_1)\varphi(x_2)]$
subject to the following constraints on the $\rho(x_{1},x_{2})\,$:
\begin{equation}
\int dx_{2}\,\rho(x_{1},x_{2})=P_{1}(x_{1}).\label{margprob1}
\end{equation}%
Notice that the probability of each $x_{1}\in\mathcal{X}_1$ within $\rho(x_1,x_2)$ is being constrained to $P_{1}(x_{1})$ in the marginal. We therefore need one Lagrange multiplier $\lambda _{1}(x_{1})$ for each $x_1\in\mathcal{X}_1$ to tie each value of $\int dx_2 \,\rho(x_1,x_2)$ to $P_1(x_1)$. Maximizing the entropy with respect to this constraint is,
\begin{equation}
\delta \left[ S-\int dx_{1}\lambda _{1}(x_{1})\left( \int
dx_{2}\,\rho(x_{1},x_{2})-P_{1}(x_{1})\right) \right] =0,
\end{equation}%
which requires that
\begin{equation}
\lambda _{1}(x_{1})=\phi _{x_{1}x_{2}}\left( \rho(x_{1},x_{2}),\varphi_{1}(x_{1})\varphi_{2}(x_{2})\right),
\end{equation}%
for arbitrary variations of $\rho(x_1,x_2)$. By design, DC2 is implemented by requiring $\varphi_{1}\varphi_{2}\rightarrow P_{1}\varphi_{2}$ in this case, therefore,
\begin{equation}
\lambda _{1}(x_{1})=\phi _{x_{1}x_{2}}\left(
P_{1}(x_{1})\varphi_{2}(x_{2}),\varphi_{1}(x_{1})\varphi_{2}(x_{2})\right).
\label{lambda1 a}
\end{equation}%
This equation must hold for all choices of $%
x_{2}$ and all choices of the prior $\varphi_{2}(x_{2})$ as $\lambda _{1}(x_{1})$ is independent of $x_{2}$. Suppose we had chosen a different prior $\varphi_{2}^{\prime }(x_{2})=\varphi_{2}(x_{2})+\delta \varphi_{2}(x_{2})$ that
disagrees with $\varphi_{2}(x_{2})$. For all $x_2$ and $\delta \varphi_{2}(x_{2})$, the multiplier $\lambda _{1}(x_{1})$ remains unchanged as it constrains the independent $\rho(x_1)\rightarrow P_1(x_1)$. This means that any dependence that the right-hand side might potentially have had on $x_{2}$ and on the
prior $\varphi_{2}(x_{2})$ \emph{must cancel out}. This means that
\begin{equation}
\phi _{x_{1}x_{2}}\left(P_{1}(x_{1})\varphi_{2}(x_{2}),\varphi_{1}(x_{1})\varphi_{2}(x_{2})\right)=f_{x_1}(P_{1}(x_{1}),\varphi_{1}(x_{1})).
\end{equation}
Since any value of $\varphi_{2}$ gives the same $f_{x_1}$,
we may choose a convenient constant prior set equal to one, $\varphi_{2}(x_{2})=1$, therefore
\begin{equation}
f_{x_1}(P_{1}(x_{1}),\varphi_{1}(x_{1}))=\phi _{x_{1}x_2}\left(
P_{1}(x_{1})\cdot 1,\varphi_{1}(x_{1})\cdot 1\right)=\phi _{x_{1}}\left(
P_{1}(x_{1}),\varphi_{1}(x_{1})\right)
\end{equation}
 in general. This gives
\begin{equation}
\lambda _{1}(x_{1})=\phi _{x_{1}}\left(
P_{1}(x_{1}),\varphi_{1}(x_{1})\right) .  \label{lambda1 b}
\end{equation}%
The left-hand side above does not depend on $x_{2}$, and therefore neither does
the right-hand side. An argument exchanging systems $1$ and $2$ gives a similar result.

\noindent \textbf{Case 1---Conclusion:} When system 2 is not updated the dependence on $\varphi_{2}$ and $x_2$
drops out,     
\begin{equation}
\phi_{x_1x_2} \left( P_{1}(x_1)\varphi_{2}(x_2),\varphi_{1}(x_1)\varphi_{2}(x_2)\right) =\phi_{x_1} \left( P_{1}(x_1),\varphi_{1}(x_1)\right),
\label{lambda1 c}
\end{equation}
and vice-versa when system 1 is not updated,
\begin{equation}
\phi_{x_1x_2} \left( \varphi_{1}(x_1)P_{2}(x_2),\varphi_{1}(x_1)\varphi_{2}(x_2)\right) =\phi_{x_2} \left( P_{2}(x_2),\varphi_{2}(x_2)\right) .\label{lambda1 d}
\end{equation}
As we seek the general functional form of $\phi_{x_1x_2}$, and because the $x_2$ dependence drops out of (\ref{lambda1 c}) and the $x_1$ dependence drops out of (\ref{lambda1 d}) for arbitrary $\varphi_1,\varphi_2$ and $\varphi_{12}=\varphi_1\varphi_2$, the explicit coordinate dependence in $\phi$ consequently drops out of both such that,
\begin{eqnarray}
\phi_{x_1x_2}\rightarrow \phi.
\end{eqnarray}
Thus, $\phi=\phi(\rho(x),\varphi(x))$ must only depend on coordinates through the probability distributions themselves.

\noindent \textbf{Case 2:} Now consider a different special case in which
the marginal posterior distributions for systems $1$ and $2$ are both
completely specified to be $P_{1}(x_{1})$ and $P_{2}(x_{2})$, respectively.
Maximize the joint entropy $S[\rho(x_1,x_2),\varphi(x_1)\varphi(x_2)]$ subject to the following constraints on the $\rho(x_{1},x_{2})\,$, 
\begin{equation}
\int dx_{2}\,\rho(x_{1},x_{2})=P_{1}(x_{1})\quad \text{and}\quad \int
dx_{1}\,\rho(x_{1},x_{2})=P_{2}(x_{2})~.\label{margprob2}
\end{equation}%
Again, this is one constraint for each value of $x_{1}$ and one constraint for each value of $x_{2}$, which, therefore, require the separate Lagrange multipliers $\mu _{1}(x_{1})$ and $\mu
_{2}(x_{2})$. Maximizing $S$ with respect to these constraints is~then, 
\begin{eqnarray}
0 &=&\delta \left[ S-\int dx_{1}\mu _{1}(x_{1})\left( \int
dx_{2}\,\rho(x_{1},x_{2})-P_{1}(x_{1})\right) \right.   \nonumber \\
&&-\left. \int dx_{2}\mu _{2}(x_{2})\left( \int
dx_{1}\,\rho(x_{1},x_{2})-P_{2}(x_{2})\right) \right] \,,
\end{eqnarray}%
leading to

\begin{equation}
\mu
_{1}(x_{1})+\mu _{2}(x_{2})=\phi \left( \rho(x_{1},x_{2}),\varphi_{1}(x_{1})\varphi_{2}(x_{2})\right).
\end{equation}%
The updating is being designed so that $\varphi_{1}\varphi_{2}\rightarrow P_{1}P_{2}$, as the independent subsystems are being updated based on expectation values that are silent about correlations. DC2 thus imposes,
\begin{equation}
\mu _{1}(x_{1})+\mu _{2}(x_{2})=\phi \left(
P_{1}(x_{1})P_{2}(x_{2}),\varphi_{1}(x_{1})\varphi_{2}(x_{2})\right) .  \label{mu12 a}
\end{equation}%
Write (\ref{mu12 a}) as,
\begin{equation}
\mu _{1}(x_{1})=\phi \left(
P_{1}(x_{1})P_{2}(x_{2}),\varphi_{1}(x_{1})\varphi_{2}(x_{2})\right)-\mu _{2}(x_{2}) .  \label{mu12 b}
\end{equation}%
The left-hand side is independent of $x_{2}$ so we can perform a technique
similar to before. Suppose we had chosen a different \emph{%
constraint} $P_{2}^{\prime }(x_{2})$ that differs from $P_{2}(x_{2})$ and a new prior $\varphi'_2(x_{2})$ that differs from $\varphi_2(x_{2})$ except at the point $\bar{x}_2$. At the value $\bar{x}_2$, the multiplier $\mu
_{1}(x_{1})$ remains unchanged for all $P_{2}^{\prime }(x_{2})$, $\varphi'_2(x_{2})$, and thus $x_2$. This means that any dependence that the right-hand side might potentially have had on $x_{2}$ and on the choice of $P_{2}'(x_{2})$, $\varphi'_2(x_{2})$ must cancel out, leaving $\mu _{1}(x_{1})$ unchanged. That is, the Lagrange multiplier $\mu(x_2)$ cancels out these dependences such that 
\begin{equation}
\phi \left(
P_{1}(x_{1})P_{2}(x_{2}),\varphi_{1}(x_{1})\varphi_{2}(x_{2})\right)-\mu _{2}(x_{2})=g(P_{1}(x_{1}),\varphi_{1}(x_{1})).
\end{equation}
Because $g(P_{1}(x_{1}),\varphi_{1}(x_{1}))$ is independent of arbitrary variations of $P_{2}(x_{2})$ and $\varphi_{2}(x_{2})$ on the left hand side (LHS) above---it is satisfied equally well for all choices.  The form of $g=\phi(P_{1}(x_{1}),q_{1}(x_{1}))$ is apparent if $P_{2}(x_{2})=\varphi_{2}(x_{2})=1$ as $\mu_2(x_2)=0$ similar to Case 1 as well as DC1'. Therefore, the Lagrange multiplier is 
\begin{equation}
\mu _{1}(x_{1})=\phi\left(
P_{1}(x_{1}),\varphi_{1}(x_{1})\right) .
\end{equation}%
A similar analysis carried out for $\mu _{2}(x_{2})$ leads to  
\begin{equation}
\mu _{2}(x_{2})=\phi \left( P_{2}(x_{2}),\varphi_{2}(x_{2})\right) .
\end{equation}
\noindent \textbf{Case 2---Conclusion:}
Substituting back into (\ref{mu12 a}) gives us a functional equation for $%
\phi \,$, 
\begin{equation}
\phi \left( P_{1}P_{2},\varphi_{1}\varphi_{2}\right) =\phi \left( P_{1},\varphi_{1}\right)
+\phi \left( P_{2},\varphi_{2}\right) .\label{funcequationprob}
\end{equation}%
The general solution for this functional equation is derived in the Appendix \ref{appendix1}, and is
\begin{eqnarray}
\phi(\rho,\varphi)=a_1\ln(\rho(x))+a_2\ln(\varphi(x)),
\end{eqnarray}
where $a_1,a_2$ are constants. The constants are fixed by using DC1'. Letting $\rho_1(x_1)=\varphi_1(x_1)=\varphi_1$ gives $\phi(\varphi,\varphi)=0$ by DC1', and, therefore,
\begin{eqnarray}
\phi(\varphi,\varphi)=(a_1+a_2)\ln(\varphi)=0,
\end{eqnarray}
so we are forced to conclude $a_1=-a_2$ for arbitrary $\varphi$.
Letting $a_1\equiv A=-|A|$ such that we are really maximizing the entropy (although this is purely convention) gives the general form of $\phi$ to be
\begin{eqnarray}
\phi(\rho,\varphi)=-|A|\ln\Big(\frac{\rho(x)}{\varphi(x)}\Big).
\end{eqnarray}
As long as $A\neq0$, the value of $A$ is arbitrary. The general form of the entropy designed for the purpose of inference of $\rho$ is found by integrating $\phi$, and, therefore,
\begin{eqnarray}
S[\rho(x),\varphi(x)]=-|A|\int dx\, \Big(\rho(x)\ln\Big(\frac{\rho(x)}{\varphi(x)}\Big)-\rho(x)\Big)+C[\varphi].\label{theentropy}
\end{eqnarray}
The constant in $\rho$, $C[\varphi]$, will always drop out when varying $\rho$.
The apparent extra term ($|A|\int\rho(x)dx$) from integration cannot be dropped while simultaneously satisfying DC1', which requires $\rho(x)=\varphi(x)$ in the absence of constraints or when there is no change to one's information. In previous versions where the integration term ($|A|\int\rho(x)dx$) is dropped, one obtains solutions like $\rho(x)=e^{-1}\varphi(x)$ (independent of whether $\varphi(x)$ was previously normalized or not) in the absence of new information. Obviously, this factor can be taken care of by normalization, and, in this way, both forms of the entropy are equally valid; however, this form of the entropy better adheres to the PMU through DC1' when normalization is not applied. Given that we may regularly impose normalization, we may drop the extra $\int \rho(x) dx$ term and $C[\varphi]$. For convenience then, (\ref{theentropy}) becomes 
\begin{eqnarray}
S[\rho(x),\varphi(x)]\rightarrow S^*[\rho(x),\varphi(x)]=-|A|\int dx\, \rho(x)\ln\Big(\frac{\rho(x)}{\varphi(x)}\Big).
\end{eqnarray}
Given that normalization is applied regularly, the same selected posterior $\rho(x)$ maximizes both $S[\rho(x),\varphi(x)]$ and $S^*[\rho(x),\varphi(x)]$, so in future use, the star notation will be dropped.

\paragraph{Remarks}
It can be seen that the relative entropy is invariant under coordinate transformations. This implies that a system of coordinates carry no information, which is to say that changing coordinates is simply a change in the \emph{label} of the proposition, while the set of propositions ultimately stay the same \cite{book}.

The general solution to the maximum entropy procedure with respect to $N$ linear constraints in $\rho$, $\expt{A_i(x)}$, and normalization gives a canonical-like selected posterior probability distribution,
\begin{eqnarray}
\rho(x)=\frac{\varphi(x)}{Z}\exp\Big(\sum_i\alpha_iA_i(x)\Big).\label{gensolrho}
\end{eqnarray}
 The positive constant $|A|$ may always be absorbed into the Lagrange multipliers so we may let it equal unity without loss of generality. DC1' is fully realized when we maximize with respect to a set of constraints on $\rho(x)$ that are already satisfied by $\varphi(x)$. If all of the constraints are already held by $\varphi(x)$, their corresponding Lagrange multipliers are forcibly zero (as can be seen in (\ref{gensolrho}) using (\ref{theentropy})), in agreement with Jaynes, as the expectation values $\expt{A}=\expt{A}(\alpha)$ are monotonic in their Lagrange multipliers and are satisfied when set equal to zero. This gives the expected result $\rho(x)=\varphi(x)$ as there is no new information. Our design has arrived at a refined maximum entropy method \cite{Jaynes1}, whose universality can be seen by following \cite{GiffinBayes,Giffin}. We will follow \cite{GiffinBayes,Giffin} using density matrices in the next chapter.

\section{The Design of the Quantum Relative Entropy}
 In the last section, we assumed that the universe of discourse (the set of relevant propositions or microstates) $\mathcal{X}=\mathcal{A}\times\mathcal{B}\times...$ was known. In quantum physics, things are a bit more ambiguous because many probability distributions, or many experiments, can be associated with a given density matrix. As any probability distribution from a given density matrix, $\rho(a)=\mbox{Tr}(\ket{a}\bra{a}\hat{\rho})=\sum_n\braket{n}{a}\bra{a}\hat{\rho}\ket{n}$,\footnote{$\mbox{Tr}(...)$ is the trace.} may be ranked using the standard relative entropy, it is unclear why we would choose one universe of discourse over another. In lieu of this, so that one universe of discourse is not given preferential treatment, we consider ranking entire density matrices against one another. Probability distributions of interest may be found from the selected posterior density matrix. This moves our universe of discourse from sets of propositions $\mathcal{X}\rightarrow\mathcal{H}$ to Hilbert~ space(s).

When the objects of study are quantum systems, we desire an objective procedure to update from a prior density matrix $\hat{\varphi}$ to a posterior density matrix $\hat{\rho}$. We will apply the same intuition used for ranking probability distributions (Section \ref{s2}) and implement the PMU, PI, and design criteria to the ranking of density matrices. 



\subsection{Designing the Quantum Relative Entropy}
In this section, we design the quantum relative entropy using the same inferentially guided \emph{design criteria} as were used in the standard relative entropy. 
\paragraph{DC1: Subdomain Independence}

The goal is to design a function $S(\hat{\rho},\hat{\varphi})$ that is able to rank density matrices. This insists that $S(\hat{\rho},\hat{\varphi})$ be a real scalar valued function of the posterior $\hat{\rho}$, and prior $\hat{\varphi}$ density matrices, which we will call the quantum relative entropy or simply the entropy in this section. An arbitrary variation of the entropy with respect to $\hat{\rho}$ is,
\begin{eqnarray}
\delta S(\hat{\rho},\hat{\varphi})&=&\sum_{ij}\frac{\delta S(\hat{\rho},\hat{\varphi})}{\delta \rho_{ij}}\delta\rho_{ij}=\sum_{ij}\Big(\frac{\delta S(\hat{\rho},\hat{\varphi})}{\delta \hat{\rho}}\Big)_{ij}\delta(\hat{\rho})_{ij}=\sum_{ij}\Big(\frac{\delta S(\hat{\rho},\hat{\varphi})}{\delta \hat{\rho}^T}\Big)_{ji}\delta(\hat{\rho})_{ij}\nonumber\\
&=&\mbox{Tr}\Big(\frac{\delta S(\hat{\rho},\hat{\varphi})}{\delta \hat{\rho}^T}\delta\hat{\rho}\Big),
\end{eqnarray}
where $\hat{\rho}^T$ is the transpose of $\hat{\rho}$. We wish to maximize this entropy with respect to expectation value constraints, such as $\expt{A}=\mbox{Tr}(\hat{A}\hat{\rho})$ on $\hat{\rho}$. Using the Lagrange multiplier method to maximize the entropy with respect to $\expt{A}$ and normalization, is setting the variation equal to zero,
\begin{eqnarray}
\delta\Big(S(\hat{\rho},\hat{\varphi})-\lambda[\mbox{Tr}(\hat{\rho})-1]-\alpha[\mbox{Tr}(\hat{A}\hat{\rho})-\expt{A}]\Big)=0,
\end{eqnarray}
 where $\lambda$ and $\alpha$ are the Lagrange multipliers for the respective constraints. Because $S(\hat{\rho},\hat{\varphi})$ is a real number, we require $\delta S$ to be real, and because $\hat{\rho}$ is Hermitian, $\delta\hat{\rho}$ and $\frac{\delta S(\hat{\rho},\hat{\varphi})}{\delta \hat{\rho}^T}$ are Hermitian. Furthermore, because expectation values are real, $\hat{A}$'s are Hermitian and their corresponding Lagrange multipliers are real too. Arbitrary variations of $\hat{\rho}$ give,
\begin{eqnarray}
\mbox{Tr}\Big(\Big(\frac{\delta S(\hat{\rho},\hat{\varphi})}{\delta \hat{\rho}^T}-\lambda\hat{1}-\alpha\hat{A}\Big)\delta\hat{\rho}\Big)=0,\label{2.39}
\end{eqnarray}
where $\hat{1}$ is the identity matrix. For these arbitrary variations, the variational derivative of $S$ satisfies,
\begin{eqnarray}
\frac{\delta S(\hat{\rho},\hat{\varphi})}{\delta \hat{\rho}^T}=\lambda\hat{1}+\alpha\hat{A}\label{52},
\end{eqnarray}
at the maximum; however, forcing $\delta\hat{\rho}$ to be Hermitian gives the same result (after a bit more algebra) as is demonstrated in Appendix \ref{deltarhos}. As in the remark earlier, \emph{all} forms of $S$ that give the correct form of $\frac{\delta S(\hat{\rho},\hat{\varphi})}{\delta \hat{\rho}^T}$ under variation are \emph{equally valid} for the purpose of inference. For notational convenience, we let
\begin{eqnarray}
\frac{\delta S(\hat{\rho},\hat{\varphi})}{\delta \hat{\rho}^T}\equiv\phi(\hat{\rho},\hat{\varphi}),
\end{eqnarray}
which is a matrix valued function of the posterior and prior density matrices. The form of $\phi(\hat{\rho},\hat{\varphi})$ is already ``local'' in $\hat{\rho}$ (the variational derivative is with respect to the whole density matrix and the RHS is a function of the density matrices), so we don't need to constrain it further as we did in the original DC1. 

\paragraph{DC1':}  \emph{In the absence of new information, the new state $\hat{\rho}$ is equal to the old state $\hat{\varphi}$.}\\

 Applied to the ranking of density matrices, in the absence of new information, the density matrix $\hat{\varphi}$ should not change, that is, the posterior density matrix $\hat{\rho}=\hat{\varphi}$ is equal to the prior density matrix. Maximizing the entropy without applying any constraints gives,
\begin{eqnarray}
\frac{\delta S(\hat{\rho},\hat{\varphi})}{\delta \hat{\rho}^T}=\hat{0},
\end{eqnarray}
and, therefore, DC1' imposes the following condition in this case:
\begin{eqnarray}
\frac{\delta S(\hat{\rho},\hat{\varphi})}{\delta \hat{\rho}^T}=\phi(\hat{\rho},\hat{\varphi})=\phi(\hat{\varphi},\hat{\varphi})=\hat{0}.
\end{eqnarray}

As in the original DC1', if $\hat{\varphi}$ is known to obey some expectation value $\expt{\hat{A}}$, and then if one goes out of their way to constrain $\hat{\rho}$ to that expectation value and nothing else, it follows from the PMU that $\hat{\rho}=\hat{\varphi}$, as no information has been gained. This is not imposed directly but can be verified later due to the monotonocity of expectation values in their corresponding Lagrange multipliers. 

\paragraph{DC2: Subsystem Independence} 

The discussion of DC2 is the same as the standard relative entropy DC2 -- it is not a consistency requirement, and the updating is \emph{designed} so that the independence reflected in the prior is maintained in the posterior by default via the PMU when the information provided is silent about correlations.

\paragraph{DC2 Implementation} 

Consider a composite system living in the Hilbert space $\mathcal{H}=\mathcal{H}%
_{1}\otimes \mathcal{H}_{2}$. Assume that all prior evidence led us to
believe the systems were independent. This is reflected in the prior
density matrix: if the individual system priors are $\hat{\varphi}_{1}$ and $%
\hat{\varphi}_{2}$, then the joint prior for the whole system is $%
\hat{\varphi}_{1}\otimes\hat{\varphi}_{2}$. Further suppose that new information is acquired
such that $\hat{\varphi}_{1}$ would itself be updated to $\hat{\rho}_1$ and
that $\hat{\varphi}_{2}$ would be itself be updated to $\hat{\rho}_{2}$. By design, the implementation of DC2
constrains the entropy functional such that in this case, the joint product prior density matrix $%
\hat{\varphi}_{1}\otimes\hat{\varphi}_{2} $ updates to the product posterior $\hat{\rho}_{1}\otimes\hat{\rho}_{2} $
so that inferences about one do not affect inferences about the other.

The argument below is considerably simplified if we expand the space of
density matrices to include density matrices that are not necessarily normalized. This does
not represent any limitation because normalization can always be imposed as one additional constraint. We consider a few special cases below: \\

\noindent \textbf{Case 1:} We receive the extremely constraining information
that the posterior distribution for system $1$ is completely specified to be 
$\hat{\rho}_1$ while we receive no information about system $2$. We
treat the two systems jointly. Maximize the joint entropy $S(\hat{\rho}_{12},\hat{\varphi}_{1}\otimes\hat{\varphi}_{2})$, subject to the following constraints on the~$\hat{\rho}_{12}\,$, 
\begin{equation}
\mbox{Tr}_2(\hat{\rho}_{12})=\hat{\rho}_{1},
\end{equation}%
where $\mbox{Tr}_i(...)$ is the partial trace over the vectors in $\mathcal{H}_i$. Notice that all of the $N^2$ elements in $\mathcal{H}_1$ of $\hat{\rho}_{12}$ are being constrained. We therefore need a Lagrange multiplier which spans $\mathcal{H}_1$ and has $N^2$ components -- it is a square Lagrange multiplier matrix $\hat{\lambda}_1$. This is readily seen by observing the component form expressions of the Lagrange multipliers $(\hat{\lambda}_1)_{ij}=\lambda_{ij}$. Maximizing the entropy with respect to this $\mathcal{H}_2$ independent constraint is
\begin{eqnarray}
0=\delta\Big(S-\sum_{ij}\lambda_{ij}\Big(\mbox{Tr}_2(\hat{\rho}_{12}) -\hat{\rho}_{1}\Big)_{ij}\Big),
\end{eqnarray}
but reexpressing this with its transpose $(\hat{\lambda}_1)_{ij}=(\hat{\lambda}_1^T)_{ji}$, gives
\begin{eqnarray}
0=\delta\Big(S-\mbox{Tr}_1(\hat{\lambda}_1[\mbox{Tr}_2(\hat{\rho}_{12}) -\hat{\rho}_{1}])\Big),
\end{eqnarray}
where we have relabeled $\hat{\lambda}_1^T\rightarrow \hat{\lambda}_1$, for convenience, as the name of the Lagrange multipliers are arbitrary. This variation is,
\begin{eqnarray}
0=\mbox{Tr}\Big(\Big(\frac{\delta S}{\delta\hat{\rho}_{12}^T}-\hat{\lambda}_1\otimes\hat{1}_2\Big)\delta\hat{\rho}_{12}\Big),
\end{eqnarray}
where $\otimes$ denotes, equally well, the tensor, direct, or Kronecker product. Similar to Appendix \ref{deltarhos}, but without assuming $\hat{\lambda}$ is Hermitian, one finds,
\begin{equation}
\hat{\lambda}_1\otimes \hat{1}_2=\phi \left( \hat{\rho}_{12},\hat{\varphi}_{1}\otimes\hat{\varphi}_{2}\right),\,
\end{equation}%
and therefore one concludes $\hat{\lambda}$ must be Hermitian because the RHS is Hermitian. DC2 is implemented by requiring $\hat{\varphi}_{1}\otimes\hat{\varphi}_{2}\rightarrow \hat{\rho}_{1}\otimes\hat{\varphi}_{2}$, such that the function $\phi$ is designed to reflect subsystem independence in this case; therefore, we have
\begin{equation}
\hat{\lambda}_1\otimes \hat{1}_2=\phi \left( \hat{\rho}_{1}\otimes\hat{\varphi}_{2},\hat{\varphi}_{1}\otimes\hat{\varphi}_{2}\right).\label{qlambda1 a}
\end{equation}
Had we chosen a different  prior $\hat{\varphi}_{2}^{\prime }=\hat{\varphi}_{2}+\delta \hat{\varphi}_{2}$, for all $\delta \hat{\varphi}_{2}$ the LHS $\hat{\lambda}_1\otimes \hat{1}_2$ remains unchanged given that $\phi$ is independent of scalar functions of $\hat{\varphi}_2$, e.g., $h(\hat{\varphi}_2)=\sum_{ij}\varphi_{ij}$. These scalar functions could be simultaniously lumped into $\hat{\lambda}_1$ and $\phi$ and keep $\hat{\rho}_{1}$ fixed. The potential dependence on scalar functions of $\hat{\varphi}_2$ can be removed by imposing DC2 in a subsystem independent situation where $\hat{\rho}_{1}'$ in $\phi$ need not be fixed under variations of $\hat{\varphi}_2$, and then use DC2 to impose that it is fixed.\footnote{The resulting equation from such a situation, when for instance maximizing the entropy of an independent joint prior while imposing a constraint in $\mathcal{H}_1$, $\mbox{Tr}(\hat{A}_1\otimes\hat{1}_2\cdot\hat{\rho}_{12})=\expt{A_1}$, facilitated by a scalar Lagrange multiplier $\lambda$, is,  
\begin{equation}
\lambda\hat{A}_1\otimes \hat{1}_2=\phi \left( \hat{\rho}_{1}'\otimes\hat{\varphi}_{2},\hat{\varphi}_{1}\otimes\hat{\varphi}_{2}\right),\nonumber
\end{equation}
after imposing DC2. For subsystem independence to be imposed here, $\hat{\rho}_{1}'$ must be independent of variations in $\hat{\varphi}_2$, and, therefore, in a general subsystem independent case, $\phi$ is independent of scalar functions of $\hat{\varphi}_2$.} This means that any dependence that the right-hand side of (\ref{qlambda1 a}) might potentially have had on $\hat{\varphi}_{2}$ \emph{must drop out}, meaning,
\begin{equation}
\phi \left( \hat{\rho}_{1}\otimes\hat{\varphi}_{2},\hat{\varphi}_{1}\otimes\hat{\varphi}_{2}\right)=f( \hat{\rho}_{1}, \hat{\varphi}_{1})\otimes\hat{1}_2.\label{1234}
\end{equation}
Since $\hat{\varphi}_{2}$ is arbitrary, suppose further that we choose a unit prior, $%
\hat{\varphi}_{2}=\hat{1}_2\,$, and note that $\hat{\rho}_{1}\otimes\hat{1}_2$, $\hat{\varphi}_{1}\otimes\hat{1}_2$, and $f( \hat{\rho}_{1}, \hat{\varphi}_{1})\otimes\hat{1}_2$ are block diagonal in $\mathcal{H}_2$.\footnote{ By being ``block diagonal in $\mathcal{H}_2$" we mean that:
\begin{eqnarray} \hat{A}_1\otimes\hat{1}_2=\left( \begin{array}{ccc}
A_{11}\hat{1}_2& \dots & A_{1N}\hat{1}_2\\
  \vdots&\ddots&\vdots\\
A_{N1}\hat{1}_2& \dots&A_{NN}\hat{1}_2
\end{array} \right)_1=\left( \begin{array}{ccc}
\hat{A}_1& \dots & 0\\
  \vdots&\hat{A}_1&\vdots\\
0& \dots&\hat{A}_1
\end{array} \right)_2.
\end{eqnarray}
}
This gives
\begin{eqnarray}
f( \hat{\rho}_{1}, \hat{\varphi}_{1})\otimes\hat{1}_2=\phi \left( \hat{\rho}_{1}\otimes\hat{1}_{2},\hat{\varphi}_{1}\otimes\hat{1}_{2}\right).
\end{eqnarray}
 Because the LHS of the above equation is block diagonal in $\mathcal{H}_2$, the RHS is block diagonal in $\mathcal{H}_2$ and, because the function $\phi$ is understood to be a power series expansion in its arguments, 
\begin{eqnarray}
f( \hat{\rho}_{1}, \hat{\varphi}_{1})\otimes\hat{1}_2=\phi \left( \hat{\rho}_{1}\otimes\hat{1}_{2},\hat{\varphi}_{1}\otimes\hat{1}_{2}\right)=\phi \left( \hat{\rho}_{1},\hat{\varphi}_{1}\right)\otimes\hat{1}_{2}.
\end{eqnarray}
This gives
\begin{equation}
\hat{\lambda}_1\otimes \hat{1}_2=\phi \left( \hat{\rho}_{1},\hat{\varphi}_{1}\right)\otimes\hat{1}_{2},  \label{qlambda1 b}
\end{equation}
and, therefore, the $\hat{1}_2$ factors out and $\hat{\lambda}_1=\phi \left( \hat{\rho}_{1},\hat{\varphi}_{1}\right)$. %
A similar argument exchanging systems $1$ and $2$ shows  $\hat{\lambda}_2=\phi \left( \hat{\rho}_{2},\hat{\varphi}_{2}\right)$.

\noindent \textbf{Case 1---Conclusion:} The analysis leads us to
conclude that when the system 2 is not updated, the dependence on $\hat{\varphi}_{2}$ drops out,    
\begin{equation}
\phi \left( \hat{\rho}_{1}\otimes\hat{\varphi}_{2},\hat{\varphi}_{1}\otimes\hat{\varphi}_{2}\right)=\phi \left( \hat{\rho}_{1},\hat{\varphi}_{1}\right)\otimes\hat{1}_2\label{qlambda1 c},
\end{equation} 
and, similarly,
\begin{equation}
\phi \left( \hat{\varphi}_{1}\otimes\hat{\rho}_{2},\hat{\varphi}_{1}\otimes\hat{\varphi}_{2}\right)=\hat{1}_1\otimes\phi \left( \hat{\rho}_{2},\hat{\varphi}_{2}\right)\label{qlambda1 d}.
\end{equation} \\

\noindent \textbf{Case 2:} Now consider a different special case in which
the marginal posterior density matrices for systems $1$ and $2$ are both
completely specified to be $\hat{\rho}_1$ and $\hat{\rho}_2$, respectively.
Maximize the joint entropy, $S(\hat{\rho}_{12},\hat{\varphi}_{1}\otimes\hat{\varphi}_{2})$,
subject to the following constraints on the $\hat{\rho}_{12}\,$, 
\begin{equation}
\mbox{Tr}_2(\hat{\rho}_{12})=\hat{\rho}_{1}\quad \text{and}\quad \mbox{Tr}_1(\hat{\rho}_{12})=\hat{\rho}_{2}.
\end{equation}%
 Here, each expectation value constrains the entire space $\mathcal{H}_i$, where $\hat{\rho}_{i}$ lives. The Lagrange multipliers must span their respective spaces, so we implement the constraint with the Lagrange multiplier operators $\hat{\mu}_i$, and,
\begin{eqnarray}
0=\delta\Big(S-\mbox{Tr}_1(\hat{\mu}_1[\mbox{Tr}_2(\hat{\rho}_{12}) -\hat{\rho}_{1}])-\mbox{Tr}_2(\hat{\mu}_2[\mbox{Tr}_1(\hat{\rho}_{12}) -\hat{\rho}_{2}])\Big).
\end{eqnarray}
For arbitrary variations of $\hat{\rho}_{12}$, we have
\begin{equation}
\hat{\mu}_1\otimes \hat{1}_2+\hat{1}_1\otimes \hat{\mu}_2=\phi \left( \hat{\rho}_{12},\hat{\varphi}_{1}\otimes\hat{\varphi}_{2}\right).\,
\end{equation}%
By design, DC2 is implemented by requiring $\hat{\varphi}_{1}\otimes\hat{\varphi}_{2}\rightarrow \hat{\rho}_{1}\otimes\hat{\rho}_{2}$ in this case; therefore, we have
\begin{equation}
\hat{\mu}_1\otimes \hat{1}_2+\hat{1}_1\otimes \hat{\mu}_2=\phi \left( \hat{\rho}_{1}\otimes\hat{\rho}_{2},\hat{\varphi}_{1}\otimes\hat{\varphi}_{2}\right).\,\label{qmu12 a}
\end{equation}
 Write (\ref{qmu12 a}) as 
\begin{equation}
\hat{\mu}_1\otimes \hat{1}_2=\phi \left( \hat{\rho}_{1}\otimes\hat{\rho}_{2},\hat{\varphi}_{1}\otimes\hat{\varphi}_{2}\right)-\hat{1}_1\otimes \hat{\mu}_2 ~.  \label{1mu12 b}
\end{equation}%
The LHS is independent of changes that might occur in $\mathcal{H}_2$ on the RHS of (\ref{1mu12 b}). This means that any variation of $\hat{\rho}_{2}$ and $\hat{\varphi}_{2}$ must be canceled out by $\hat{\mu}_2$ -- it removes the dependence of $\hat{\rho}_{2}$ and $\hat{\varphi}_{2}$ in $\phi$.  Therefore, any dependence that the RHS might potentially have had on $\hat{\rho}_{2}$, $\hat{\varphi}_{2}$ must drop out in a general subsystem independent case, leaving $\hat{\mu}_{1}$ unchanged. Consequently,
\vspace{12pt}
\begin{equation}
\phi \left(
\hat{\rho}_1\otimes \hat{\rho}_2,\hat{\varphi}_{1}\otimes \hat{\varphi}_{2}\right)-\hat{1}_1\otimes\hat{\mu}_{2}=g(\hat{\rho}_{1},\hat{\varphi}_{1})\otimes \hat{1}_2.
\end{equation}
Because $g(\hat{\rho}_{1},\hat{\varphi}_{1})$ is independent of arbitrary variations of $\hat{\rho}_{2}$ and $\hat{\varphi}_{2}$ on the LHS above---it is satisfied equally well for all choices.  The form of $g(\hat{\rho}_{1},\hat{\varphi}_{1})$ reduces to the form of $f(\hat{\rho}_{1},\hat{\varphi}_{1})$ from Case 1 when $\hat{\rho}_{2}=\hat{\varphi}_{2}=\hat{1}_2$ and, similarly, DC1' gives $\hat{\mu}_2=0$. Therefore, the Lagrange multiplier is 
\begin{equation}
\hat{\mu}_1\otimes \hat{1}_2=\phi(\hat{\rho}_{1},\hat{\varphi}_{1})\otimes \hat{1}_2.
\end{equation}%
A similar analysis is carried out for $\hat{\mu}_{2}$ leading to  
\begin{equation}
\hat{1}_1\otimes \hat{\mu}_2=\hat{1}_1\otimes \phi(\hat{\rho}_{2},\hat{\varphi}_{2}).
\end{equation}
\noindent \textbf{Case 2---Conclusion:}
Substituting back into (\ref{qmu12 a}) gives us a functional equation for $%
\phi \,$,     
\begin{eqnarray}
\phi(\hat{\rho}_{1}\otimes\hat{\rho}_2,\hat{\varphi}_1\otimes\hat{\varphi}_2)=\phi(\hat{\rho}_{1},\hat{\varphi}_1)\otimes\hat{1}_2+\hat{1}_1\otimes\phi(\hat{\rho}_2,\hat{\varphi}_2),\label{77}
\end{eqnarray}
which is
\begin{eqnarray}
\phi(\hat{\rho}_{1}\otimes\hat{\rho}_2,\hat{\varphi}_1\otimes\hat{\varphi}_2)=\phi(\hat{\rho}_{1}\otimes\hat{1}_2,\hat{\varphi}_1\otimes\hat{1}_2)+\phi(\hat{1}_1\otimes\hat{\rho}_2,\hat{1}_1\otimes\hat{\varphi}_2).\label{55}
\end{eqnarray}
The general solution to this matrix valued functional equation is derived in Appendix \ref{appendix2} and is
\begin{eqnarray}
\phi(\hat{\rho},\hat{\varphi})=\stackrel{\,\,\sim}{A}\ln(\hat{\rho})+\stackrel{\,\,\sim}{B}\ln(\hat{\varphi}),
\end{eqnarray}
where tilde $\stackrel{\,\,\sim}{A}$ is a ``super-operator'' having constant coefficients and twice the number of indicies as $\hat{\rho}$ and $\hat{\varphi}$ as discussed in the Appendix (i.e., $\left(\stackrel{\,\,\sim}{A}\ln(\hat{\rho})\right)_{ij}=\sum_{k\ell}A_{ijk\ell}(\log(\hat{\rho}))_{k\ell}$ and similarly for $\stackrel{\,\,\sim}{B}\ln(\hat{\varphi})$). DC1' imposes
\begin{eqnarray}
\phi(\hat{\varphi},\hat{\varphi})=\stackrel{\,\,\sim}{A}\ln(\hat{\varphi})+\stackrel{\,\,\sim}{B}\ln(\hat{\varphi})=\hat{0},
\end{eqnarray}
which is satisfied in general when $\stackrel{\,\,\sim}{A}=-\stackrel{\,\,\sim}{B}$, and, now,
\begin{eqnarray}
\phi(\hat{\rho},\hat{\varphi})=\stackrel{\,\,\sim}{A}\Big(\ln(\hat{\rho})-\ln(\hat{\varphi})\Big).\label{supersol}
\end{eqnarray}
Recall that the RHS of (\ref{77}) is equal to the RHS of (\ref{55}), which is stated here:
\begin{eqnarray}
\phi(\hat{\rho}_{1},\hat{\varphi}_1)\otimes\hat{1}_2+\hat{1}_1\otimes\phi(\hat{\rho}_2,\hat{\varphi}_2)=\phi(\hat{\rho}_{1}\otimes\hat{1}_2,\hat{\varphi}_1\otimes\hat{1}_2)+\phi(\hat{1}_1\otimes\hat{\rho}_2,\hat{1}_1\otimes\hat{\varphi}_2).
\end{eqnarray}
We may fix the constant $\stackrel{\,\,\sim}{A}$ by substituting (\ref{supersol}) into both sides of the above equation. This gives,
\begin{eqnarray}
\Big(\stackrel{\,\,\sim}{A}_{1}\Big(\ln(\hat{\rho}_{1})-\ln(\hat{\varphi}_1)\Big)\Big)\otimes\hat{1}_2+\hat{1}_1\otimes\Big(\stackrel{\,\,\sim}{A}_{2}\Big(\ln(\hat{\rho}_{2})-\ln(\hat{\varphi}_2)\Big)\Big)\nonumber
\end{eqnarray}
\begin{eqnarray}
=\stackrel{\,\,\sim}{A}_{12}\Big(\ln(\hat{\rho}_{1}\otimes\hat{1}_2)-\ln(\hat{\varphi}_1\otimes\hat{1}_2)\Big)+\stackrel{\,\,\sim}{A}_{12}\Big(\ln(\hat{1}_{1}\otimes\hat{\rho}_2)-\ln(\hat{1}_1\otimes\hat{\varphi}_2)\Big),\label{59}
\end{eqnarray}
where $\stackrel{\,\,\sim}{A}_{12}$ acts on the joint space of $1$ and $2$ and $\stackrel{\,\,\sim}{A}_{1}$, $\stackrel{\,\,\sim}{A}_{2}$ acts on single subspaces $1$, $2$, respectively. Using the log tensor product identity, $\ln(\hat{\rho}_{1}\otimes\hat{1}_2)=\ln(\hat{\rho}_{1})\otimes\hat{1}_2$, which can be seen by series expansion, the RHS of (\ref{59}) becomes
\begin{eqnarray}
=\stackrel{\,\,\sim}{A}_{12}\Big(\ln(\hat{\rho}_{1})\otimes\hat{1}_2-\ln(\hat{\varphi}_1)\otimes\hat{1}_2\Big)+\stackrel{\,\,\sim}{A}_{12}\Big(\hat{1}_{1}\otimes\ln(\hat{\rho}_2)-\hat{1}_1\otimes\ln(\hat{\varphi}_2)\Big).\label{60}
\end{eqnarray}
Note that arbitrarily letting $\hat{\rho}_2=\hat{\varphi}_2$ in (\ref{59}), (\ref{60}) gives
\begin{eqnarray}
\Big(\stackrel{\,\,\sim}{A}_{1}\Big(\ln(\hat{\rho}_{1})-\ln(\hat{\varphi}_1)\Big)\Big)\otimes\hat{1}_2
=\stackrel{\,\,\sim}{A}_{12}\Big(\ln(\hat{\rho}_{1})\otimes\hat{1}_2-\ln(\hat{\varphi}_1)\otimes\hat{1}_2\Big),
\end{eqnarray}
or arbitrarily letting $\hat{\rho}_1=\hat{\varphi}_1$ gives
\begin{eqnarray}
\hat{1}_1\otimes\Big(\stackrel{\,\,\sim}{A}_{2}\Big(\ln(\hat{\rho}_{2})-\ln(\hat{\varphi}_2)\Big)\Big)=\stackrel{\,\,\sim}{A}_{12}\Big(\hat{1}_{1}\otimes\ln(\hat{\rho}_2)-\hat{1}_1\otimes\ln(\hat{\varphi}_2)\Big).
\end{eqnarray}
As $\stackrel{\,\,\sim}{A}_{12}$, $\stackrel{\,\,\sim}{A}_{1}$, and $\stackrel{\,\,\sim}{A}_{2}$ are constant tensors, inspecting the above equalities determines the form of the tensor to be $\stackrel{\,\,\sim}{A}\,=\stackrel{\,\,}{A}\stackrel{\,\,\sim}{1}$ where $A$ is a scalar constant and $\stackrel{\,\,\sim}{1}$ is the super-operator identity over the appropriate (joint) Hilbert space.

Because our goal is to maximize the entropy function, we let the arbitrary constant $A=-|A|$ and distribute $\stackrel{\,\,\sim}{1}$ identically, which gives the final functional form,
\begin{eqnarray}
\phi(\hat{\rho},\hat{\varphi})=-|A|\Big(\ln(\hat{\rho})-\ln(\hat{\varphi})\Big).
\end{eqnarray}
 ``Integrating'' $\phi$ gives a general form for the quantum relative entropy,
\begin{eqnarray}
S(\hat{\rho},\hat{\varphi} )=-|A|\mbox{Tr}(\hat{\rho} \log \hat{\rho} -\hat{\rho}\log \hat{\varphi} -\hat{\rho})+C[\hat{\varphi}]=-|A|S_U(\hat{\rho},\hat{\varphi})+|A|\mbox{Tr}(\hat{\rho})+C[\hat{\varphi}], 
\end{eqnarray}
where $S_U(\hat{\rho},\hat{\varphi})$ is Umegaki's form of the relative entropy \cite{Umegaki,Uhlmann,Schumacher}, the extra $|A|\mbox{Tr}(\hat{\rho})$ from integration is an artifact present for the preservation of DC1', and $C[\hat{\varphi}]$ is a constant in the sense that it drops out under arbitrary variations of $\hat{\rho}$. This entropy leads to the same inferences as Umegaki's form of the entropy with an added bonus that $\hat{\rho}=\hat{\varphi}$ in the absence of constraints or changes in information---rather than $\hat{\rho}=e^{-1}\hat{\varphi},$ which would be given by maximizing Umegaki's form of the entropy. In this sense, the extra $|A|\mbox{Tr}(\hat{\rho})$ better adheres the quantum relative entropy to the PMU (DC1'), in cases when normalization is not applied. In the spirit of this derivation, we will keep the $\mbox{Tr}(\hat{\rho})$ term there, but, for all practical purposes of inference, as long as there is a normalization constraint, it plays no role, and we find (letting $|A|=1$ and $C[\hat{\varphi}]=0$), 
\begin{eqnarray}
S(\hat{\rho},\hat{\varphi} )\rightarrow S^*(\hat{\rho},\hat{\varphi} )=-S_U(\hat{\rho},\hat{\varphi})=-\mbox{Tr}(\hat{\rho} \log \hat{\rho} -\hat{\rho}\log \hat{\varphi}),
\end{eqnarray}
which is Umegaki's form of the quantum relative entropy. $S^*(\hat{\rho},\hat{\varphi})$ is an equally valid entropy because, given normalization is applied, the same selected posterior $\hat{\rho}$ maximizes both $S(\hat{\rho},\hat{\varphi})$ and $S^*(\hat{\rho},\hat{\varphi})$. The tool for inferentially updating density matrices was designed from the first principles of inference, as was suggested to be possible in private communications with Caticha \cite{Private}.

\subsection{Remarks}
 Due to the universality and the equal application of the PMU (the same design criteria for both the standard and quantum case), the quantum relative entropy reduces to the standard relative entropy when $[\hat{\rho},\hat{\varphi}]=0$ or when the experiment being performed $\hat{\rho}\rightarrow \rho(a)=\mbox{Tr}(\hat{\rho}\ket{a}\bra{a})$ is known.   Because the two entropies are derived in parallel, we expect the well-known inferential results and consequences of the relative entropy to have a quantum relative entropy analog, and this is indeed what we see in the next chapter. 

Maximizing the quantum relative entropy with respect to some constraints $\expt{\hat{A}_i}$, where $\{\hat{A}_i\}$ are a set of  arbitrary Hermitian operators, and normalization $\expt{\hat{1}}=1$, gives the following general solution for the posterior density matrix:
\begin{eqnarray}
\hat{\rho}=\exp\Big(\alpha_0\hat{1}+\sum_i\alpha_i\hat{A}_i+\ln(\hat{\varphi})\Big)=\frac{1}{Z}\exp\Big(\sum_i\alpha_i\hat{A}_i+\ln(\hat{\varphi})\Big)\equiv \frac{1}{Z}\exp\Big(\hat{C}\Big),\label{rhosolution}
\end{eqnarray}
where $\alpha_i$ are the Lagrange multipliers of the respective constraints and normalization $\frac{1}{Z}\equiv e^{\alpha_0}$ may be factored out of the exponential in general because the identity matrix commutes universally. If $\hat{\varphi}\propto \hat{1}$, it is well known that the analysis arrives at the same expression for $\hat{\rho}$ after normalization, as it would if the von Neumann entropy were used, and thus one can find expressions for thermalized quantum states $\hat{\rho}=\frac{1}{Z}e^{-\beta \hat{H}}$ ($\beta=\frac{1}{kT}$ is the inverse temperature and $\hat{H}$ is the Hamiltonian operator). The remaining problem is to solve for the $N$ Lagrange multipliers using their $N$ associated expectation value constraints. In principle, their solution is found by computing $Z=\sum_ie^{\lambda_i}$, which is a sum over the eigenvalues of $\hat{C}$, and then using standard methods,
\begin{eqnarray}
\expt{\hat{A}_i}=-\frac{\d}{\d \alpha_i}\ln(Z)\label{gen},
\end{eqnarray}
and inverting to find $\alpha_i=\alpha_i(\expt{\hat{A}_i})$, which has a unique solution due to the joint concavity (convexity depending on the sign convention) of the quantum relative entropy \cite{Petz1,Petz2} when the constraints are linear in $\hat{\rho}$. The simple proof that (\ref{gen}) is monotonic in $\alpha$, and therefore that it is invertible, is that its derivative $\frac{\d}{\d\alpha}\expt{\hat{A}_i}=\expt{\hat{A}_i^2}-\expt{\hat{A}_i}^2\geq 0$. Between the Zassenhaus formula \cite{Suzuki},
\begin{eqnarray}
e^{t(\hat{A}+\hat{B})}=e^{t\hat{A}}e^{t\hat{B}}e^{-\frac{t^2}{2}[\hat{A},\hat{B}]}e^{\frac{t^3}{6}(2[\hat{B},[\hat{A},\hat{B}]]+[\hat{A},[\hat{A},\hat{B}]])}...,
\end{eqnarray}
and Horn's inequality \cite{Horn,Bhatia,Tao}, the solutions to (\ref{gen}) lack a certain calculational elegance because it is difficult to express the eigenvalues of $\hat{C}=\log(\hat{\varphi})+\sum\alpha_i\hat{A}_i$ (in the exponential) in simple terms of the eigenvalues of the $\hat{A}_i$'s and $\hat{\varphi}$, in general, when the matrices do not commute. A pedagogical exercise is, starting with a prior that is a mixture of spin-z up and down $\hat{\varphi}=a\ket{+}\bra{+}+b\ket{-}\bra{-}$ ($a,b\neq0$), to maximize the quantum relative entropy with respect to an expectation of a general Hermitian operator with which the prior density matrix does not commute. This 2 by 2 spin example is given in the Appendix \ref{spinexampleQRE}.

\chapter{Inferential Applications of the Quantum Relative Entropy\label{Entropy Applications}}

This chapter follows \cite{QBR}, which develops applications of the quantum relative entropy (QRE) to describe quantum measurement within the standard formalism of Quantum Mechanics (QM). The main advantage of our technique is the conceptual clarity it achieves by reformulating quantum measurement from the point of view of inference.

 In QM the wavefunction has two modes of evolution \cite{vonNeumann,Luders}: one is the continuous unitary evolution given by the dynamical Schr\"{o}dinger equation, while the other is the discrete collapse of the wavefunction that occurs when a detection is made. The collapse postulate is generally implemented \emph{ad hoc} to empirically represent the affect of detection on a quantum system.

 In a von Neumann measurement scheme, the goal is to make measurements on a pure state of interest $\ket{\Psi}_{\theta}=\sum_{\theta} c_{\theta}\ket{\theta}$ (in the Hilbert space $\mathcal{H}_{\theta}$). This is accomplished by first entangling $\ket{\Psi}_{\theta}$ with a pointer variable state $\ket{0}_x$ (in the Hilbert space $\mathcal{H}_x$) via a unitary time evolution, and then by making detections of $x$, to ``measure" $\theta$. The entangled states in a von Neumann measurement scheme take the form of a biorthogonal state or Schmidt decomposition, $\ket{\Psi}_{\theta}\ket{0}_x\rightarrow\ket{\Psi,\Phi}_{\theta,x}=\sum_{\theta}c_{\theta}\ket{\theta,x_{\theta}}$, such that the probability of $\theta$ and $x_{\theta'}$ is $\varphi(\theta,x_{\theta'})=|c_{\theta}|^2\delta_{\theta,\theta'}$. A detection of $x_{\theta}$ ``collapses the wavefunction", which is implemented via the \emph{ad hoc} change of state $\ket{\Psi,\Phi}_{\theta,x}\rightarrow \ket{\theta,x_{\theta}}$. Thus, by preparing the system of interest as an entangled state $\ket{\Psi}_{\theta}\rightarrow\ket{\Psi,\Phi}_{\theta,x}$, one can ``measure" the $\theta$'s by making detections of $x$'s.

A positive operator-valued measure (POVM) measurement \cite{Davies,Kraus,Holevo,Nielsen} can be thought of as a generalization of a von Neumann measurement for the purpose of measuring pure or mixed states. The density matrix of interest $\hat{\varphi}_{\theta}$, which may be a pure or mixed state, is entangled with a pointer variable $x$, which is detected for the purpose of ``measuring" the state in $\mathcal{H}_{\theta}$.  If the pointer variable was detected at $x$, the resulting density matrix,
\begin{eqnarray}
\hat{\rho}_{\theta}=\frac{A_{x}\hat{\varphi}_{\theta}A^{\dag}_{x}}{\mbox{Tr}(A_{x}\hat{\varphi}_{\theta}A^{\dag}_{x})},\label{Qbayes}
\end{eqnarray}
quantifies the collapse of $\hat{\varphi}_{\theta}$ under a POVM measurement. A POVM is a set of positive operators $\{E_x\}\in \mathcal{H}_{\theta}$ that are labeled by the pointer variable values $\{x\}$. POVMs sum to identity $\sum_xE_x=\hat{1}_{\theta}$ and are commonly decomposed $E_{x}=A_{x}^{\dag}A_x$ into the operators $A_x$ (and its adjoint $A^{\dag}_x$), which are called the ``measurement" or ``Kraus" operators, and such decompositions are not unique in general.\footnote{An example POVM for a two state system is $E_{1}=\frac{3}{4}\ket{+}\bra{+}+\frac{1}{4}\ket{-}\bra{-}$ and $E_{2}=\frac{1}{4}\ket{+}\bra{+}+\frac{3}{4}\ket{-}\bra{-}$.} In the POVM measurement above, because the $x$'s are entangled with $\hat{\varphi}_{\theta}$, a detection of $x$, leads to the relevant measurement operators, $E_x=A^{\dag}_{x}A_{x}$, to be applied to $\hat{\varphi}_{\theta}$, resulting in (\ref{Qbayes}) after normalization. The POVM measurement scheme (\ref{Qbayes}) is also known as the Quantum Bayes Rule\footnote{Note that the Quantum Bayes Rule developed here solves a different problem than was explored in \cite{Schack}. Here a single density matrix is being inferentially updated rather than the probability corresponding to the form of $N$ copies of an unknown density matrix, i.e. quantum state tomography. This type of quantum tomography may be replicated using the \emph{standard} maximum relative entropy method as it is a special case of (the standard) Bayes Rule.} \cite{Korotkov1,Korotkov2,Jordan}, or as the fundamental theorem of quantum measurement \cite{Jacobs}. For the remainder of the thesis, we will refer to (\ref{Qbayes}) as the Quantum Bayes Rule (QBR).

The QBR is a generalization of L\"{u}ders Rule \cite{Luders} in which $\hat{A}_x$'s are projectors. Here we will derive the QBR from entropic arguments, which we claim eliminates the need for \emph{ad hoc} collapse postulates in QM. Our result supports the interpretation that entropy may be used to inferentially collapse density matrices using projectors (L\"{u}ders Rule), as was derived in \cite{Hellmann,Kostecki}; however, our result is generalized to the QBR and beyond. Rather than appealing to group theoretic arguments \cite{Hellmann,Kostecki}, our derivations are seemingly simpler as they only require solving Lagrange multiplier problems. Because the quantum relative entropy was derived as the tool for density matrix inference in \cite{QRE}, our result discusses collapse from a purely inferential perspective, and thus the quantum measurement proceedure gains more clarity. The present derivation of the Quantum Bayes Rule using the quantum maximum entropy method parallels the standard (probability) maximum entropy derivation of Bayes Rule in \cite{GiffinBayes}, and so, their derivation will be reviewed for pedagogy. 

 As both forms of the standard and quantum relative entropy resemble one another and were derived in parallel \cite{QRE}, they inevitably share analogous solutions and face similar limitations; however, because we are dealing with density matrices, these limitations have consequences in quantum mechanical experiments.  In standard probability theory, there is a phrase, ``The maximum entropy method cannot fix flawed information" \cite{book}, and a similar theme permeates the inference procedure for density matrices. Because the entropy was designed to update from a prior density matrix $\hat{\varphi}$ to a posterior density matrix $\hat{\rho}$, the form of $\hat{\varphi}$ must accurately describe the prior state of knowledge of the system if $\hat{\rho}$ is going to objectively represent the updated state of knowledge for that quantum system. For instance, if our prior knowledge tells us that a particle is located within a certain interval, it makes no sense to impose that the particle has an average position anywhere but within that interval. We derive this type of logical compatibility in the quantum maximum entropy method, which we name the Prior Density Matrix Theorem (PDMT). The PDMT states that a prior density matrix can only be updated in the Hilbert space it originally spans. The PDMT is purely mathematical and follows from logic analogous to that of Bayes Theorem -- if a prior state assigns zero probability to some state, $\varphi(\theta_0)=0$, Bayes Theorem gives $\varphi(\theta_0)\stackrel{*}{\rightarrow}\varphi(\theta_0|x)=\frac{\varphi(x|\theta_0)}{\varphi(x)}\varphi(\theta_0)=0$. The PDMT is a purely inferential consequence of entropic updating, yet it sheds light on some of the nontrivial notions of quantum measurement and QM in general. 

 A special case of the PDMT implies that if the prior state is a pure state $\ket{\Psi}=\sum_{\theta}c_{\theta}\ket{\theta}$, then the only inferential update one can make is normalization using the quantum maximum entropy method. Thus, to be able to reproduce standard quantum mechanical projection measurements,  one must first \emph{ad hocly} allow the pure state to decohere (or partially decohere) within the measurement device such that it can be updated nontrivially. This is a reformulation of L\"uders notion \cite{Luders} that the function of a measurement device must be to project the pure state into a mixed state $\hat{\rho}\rightarrow \sum \hat{P}_i\hat{\rho}\hat{P}_i$ (where $\hat{P}_i$ are rank-1 orthonormal projectors), except our argument is formulated purely in terms of entropy and inference. This concept is not as foreign or as objectionable as it may seem if we consider the well known results of the quantum two slit experiment. If a ``which slit" detection of the particle is made, then the resulting probability distribution is a decohered sum of distributions on the screen (after many trials), whereas omitting this detection allows for interference effects. Decoherence of the pure state was required for a which slit inference. Once the particle hits the screen, to detect its state, it decoheres (potentially again) on the detection screen. This imprints a mixed state realization of the incoming pure or decohered state on the screen $\hat{\rho}\rightarrow \sum \hat{P}_x\hat{\rho}\hat{P}_x$, which may be detected and collapse the state.\footnote{Note that both $\hat{\rho}$ and $\sum \hat{P}_x\hat{\rho}\hat{P}_x$ have the same positional probability distribution $p(x)$, but in general they evolve differently in time.} When making measurements in QM, it is important to include both the system of interest and the auxiliary pointer variable (and/or the measurement device), such that the appropriate prior density matrix is generated. In this sense, ``collapse of the wavefunction" is better stated as the ``collapse of the mixed state after decoherence" -- which then, as we will see, is nothing more than standard probability updating. In conclusion, it is illogical for a prior pure state to directly collapse into a different pure state because, in some sense, we are already maximally informed about the prior (pure) state of the system, which may be viewed as a (priorly) collapsed state itself.

In preparation for the derivation of the Quantum Bayes Rule using the quantum maximum entropy method, the derivation of Bayes Rule using the standard maximum entropy method is reviewed below \cite{GiffinBayes}. We will introduce the PDMT and apply the quantum maximum entropy method to derive the aforementioned cases of interest. 

\paragraph{Maximum Entropy and Bayes}
  When the information provided is in the form of data, entropic updating is consistent with Bayes Rule,
\begin{eqnarray}
\rho(\theta)\stackrel{1}{=}\varphi(\theta|x)\stackrel{2}{=}\frac{\varphi(x|\theta)\varphi(\theta)}{\varphi(x)},
\end{eqnarray}
where Bayes Rule is the first equal sign and Bayes Theorem is the second equal sign \cite{book}. This leads to the realization that Bayesian and entropic inference methods are consistent with one another \cite{GiffinBayes}. 

The posterior distribution $\rho(\theta)$ can only be realized once the data about $x$'s has been processed. This implies the state space of interest is the product space of $\mathcal{X}\times\Theta$ with a joint prior $\varphi(x,\theta)$. Suppose we collect data and observe the value $x'$. The data constrains the joint posterior distribution $\rho(x,\theta)$ to reflect the fact that the value of $x$ is known to be $x'$, that is,
\begin{eqnarray}
\rho(x)=\int d\theta \rho(x,\theta)=\delta(x-x'),\label{data}
\end{eqnarray}
however; this data constraint is not enough to specify the full joint posterior distribution, 
\begin{eqnarray}
\rho(x,\theta)=\rho(\theta|x)\rho(x)=\rho(\theta|x)\delta(x-x'),\label{data+1}
\end{eqnarray}
because $\rho(\theta|x)$ is not determined. The above equation is known as Bayesian conditionalization.

 As there are many distributions that satisfy this data constraint, we rank candidate distributions using the maximum entropy method. Note that the data constraint (\ref{data}) in principle constrains \emph{each} $x$ in $\rho(x,\theta)$. This means a Lagrange multiplier $\alpha(x)$ is required to tie down each  $x\in\mathcal{X}$ of the marginal distribution $\rho(x)$. Maximizing the entropy with respect to these constraints and normalization is,
\begin{eqnarray}
0=\delta\Big(S-\lambda[\int \rho(x,\theta)\,dxd\theta-1]-(\int \alpha(x)[\int\rho(x,\theta)\,d\theta-\delta(x-x')]\,dx)\Big)
\end{eqnarray}
where $S=S[\rho(x,\theta),\varphi(x,\theta)]$ is the standard relative entropy and $\lambda$ is the Lagrange multiplier that imposes normalization. This leads to the following joint posterior distribution,
\begin{eqnarray}
\rho(x,\theta)=\varphi(x,\theta)\frac{e^{\alpha(x)}}{Z}.
\end{eqnarray}
The Lagrange multiplier $Z$ is found by imposing normalization,
\begin{eqnarray}
1&=&\int\rho(x,\theta)\,dxd\theta=\frac{1}{Z}\int\varphi(x,\theta)e^{\alpha(x)} \,dxd\theta\nonumber\\
&\rightarrow&Z=\int\varphi(x,\theta)e^{\alpha(x)} \,dxd\theta.
\end{eqnarray}
The Lagrange multiplier $\alpha(x)$ is found by considering the data constraint (\ref{data}),
\begin{eqnarray}
\delta(x-x')=\int\rho(x,\theta)d\theta=\frac{e^{\alpha(x)}}{Z}\int\,\varphi(x,\theta)d\theta=\frac{e^{\alpha(x)}}{Z}\varphi(x).
\end{eqnarray}
Substituting in the Lagrange multiplier gives the joint posterior distribution,
\begin{eqnarray}
\rho(x,\theta)=\frac{\varphi(x,\theta)}{\varphi(x)}\delta(x-x')=\varphi(\theta|x)\delta(x-x').
\end{eqnarray}
Integrating over $x$ gives the marginalized posterior distribution,
\begin{eqnarray}
\rho(\theta)=\varphi(\theta|x')=\frac{\varphi(x'|\theta)\varphi(\theta)}{\varphi(x')},
\end{eqnarray}
which is Bayes Rule. Jeffreys Rule is the generalization of Bayes Rule when the data is uncertain and it is also consistent with the entropic inference.\footnote{That is, set $\rho_D(x)=\int d\theta \rho(x,\theta)$ in place of (\ref{data}).} Further review can be found in \cite{book}. The universality of this entropic inference method is emphasized by its consistency with other forms of inference like Bayesian inference \cite{GiffinBayes,Giffin,GiffinThesis}.

\paragraph{A comment on biased priors}
Entropic inference of this nature is only as useful as we are objective about our subjectivity. One should be careful not to apply nonsensical constraints, for instance, attempting to impose impossible expectation values (like that the average roll of a six sided die be seven). In such a case, the maximum entropy method provides ``no solution" to the optimization problem due to its irrationality \cite{JaynesProb}. Consider a set of microstates $x\in D_0\subset D$, given a situation $s$, that have a prior probability $\varphi(x\in D_0|s)=0$, representing impossibility. In this subdomain $D_0$ and situation $s$, the values of $x$ are believed to be impossible, and furthermore, one can see that it is impossible to update this prior to anything but $\rho(x\in D_0|s)=0$ using the maximum entropy method for any amount of new information (as can be seen in (\ref{gensolrho})). In the same way, a delta function prior distribution $\varphi(x|s)\sim\delta(x-x_0)$, which claims complete certainty at $x_0$, cannot be updated. We shall call priors that have domains of unupdatablity $\varphi(x\in D_0|s)\stackrel{*}{\rightarrow}\rho(x\in D_0|s)=\varphi(x\in D_0|s)$ ``biased", whether or not the prior was attained by objective evidence or personal bias\footnote{`` I'm not biased if I'm right!" - N. Carrara}
 as the philosophical divide between the two is, in all but the most extreme cases, a bit subjective.\footnote{For instance, can an objective experimentalist be \emph{completely certain} that their measurement device hasn't misfired?} 
 
A biased state of knowledge pertaining to a situation $s$ does not imply bias for a new situation $s'$, so a realization that a nonbiased probability should be assigned to the region $D_0$ admits the system is now in a new situation $s'$. An example of this from Statistical Mechanics (and also QM) occurs if the distance between the walls of an infinite potential box is enlarged such that previous zero probability regions now gain possibility. In this sense, and others, that entropic updates are purely epistemic. If the physical situation has changed, the situation $s\rightarrow s'$ should be updated as well.\footnote{In the next chapter we see this kind of ``situational" entropic updating in Entropic Dynamics. Differences in situations $s'-s$ in Entropic Dynamics are labeled by differences in time $t'-t$.} 

\section{Quantum Entropic Inference}

Before deriving the QBR, we must first discuss the ramifications of ``biased" prior density matrices. 

\subsection{Prior density matrices}
If the prior density matrix $\hat{\varphi}=\ket{\varphi}\bra{\varphi}$ is a pure state, then we consider it to be a completely ``biased" density matrix because no amount of information can update it, i.e., $\hat{\varphi}\stackrel{*}{\rightarrow}\hat{\rho}=\hat{\varphi}$, without changing the situation, using entropic methods. An example using a pure spin state prior is discussed below to introduce the predicament, although the analysis holds for all prior density matrices that are 1-dimensional projectors.

Consider the biased prior density matrix $\hat{\varphi}=\ket{+}\bra{+}$ -- the positive spin-$z$ eigenstate. To perform the calculation with any rigor using this biased prior, we must unbias it slightly by allowing it to span more than 1-dimension, where we introduce $\hat{\varphi}=\lim_{\epsilon\rightarrow0}\Big(\ket{+}\bra{+}+\epsilon\ket{-}\bra{-}\Big)\equiv\lim_{\epsilon\rightarrow0}\hat{\varphi}_{\epsilon}$. We will use this $\epsilon$-prior $\hat{\varphi}_{\epsilon}$ for the prior, and then take the limit $\epsilon\rightarrow0$ when appropriate. In attempting to force the issue, consider maximizing the relative entropy subject to an expectation value constraint of a general 2 by 2 Hermitian matrix $\expt{c_\mu\sigma^{\mu}}$, the expectation value of a weighted sum of Pauli matrices and identity, such that $\hat{\rho}$ would require a nonzero component along spin down $\ket{-}\bra{-}$ to meet the expectation value constraint, in contrast to $\hat{\varphi}$. Maximizing the entropy subject to this constraint, normalization, and using the $\epsilon$-prior gives an $\epsilon$-posterior,
\begin{eqnarray}
\hat{\rho}_{\epsilon}=\frac{1}{Z}\exp\Big(\alpha c_\mu\sigma^{\mu}+\ln(\hat{\varphi}_{\epsilon})\Big)\equiv\frac{1}{Z}\exp(\hat{C}_{\epsilon}).
\end{eqnarray}
The Lagrange multiplier that imposes normalization may be found by diagonalizing the exponent $\hat{C}_{\epsilon}\rightarrow \hat{\Lambda}_{\epsilon}$,
\begin{eqnarray}
Z=\mbox{Tr} (\exp(\hat{C}_{\epsilon}))=\mbox{Tr} (\hat{U}_{\epsilon}\exp\Big(\hat{U}_{\epsilon}^{\dag}\hat{C}_{\epsilon}\hat{U}_{\epsilon}\Big)\hat{U}_{\epsilon}^{\dag})=\mbox{Tr}(e^{\hat{\Lambda}_{\epsilon}})=\sum_{\lambda_{\epsilon}} e^{\lambda_{\epsilon}},
\end{eqnarray}
which suggests a convenient representation of the posterior density matrix using $\hat{U}_{\epsilon}$,
\begin{eqnarray}
\hat{\rho}_{\epsilon}=\frac{1}{Z}\hat{U}_{\epsilon}\exp(\hat{\Lambda}_{\epsilon})\hat{U}_{\epsilon}^{\dag}.
\end{eqnarray}
In the limit $\epsilon\rightarrow 0$, the respective eigenvalues of $\hat{C}_{\epsilon}$, $\lambda_{\pm}$, approach a constant and $-\infty$ while their respective eigenvectors straighten out $\ket{\lambda_{\pm}}_{\epsilon}\rightarrow\ket{\pm}$, and $\hat{U}_{\epsilon}\rightarrow\hat{1}$. Therefore the posterior density matrix $\hat{\rho}=\lim_{\epsilon\rightarrow 0}\hat{\rho}_{\epsilon}=\hat{\varphi}$ is equal to the biased prior density matrix and has not been updated. Because the pure state fails to update, it is biased, analogous to a delta function probability distribution in the standard maximum entropy case. One might expect an infinitely large Lagrange multiplier $\alpha$ to be able to overcome the negative infinities one obtains when taking $\epsilon\rightarrow 0$. However, because $\expt{c_{\mu}\sigma^{\mu}}=f(\alpha)$ is monotonic in $\alpha$, an infinitely large Lagrange multiplier would imply the system is being constrained to a maximally large expectation value (which represents a very small subset of expectations) and usually implies that the posterior is actually known with certainty (in which case there is no need to use the maximum entropy method).  The mixed spin state example in Appendix \ref{spinexampleQRE}, is not biased, and therefore it avoided issues surrounding biased prior density matrices. Below we will discuss the general case and its implications.

Consider an $M$th order biased prior represented in its eigenbasis $\hat{\varphi}=\sum_{n=1}^M\varphi_n\ket{\varphi_n}\bra{\varphi_n}+\sum_{n=M+1}^N0_n\ket{\varphi_n}\bra{\varphi_n}$ in an $N$ dimensional Hilbert space ($M=1$ is a purestate). Given an $N\times N$ dimensional constraint $\hat{A}$ (however the analysis holds for $\hat{A}$ of any rank),  the prescription is to add some $\epsilon$'s to $\hat{\varphi}$ such that $\hat{\varphi}\rightarrow\hat{\varphi}_{\epsilon}$ spans $N$, and,
\begin{eqnarray} \log(\hat{\varphi}_{\epsilon})=\left( \begin{array}{cccccc}
\log(\varphi_1)& \dots & 0&0 &\dots&0\\
  \vdots&\ddots&0&0&\dots&0\\
0 & 0&\log(\varphi_M)&0&\dots&0\\
0 &0 &0 &\log(\epsilon)&&0\\
\vdots &\vdots & \vdots&&\ddots&0\\
0 &0 & 0&0&0&\log(\epsilon)
\end{array} \right),
\end{eqnarray}  
 has $N-M$ diagonal $\log(\epsilon)$ terms. Because density matrices are Hermitian, and have a sum representation, $\hat{\rho}=\sum_{ij}\rho_{ij}\ket{i}\bra{j}$, they may always be rearranged and relabeled into the form above without loss of generality. Thus, $\log(\hat{\varphi}_{\epsilon})$ may always be written as a direct sum $\log(\hat{\varphi}_{\epsilon})=\log(\hat{\varphi}_M)\oplus\log(\epsilon)\hat{1}_{N-M}$, where $\log(\hat{\varphi}_M)$ is the first $M\times M$ block of $\log(\hat{\varphi}_{\epsilon})$ and $\log(\epsilon)\hat{1}_{N-M}$ is the remaining block proportional to $\log(\epsilon)$.  Expressing the $N\times N$ constraint matrices $\hat{A}=\sum_i\alpha_i\hat{A}_i$ in the eigenbasis of $\hat{\varphi}_{\epsilon}$, and summing it, is,
\begin{eqnarray}
\hat{C}_{\epsilon}=\hat{A}+\log(\hat{\varphi}_{\epsilon}),
\end{eqnarray}
which is a general representation of the matrix that resides in the exponential of a posterior density matrix, $\hat{\rho}_{\epsilon}=\frac{1}{Z}\exp(\hat{C}_{\epsilon})$, having an $\epsilon$-prior density matrix $\hat{\varphi}_{\epsilon}$. Similarly partitioning $\hat{C}_{\epsilon}$ by letting $\hat{C}_{M}$ be its first $M\times M$ block, the characteristic polynomial equation of $\hat{C}_{\epsilon}$ may be written in the following form,
 \begin{eqnarray}
0=\det|\hat{C}_{\epsilon}-\lambda\hat{1}|&=&\det|\hat{C}_{M}-\lambda\hat{1}_M|(\log(\epsilon)-\lambda)^{N-M}+c_{1}(\log(\epsilon)-\lambda)^{N-M-1}\nonumber\\
&+&c_{2}(\log(\epsilon)-\lambda)^{N-M-2}+...+c_{N-M},
\end{eqnarray}
where the $c_{n}$'s are the remaining $\epsilon$ independent coefficients. For any finite $\lambda$, we may divide the characteristic equation by the leading $(\log(\epsilon)-\lambda)^{N-M}\approx \log(\epsilon)^{N-M}$ term, which in the limit of $\epsilon\rightarrow 0$, reduces the characteristic equation to the $M\times M$ block characteristic equation, 
 \begin{eqnarray}
0=\det|\hat{C}_{\epsilon}-\lambda\hat{1}|\rightarrow\det|\hat{C}_{M}-\lambda\hat{1}_M|,
\end{eqnarray}
for all finite $\lambda$. The eigenvectors associated to these $M$-finite eigenvalues span the $M\times M$ vector space. As this is true for all finite eigenvalues, the remaining $N-M$ eigenvalues are not finite and indeed are all equal to negative infinity, due to the $\log(\epsilon)$'s as $\epsilon\rightarrow0$. The remaining eigenvectors with the associated infinite eigenvalues therefore span the remaining $(N-M)\times (N-M)$ vector space, but are not unique because they have degenerate eigenvalues. The eigenvectors for the finite and infinite eigenvalues span disjoint subspaces and therefore so do the unitary matrices which diagonalize them $\hat{U}=\hat{U}_{M}\oplus\hat{U}_{N-M}$ as the unitary operators consist of columns of their associated eigenvectors. This disjointness is independence in the sense that the unitary operator $\hat{U}$ is block diagonal, and therefore $C_{\epsilon}=C_M\oplus C_{N-M}=\hat{U}_{M}\Lambda_{M}\hat{U}^{\dag}_{M}\oplus\hat{U}_{N-M}\Lambda_{N-M}\hat{U}^{\dag}_{N-M}$. The posterior density matrix is therefore,
\begin{eqnarray}
\hat{\rho}=\frac{1}{Z}e^{\hat{C}_M}\oplus\hat{0}_{N-M}=\frac{1}{Z}e^{\hat{A}_M+\log(\hat{\varphi}_M)}\label{biasedprior},
\end{eqnarray}
completely independent of the $\hat{A}-\hat{A}_M$ pieces of the constraints in $\hat{A}$, and $\hat{\varphi}_M=\hat{\varphi}$ is the original $M$th order biased prior.  The lack of an ability to update biased priors in the $N-M$ region is not a failure of the method of maximum entropy, but rather a failure to choose appropriate constraints given an $M$th order biased prior density matrix. 

In general, any prior density matrix that does not span the entire Hilbert space is an $M$th order biased prior density matrix. This insists the following, which we state as a theorem:
\paragraph{Prior Density Matrix Theorem (PDMT):} \emph{An $M$th order biased prior density matrix $\hat{\varphi}$ can only be inferentially updated in the eigenspace that it spans.}\\

 The immediate consequence of the PDMT is that entropic updating can only cause \emph{epistemic and inferential} changes to $\hat{\rho}$. The inability to update a pure state nontrivially\footnote{The trivial update being pure state normalization.}, like in the pure state spin $\ket{+}\bra{+}$ example, shows just this. The only way to change the state of a 1-dimensional projector (pure state) prior density matrix is to physically rotate the state by applying dynamical unitary operators $U(t',t)$ via the Schr\"{o}dinger equation because no inferential entropic update is possible.\footnote{Entropic Dynamics, however, is an application of the standard maximum entropy method and information geometry that does yield Schr\"{o}dinger evolution.} However, if $\hat{\varphi}$ is a pure state and one knows that it will evolve unitarily to $\hat{\rho}=U\hat{\varphi}U^{\dag}$, then one may evolve it. In this instance the quantum maximum entropy method is not needed because the posterior state is known; however, the quantum maximum entropy method has not failed because the PDMT indeed gives $\hat{\varphi}$ as its solution. 

Once the Quantum Bayes Rule is derived using entropy, we will see that the Schr\"{o}dinger equation and the quantum maximum entropy method complement one another in QM -- the first being responsible for continuous dynamical ``physical" changes to the system and the second being discontinuous inferential updates within the space originally spanned by $\hat{\varphi}$, all being part of the measurement process. This is not to say that the quantum maximum entropy method cannot in general mimic the results of unitary state evolution, it could very well be the special case that a unitary operator acting on an $M>1$ $\hat{\varphi}$ evolves $\hat{\varphi}$ only within $M$ (rather than rotating the eigenvectors off their hyperplane), and thus the PDMT does not prevent this update.

If one is serious about the assignment of a biased prior density matrix then the following realization is needed, ``Because pure states are completely biased, the quantum maximum entropy method cannot update to a new posterior \emph{at that time}". If however the pure state is changed ``physically" by the addition of new microstates via interaction, allowing it to decohere \cite{Zurek2003b,Schlosshauer} (and the references therein), or change by some other process, then at a later time one could employ a method similar to \cite{Wooters1,Wooters2}, that is, apply $\hat{\varphi}^{-1/2}\hat{\varphi}^{'1/2}(t)$ and its transpose on either side of $\hat{\varphi}\equiv\hat{\varphi}(0)$,
\begin{eqnarray}
\hat{\varphi}'(t)=\Big(\hat{\varphi}^{'1/2}(t)\hat{\varphi}^{-1/2}\Big)\,\hat{\varphi}\,\Big(\hat{\varphi}^{-1/2}\hat{\varphi}^{'1/2}(t)\Big),\label{changeofstate}
\end{eqnarray}
to represent a new prior density matrix that, decohered, evolved unitarily, or is part of a new experimental configuration. Now, if the prior is an $M>1$ mixed state, it is possible to inferentially update it non-trivially.  If the quantum relative entropy is going to be used for making inferences that pertain to pure state quantum measurement, the function of a measurement device must be to project the system into a basis $\hat{\rho}\rightarrow \sum_i\hat{P}_i\hat{\rho}\hat{P}_i$, which evades the potential trivial results of the PDMT. The proper prior for quantum measurement thus depends on both the system of interest \emph{and} the measurement device. This is a reformulation of L\"{u}ders' notion, except here, it has been motivated and expressed in the language of entropic updating.

There are a few things to take away from this section. The quantum maximum entropy method only updates a density matrix inferentially, as can be seen by its lack of ability to rotate biased priors into non-biased states or other biased priors states. This is exactly what we expect as the unupdatability of biased priors exists in standard probability theory.   The solution to the biased prior problem (in the standard and quantum maximum entropy method) is, if appropriate: to  change the constraint(s), the prior, both, or neither and accept the consequences of its solution. This reasoning guides us in choosing appropriate priors in subsequent derivations throughout this chapter. 

\section{The Quantum Bayes Rule}
Notationally, we will let density matrices living in a Hilbert space $\mathcal{H}_{x}\otimes\mathcal{H}_{\theta}$ to be denoted $\hat{\rho}_{x,\theta}$. Density matrices may of course be expressed in any basis within these Hilbert spaces. Below, $x$'s are the arbitrary variables (positions, momenta, ect.) that will be detected and $\theta$'s are the arbitrary variables that will be inferred.  We find it convenient to denote the diagonal $(x',x')$ block matrix of $\hat{\rho}_{x,\theta}$ with an equal sign such that $\bra{x'}\hat{\rho}_{x,\theta}\ket{x'}\equiv\hat{\rho}_{x=x',\theta}$ and similarly $\bra{\theta'}\hat{\rho}_{x,\theta}\ket{\theta'}\equiv\hat{\rho}_{x,\theta=\theta'}$, and on occasion probabilities $\rho(\theta')=\bra{\theta'}\hat{\rho}_{\theta}\ket{\theta'}\equiv\hat{\rho}_{\theta=\theta'}$. Also, a tilde above a density matrix will represent a mixed representation of the density matrix in question $\hat{\varphi}_{\theta}\rightarrow\stackrel{\sim}{\varphi}_{\theta}=\sum_{\alpha}\hat{P}_{\alpha}\hat{\varphi}_{\theta}\hat{P}_{\alpha}$, with $\hat{P}_{\theta}=\ket{\alpha}\bra{\alpha}$. Here we will review the standard POVM measurement scheme that leads to the QBR.

\paragraph{Introduction -- Quantum Bayes Rule:}
Following \cite{Jacobs}, consider a prior density matrix $\hat{\varphi}_{\theta}$ which is entangled with a pointer variable such that $\hat{\varphi}_{\theta}\rightarrow \hat{\varphi}_{x,\theta}$. The system and the pointer variable are entangled in the following way: given an initial state of the pointer variable $\ket{0}_x\bra{0}$, the joint system is entangled with a unitary operator $U$,
\begin{eqnarray}
\hat{\varphi}_{x,\theta}=U( \ket{0}_x\bra{0}\otimes\hat{\varphi}_{\theta})U^{\dag}
\end{eqnarray}
where,
\begin{eqnarray}
U=\sum_{\theta,\theta',x,x'}u_{\theta,\theta',x,x'}\ket{x'}\ket{\theta'}\bra{x}\bra{\theta}=\sum_{x,x'}\ket{x'}_x\bra{x}\otimes A_{x'x},
\end{eqnarray}
and where,
\begin{eqnarray}
 A_{x'x}=\sum_{\theta,\theta'}u_{\theta,\theta',x,x'}\ket{\theta}\bra{\theta'},
\end{eqnarray}
is the $x',x$ sub-block matrix \cite{Jacobs}. The prior density matrix of the joint system is therefore,
\begin{eqnarray}
\hat{\varphi}_{x,\theta}=\sum_{x,x'}\ket{x}_x\bra{x'}\otimes (A_{x}\hat{\varphi}_{\theta}A^{\dag}_{x'}),\label{entangle3.23}
\end{eqnarray}
where $A_{x0}\equiv A_{x}$ are the measurement operators of the POVM $\hat{E}_{x}=\hat{A}_{x}\hat{A}^{\dag}_{x}$. Due to Naimark's theorem \cite{Naimark1,Naimark2}, making a projective measurement of the pointer variable $x$ can be used to perform a POVM measurement on $\hat{\varphi}_{\theta}$. Projecting the pointer variable requires the following action on $\hat{\varphi}_{x,\theta}$,
\begin{eqnarray}
\hat{\varphi}_{x,\theta}\rightarrow (\ket{x'}\bra{x'}\otimes \hat{1}_{\theta})\hat{\varphi}_{x,\theta}(\ket{x'}\bra{x'}\otimes \hat{1}_{\theta})=\ket{x'}\bra{x'}\otimes (A_{x'}\hat{\varphi}_{\theta}A^{\dag}_{x'}),
\end{eqnarray}
which collapses the state in $x$ to $x'$ and implies the new state of the system is,
\begin{eqnarray}
\hat{\rho}_{\theta}=\frac{A_{x'}\hat{\varphi}_{\theta}A^{\dag}_{x'}}{\mbox{Tr}(A_{x'}\hat{\varphi}_{\theta}A^{\dag}_{x'})},\label{Qbayes2}
\end{eqnarray}
after normalization. Again this is known as the Quantum Bayes Rule (QBR) \cite{Korotkov1,Korotkov2,Jordan}, the fundamental theorem of quantum measurement \cite{Jacobs}, or the POVM measurement formalism \cite{Davies,Kraus,Holevo,Nielsen}. 

In the remainder of this section we will derive inference rules using the quantum maximum entropy method in order of increasing generality. In each case, the relevant prior density matrix will be generated using standard methods of decoherence in QM. First, we will consider the simplest case -- it is the inference of a single system collapsing into one of its eigenstates. We will then reconsider this simple case in the presence of measurement uncertainty -- we call this case a ``simple partial (or uncertain) collapse". The next level of generality is to consider inferences on systems that are jointly coupled or entangled, i.e., a system of interest entangled with a pointer variable. In such a case, one must obtain the appropriate joint prior density matrix for inference from QM. Inferences that lead the pointer variable to collapse to a definite state yield a derivation of the QBR from the quantum maximum entropy method. As before, if there is measurement uncertainty in the pointer variable, one has a partial (or uncertain) collapse of the pointer variable state. This yields a posterior state that we call the Quantum Jeffreys Rule (QJR). 

In the joint entangled system setting, we may wish to make detections of both the system of interest and the pointer variable. We find that the order of detection does not affect the final joint probability $\rho(x,\theta)$, which is consistent with the results of the delayed choice experiment -- a similar argument that argues from the point of view of epistemic probability theory is given in \cite{Gaasbeek}. Finally this section is concluded with a simple example of quantum control in the quantum maximum entropy method setting. It is an instance of the QJR, and it is shown, for a joint entangled system, that capturing the pointer particle inside a thermal box with temperature $\beta$, or $\beta'$, leads to changes in the statistics of the system of interest -- thus a bit of control may be exerted by the experimentalist by choosing the temperature of the thermal box. Further generalizations are discussed in the following section.

\paragraph{Simple Collapse:}
This entropic update is a special case of (\ref{Qbayes2}) when the $A_{x}$'s are all projectors rather than a more general POVM, i.e. we are going to reproduce L\"{u}der's Rule. We must first argue for the appropriate form of a prior density matrix that is undergoing measurement.

As we are simply doing a projective measurement on $\hat{\varphi}_{x}$, ``another" pointer variable is not needed to generate the POVM. We follow the intuition that if we are going to make inferences on the basis of detection, the prior density matrix should appropriately reflect the fact that it has interacted with a measurement device. A projective measurement on the $x$'s in experiment requires entangling $\hat{\varphi}_{x}$ to detector states $\{\ket{d}\}$ and letting them decohere within the detector. This avoids the potentially trivial results of the PDMT. For concreteness we may imagine that $\hat{\varphi}_{x}=\ket{\Phi}_x\bra{\Phi}$ is the pure state density matrix of a particle that went though a two slit apparatus (no ``which slit" measurement has been made) and is impinging onto a screen, CCD array, or a similar device designed to detect $x$. Let the pure state be $\ket{\Phi}=\sum_x\sqrt{\varphi(x)}e^{i\phi(x)}\ket{x}$. The pure state evolves with the detector states,
\begin{eqnarray}
\ket{d_0}\bra{d_0}\otimes\hat{\varphi}_{x}\rightarrow\hat{\varphi}_{d,x}=\hat{U}\ket{d_0}\bra{d_0}\otimes\hat{\varphi}_{x}\hat{U}^{\dag}=\ket{\mathcal{D},\Phi}_{d,x}\bra{\mathcal{D},\Phi}.
\end{eqnarray}
The unitary operators which couple the states are,
\begin{eqnarray}
\hat{U}=\sum_{d_y,d_{y'},x,x'}u_{d_yd_{y}xx'}\ket{d_y}\ket{x}\bra{d_{y'}}\bra{x'}=\sum_{d_y,d_{y'}}\ket{d_y}\bra{d_{y'}}\otimes B_{d_yd_{y'}},
\end{eqnarray}
with the sub-block matrices,
\begin{eqnarray}
B_{d_yd_{y'}}=\sum_{x,x'}u_{d_yd_{y'}xx'}\ket{x}\bra{x'}.
\end{eqnarray}
We define a good detector as one in which the $\ket{x}$th pointer variable state only entangles with the local state of the detector $\ket{d_x}$, which is an argument for the sub-block matrix to take a simple form,
\begin{eqnarray}
B_{d_y0}=\sum_{x,x'}\delta_{y,x}\delta_{y,x'}\ket{x}\bra{x'}=\ket{y}\bra{y}.
\end{eqnarray}
This then gives a fully entangled (von Neumann measurement) state,
\begin{eqnarray}
\ket{\mathcal{D},\Phi}=\sum_x\sqrt{\varphi(x)}e^{i\phi(x)}\ket{d_x,x}.\nonumber
\end{eqnarray} 
Tracing over the detector states $\{\ket{d_x}\}$ represents that the system has interacted with the measurement device (in this case the screen), but its value has yet to be registered by either the device or the observer, and that we have neglected to keep track of the detector states, 
\begin{eqnarray}
\hat{\varphi}_x\rightarrow\stackrel{\sim}{\varphi}_{x}(t)=\mbox{Tr}_d(\hat{\varphi}_{d,x})=\sum_{x}\varphi(x)\ket{x}\bra{x}.
\end{eqnarray}
This is a mixed state realization of the original two slit pure state probability $\varphi(x)=|\braket{x}{\Phi}|^2=\,\stackrel{\sim}{\varphi}(x)$ on the screen. This is not original and may be obtained following \cite{Luders} using projectors $\hat{\varphi}\rightarrow \stackrel{\sim}{\varphi}_{x}(t)=\sum_{x}\hat{P}_x\hat{\varphi}_x\hat{P}_x$ or more directly \cite{Zurek2003b, Schlosshauer}. Now that the form of the prior density matrix has been argued for, we may utilize the quantum maximum entropy method.

In principle, when the result of a projective measurement ($\stackrel{\sim}{\varphi}_{x}$) is registered, the state of the system is known with certainty. This is represented by the following constraint on the posterior probability distribution, 
\begin{eqnarray}
\mbox{Tr}(\ket{x}\bra{x}\hat{\rho}_{x})=\rho(x)=\delta_{xx'},\label{delta}
\end{eqnarray}
which is an expectation value on the posterior density matrix $\hat{\rho}_{x}$, stating that the system was detected in the state $x'$ with certainty. Because this constraint must be imposed for every $x$, there is one Lagrange multiplier $\alpha_x$ for each $x$. Maximizing the quantum relative entropy with respect to this constraint and normalization is setting
\begin{eqnarray}
0=\delta\Big( S-\lambda[\mbox{Tr}(\hat{\rho}_{x})-1]-\sum_x\alpha_x[\mbox{Tr}(\ket{x}\bra{x}\hat{\rho}_{x})-\delta_{xx'}]\Big),\label{47}
\end{eqnarray}
where $S=S(\hat{\rho}_x,\stackrel{\sim}{\varphi}_{x})$ is the quantum relative entropy. The posterior which maximizes the quantum relative entropy subject to these constraints is,
 \begin{eqnarray}
\hat{\rho}_{x}=\frac{1}{Z}\exp\Big(\sum_x\alpha_x\ket{x}\bra{x}+\log(\stackrel{\sim}{\varphi}_{x})\Big).
\end{eqnarray}
Because the constraint and prior commute, the posterior density matrix takes a simple form,
\begin{eqnarray}
\hat{\rho}_{x}=\frac{1}{Z}\sum_x\varphi(x)e^{\alpha_x}\ket{x}\bra{x}.
\end{eqnarray}
The normalization constraint gives,
\begin{eqnarray}
Z=\sum_x\varphi(x)e^{\alpha_x},
\end{eqnarray}
and the expectation value constraint (\ref{delta}) gives,
\begin{eqnarray}
\delta_{xx'}=\mbox{Tr}(\ket{x}\bra{x}\hat{\rho}_{x})=\frac{\varphi(x)e^{\alpha_x}}{Z}.
\end{eqnarray}
The final form of the posterior density matrix is found by substituting for $e^{\alpha_x}$, and the result is a collapsed state,
\begin{eqnarray}
\hat{\rho}_{x}=\sum_x\delta_{xx'}\ket{x}\bra{x}=\ket{x'}\bra{x'},
\end{eqnarray}
as expected.  Written in a suggestive ``Bayes update", or ``projective collapse" form,
\begin{eqnarray}
\hat{\rho}_{x}\equiv\,\,\stackrel{\sim}{\varphi}_{x|x'}\,\,=\frac{\ket{x'}\bra{x'}\stackrel{\sim}{\varphi}_{x}\ket{x'}\bra{x'}}{\varphi(x')}
=\frac{\ket{x'}\stackrel{\sim}{\varphi}_{x=x'}\bra{x'}}{\varphi(x')},\label{simplecollapse}
\end{eqnarray}
the result is more easily comparable to L\"{u}ders strong collapse rule and the QBR. Note the tilde on $\stackrel{\sim}{\varphi}_{x}$ indicates that it is the appropriate prior for inference as the state has decohered in the detector. Although $\stackrel{\sim}{\varphi}_{x=x'}=\hat{\varphi}_{x=x'}=\varphi(x')$ are numerically equal probabilities, substitution of this above is incorrect because $\hat{\varphi}_{x}$ is the state at the time prior to interacting with the measurement device. Although this is perhaps a bit fussy, it makes explicit why secure channels exist in quantum cryptography -- the statistics and dynamics of a quantum system change when it is measured ($\hat{\varphi}_x\rightarrow\stackrel{\sim}{\varphi}_{x}$) because the state must decohere before it is inferentially updated ($\stackrel{\sim}{\varphi}_{x}\rightarrow \hat{\rho}_{x}$) due to the measurement. Thus, ``wavefunction collapse" is really decoherence and then collapse of the mixed state. Note that this result is completely consistent with the use of Bayesian inference to update probability distributions upon the observation of data in QM, i.e., $\varphi(x)=|\psi(x)|^2\rightarrow \rho(x)=\delta_{xx'}$.

Equation (\ref{simplecollapse}) is the special case of the QBR (\ref{Qbayes}) when the measurements are projective. We call this inference a ``simple collapse" and reserve the title of QBR for the general result that will be derived later. Note that this derivation does not require first solving for the ``weak" collapse and taking the limit \cite{Hellmann}. This is because \cite{QRE} gives the general solution to $\hat{\rho}$ (equation (\ref{rhosolution})) when the constraints are linear in $\hat{\rho}$.

\paragraph{Simple Partial (or Uncertain) Collapse:}
A state may partially collapse if after detection the state still has a certain probability of being in one state or another, due to measurement uncertainty. This leads to a quantum analog of Jeffreys rule; however, we shall reserve the title of the ``Quantum Jeffreys Rule" for the general result derived later. Given the similarly prepared prior density matrix $\stackrel{\sim}{\varphi}_{x}$, we maximize the entropy with respect to a set of constraints $\{\rho_D(x)=\mbox{Tr}(\ket{x}\bra{x}\hat{\rho}_{x})\}$ to codifying a lack of certainty in the final state outcome (perhaps a narrow Gaussian distribution rather than exact knowledge in (\ref{delta})).  Maximizing the entropy with respect to these constraints and normalization again gives the posterior,
\begin{eqnarray}
\hat{\rho}_{x}=\frac{1}{Z}\sum_{x}\varphi(x)e^{\alpha_{x}}\ket{x}\bra{x},
\end{eqnarray}
and normalization implies,
\begin{eqnarray}
Z=\mbox{Tr}(\hat{\rho}_{x})=\sum_{x}\varphi(x)e^{\alpha_{x}}.
\end{eqnarray}
Satisfying the remaining expectation value constraint ($\rho_D(x)=\mbox{Tr}(\ket{x}\bra{x}\hat{\rho}_{x})$) gives $e^{\alpha_{x}}=Z\frac{\rho_D(x)}{\varphi(x)}$ for each $x$, and therefore,
\begin{eqnarray}
\hat{\rho}_{x}=\sum_{x}\rho_D(x)\ket{x}\bra{x}=\sum_{x'}\rho_D(x')\frac{\ket{x'}\stackrel{\sim}{\varphi}_{x=x'}\bra{x'}}{\varphi(x')}=\sum_{x'}\rho_D(x')\stackrel{\sim}{\varphi}_{x|x'},
\end{eqnarray}
which is a ``partial" or ``uncertain"  collapse, reproducing \cite{Hellmann,Kostecki}. 

\paragraph{The Appropriate Joint Prior for the QBR:}
The previous sections derive the collapse of a single simple system from the quantum maximum entropy method. There, the appropriate prior density matrix was generated from standard decoherence arguments in QM. Here, the inference is over a joint entangled system, and therefore, the appropriate prior density matrix is itself a joint prior density matrix. The form of the joint prior density matrix for QBR depends on the physical interactions between the system of interest, the pointer variable, and later the detector used on the pointer variable. Thus, the appropriate joint prior density matrix is generated from standard methods in QM.

 Notice that if $\hat{\varphi}_{\theta}$ is an $M$th order biased prior, then,
\begin{eqnarray}
\hat{\varphi}_{x,\theta}=U( \ket{0}_x\bra{0}\otimes\hat{\varphi}_{\theta})U^{\dag}=\sum_{x,x'}\ket{x}\bra{x'}\otimes  (A_{x}\hat{\varphi}_{\theta}A^{\dag}_{x'}),\label{3.39}
\end{eqnarray}
is an $M$th order biased prior, meaning that $\hat{\varphi}_{x,\theta}$ can only be inferentially updated in that subspace (which may or may not be desirable). This is potentially problematic if $M=1$ because $\hat{\varphi}_{x,\theta}=\hat{\varphi}_{x,\theta}^2$ is a pure state and cannot be updated non-trivially due to the PDMT. The unitary operators above represent the interaction between the pointer variable and the system of interest, which entangles them according to (\ref{entangle3.23}), and effectively generates the statistical dependencies that appear in the joint prior probability density $\varphi(x,\theta)=\varphi(x|\theta)\varphi(\theta)$ of $\hat{\varphi}_{x,\theta}$, as well as the likelihood function $\varphi(x|\theta)$.

We follow the intuition that if we are going to make inferences on the basis of detection, the prior density matrix should appropriately reflect the fact that it has interacted with a measurement device. This interaction will be modeled by entangling the pointer variable and detector states $\{\ket{d_y}\}$, which act as a local environment states within the detector, via a unitary evolution (following \cite{Schlosshauer} and the notation in \cite{Jacobs}, but a simple projection argument from L\"uders on the pointer variable states of $\hat{\varphi}_{x,\theta}$ would also suffice), 
\begin{eqnarray}
\ket{d_0}\bra{d_0}\otimes\hat{\varphi}_{x,\theta}\rightarrow \hat{\varphi}_{d,x,\theta}=(\hat{U}_{dx}\otimes \hat{1}_{\theta})(\ket{d_0}\bra{d_0}\otimes\hat{\varphi}_{x,\theta})(\hat{U}^{\dag}_{dx}\otimes \hat{1}_{\theta})
\end{eqnarray}
where,
\begin{eqnarray}
\hat{U}_{dx}=\sum_{d_y,d_{y'},x,x'}u_{d_yd_{y}xx'}\ket{d_y}\ket{x}\bra{d_{y'}}\bra{x'}=\sum_{d_y,d_{y'}}\ket{d_y}\bra{d_{y'}}\otimes B_{d_yd_{y'}},
\end{eqnarray}
and the sub-block matrices are,
\begin{eqnarray}
B_{d_yd_{y'}}=\sum_{x,x'}u_{d_yd_{y'}xx'}\ket{x}\bra{x'}.
\end{eqnarray}
We define a good detector as one in which the $\ket{x}$th pointer variable state only entangles with the local state of the detector $\ket{d_x}$, which is an argument for the sub-block matrix to take a simple form,
\begin{eqnarray}
B_{d_y0}=\sum_{x,x'}\delta_{y,x}\delta_{y,x'}\ket{x}\bra{x'}=\ket{y}\bra{y}.
\end{eqnarray}
The entangled density matrix becomes,
\begin{eqnarray}
\hat{\varphi}_{d,x,\theta}=\sum_{y,y',x,x'}\ket{d_y}\bra{d_{y'}}\otimes(B_{y'0}\otimes I_{\theta})\hat{\varphi}_{x,\theta}(B_{y'0}^{\dag}\otimes I_{\theta})\nonumber\\
=\sum_{y,y'}\ket{d_y}\bra{d_{y'}}\otimes\ket{y}\bra{y'}\otimes  (A_{y}\stackrel{\sim}{\varphi}_{\theta}A^{\dag}_{y'}).
\end{eqnarray}
The local environment of the detector states in which the pointer variable reside, are traced over, as we do not keep track of their evolution.  This is to say, a small period of time after the projective measurement has been made, the pointer variable states transition to a mixed state, which gives a standard (classical) probability distribution of the pointer variable states over the detector. The prior density matrix after projective measurement has been made is thus a block diagonal sum of states,
\begin{eqnarray}
\stackrel{\sim}{\varphi}_{x,\theta}(t)=\mbox{Tr}_{d}(\hat{\varphi}_{d,x,\theta})=\sum_x\ket{x}\bra{x}\otimes  (A_{x}\hat{\varphi}_{\theta}A^{\dag}_{x})=\sum_{x'}\ket{x'}\bra{x'}\otimes  \stackrel{\sim}{\varphi}_{x=x',\theta},
\end{eqnarray}
which we claim is the appropriate density matrix for POVM inference. This form of the prior is no longer biased, even if $\hat{\varphi}_{\theta}$ was itself biased. 


\paragraph{The constraints leading to the QBR:}
In principle, when the result of a projective measurement on the pointer variable is registered, the state of the pointer variable is known with certainty. This is represented by the following constraint on the posterior probability distribution,
\begin{eqnarray}
\rho(x)=\mbox{Tr}(\ket{x}\bra{x}\otimes\hat{1}_{\theta}\cdot\hat{\rho}_{x,\theta})=\delta_{xx'},\label{delta2}
\end{eqnarray}
which resembles (\ref{data}) (the symbol $\cdot$ is the standard matrix product). Notice that this information alone is not enough to fully constrain $\hat{\rho}_{x,\theta}$ as there are many $\hat{\rho}_{x,\theta}$ which satisfy this constraint. We therefore employ the quantum maximum entropy method and constrain the posterior subject to normalization and the data constraint, and with respect to the appropriate prior $\stackrel{\sim}{\varphi}_{x,\theta}$,
\begin{eqnarray}
0=\delta\Big( S-\lambda[\mbox{Tr}(\hat{\rho}_{x,\theta})-1]-\sum_x\alpha_x[\mbox{Tr}(\ket{x}\bra{x}\otimes\hat{1}_{\theta}\cdot\hat{\rho}_{x,\theta})-\delta_{xx'}]\Big),
\end{eqnarray}
which gives,
 \begin{eqnarray}
\hat{\rho}_{x,\theta}=\frac{1}{Z}\exp\Big(\sum_x\alpha_x\ket{x}\bra{x}\otimes\hat{1}_{\theta}+\log(\stackrel{\sim}{\varphi}_{x,\theta})\Big).
\end{eqnarray}
Because the prior density matrix is block diagonal $\log(\stackrel{\sim}{\varphi}_{x,\theta})=\sum_x\ket{x}\bra{x}\otimes\log(A_{x}\hat{\varphi}_{\theta}A^{\dag}_{x})\equiv\sum_x\ket{x}\bra{x}\otimes\log(\stackrel{\sim}{\varphi}_{x=x,\theta})$ we have,
\begin{eqnarray}
\hat{\rho}_{x,\theta}=\frac{1}{Z}\sum_x\ket{x}\bra{x}\otimes e^{\alpha_x\hat{1}_{\theta}+\log(\stackrel{\sim}{\varphi}_{x=x,\theta})}=\frac{1}{Z}\sum_xe^{\alpha_x}\ket{x}\bra{x}\otimes\stackrel{\sim}{\varphi}_{x=x,\theta}.
\end{eqnarray}
Imposing normalization gives,
\begin{eqnarray}
Z=\mbox{Tr}\Big(\sum_xe^{\alpha_x}\ket{x}\bra{x}\otimes\stackrel{\sim}{\varphi}_{x=x,\theta}\Big)=\sum_xe^{\alpha_x}\sum_{\theta}\bra{\theta}\stackrel{\sim}{\varphi}_{x=x,\theta}\ket{\theta}=\sum_xe^{\alpha_x}\stackrel{\sim}{\varphi}(x).
\end{eqnarray}
The data expectation value constraint forces,
\begin{eqnarray}
\delta_{xx'}=\frac{1}{Z}\mbox{Tr}(\ket{x}\bra{x}\otimes\hat{1}_{\theta}\,\cdot\,\sum_x e^{\alpha_x}\ket{x}\bra{x}\otimes\stackrel{\sim}{\varphi}_{x=x,\theta})\nonumber\\
=\frac{1}{Z}\mbox{Tr}(e^{\alpha_x}\ket{x}\bra{x}\otimes\stackrel{\sim}{\varphi}_{x=x,\theta})=\frac{e^{\alpha_x}}{Z}\sum_{\theta}\bra{\theta}\stackrel{\sim}{\varphi}_{x=x,\theta}\ket{\theta}=\frac{e^{\alpha_x}}{Z}\stackrel{\sim}{\varphi}(x),
\end{eqnarray}
meaning, $e^{\alpha_x}=\frac{Z\delta_{xx'}}{\stackrel{\sim}{\varphi}(x)}$. Substituting in for the Lagrange multipliers gives the final form of the posterior density matrix ,
\begin{eqnarray}
\hat{\rho}_{x,\theta}=\sum_x\frac{\delta_{xx'}}{\stackrel{\sim}{\varphi}(x)}\ket{x}\bra{x}\otimes\stackrel{\sim}{\varphi}_{x=x,\theta}=\frac{\ket{x'}\bra{x'}\otimes\stackrel{\sim}{\varphi}_{x=x',\theta}}{\stackrel{\sim}{\varphi}(x')}.\label{QBRjoint}
\end{eqnarray}
The marginal posterior is the Quantum Bayes Rule,
\begin{eqnarray}
\hat{\rho}_{\theta}=\mbox{Tr}_{x}(\hat{\rho}_{x,\theta})=\frac{\stackrel{\sim}{\varphi}_{x=x',\theta}}{\stackrel{\sim}{\varphi}(x')}\equiv \stackrel{\sim}{\varphi}_{\theta|x'},\label{QBayes3}
\end{eqnarray}
which is equivalent to the standard POVM measurement formulation, 
\begin{eqnarray}
\hat{\rho}_{\theta}\,=\frac{A_{x'}\hat{\varphi}_{\theta}A^{\dag}_{x'}}{\varphi(x')},
\end{eqnarray}
 because $\stackrel{\sim}{\varphi}_{x=x',\theta}\,=A_{x'}\hat{\varphi}_{\theta}A^{\dag}_{x'}$ and $\stackrel{\sim}{\varphi}(x')=\varphi(x')$. The posterior probability of $\theta$ indeed gives the standard Bayes Rule,
  \begin{eqnarray}
\rho(\theta)=\mbox{Tr}(\hat{\rho}_{\theta}\ket{\theta}\bra{\theta})=\mbox{Tr}(\stackrel{\sim}{\varphi}_{\theta|x'}\ket{\theta}\bra{\theta})=\varphi(\theta|x')=\frac{\varphi(\theta,x')}{\varphi(x')}.
\end{eqnarray}
As stated in \cite{Jordan}, the off diagonal elements, $\bra{\theta}\hat{\rho}_{\theta}\ket{\theta'}$, have a more exotic updating rule. One may make further inferences about $\hat{\rho}_{\theta}$ in (\ref{QBayes3}), for instance, an inference leading to its ``simple collapse" (\ref{simplecollapse}).\\
\paragraph{Quantum Jeffreys Rule (QJR):} In the same way as before, we may easily generalize this rule to cases in which the final state of the pointer variable is uncertain and encoded by a probability distribution $\rho_D(x)$ rather than one exhibiting certainty $\delta_{xx'}$. Simply replacing the expectation value constraint (\ref{delta2}) by,
\begin{eqnarray}
\mbox{Tr}(\ket{x}\bra{x}\otimes\hat{1}_{\theta}\cdot\hat{\rho}_{x,\theta})=\rho_D(x),
\end{eqnarray}
and performing the quantum maximum entropy method gives the marginal posterior,
\begin{eqnarray}
\hat{\rho}_{\theta}=\mbox{Tr}_{x}(\hat{\rho}_{x,\theta})=\sum_x\rho_D(x) \stackrel{\sim}{\varphi}_{\theta|x}=\sum_x\rho_D(x)\frac{A_{x}\hat{\varphi}_{\theta}A^{\dag}_{x}}{\varphi(x)},
\end{eqnarray}
which we call the Quantum Jeffreys Rule.

\paragraph{What about inferring $x$ from $\theta$ measurements? } 
Above we have derived the QBR and the QJR from the quantum maximum entropy method where it was assumed that a density matrix in $\mathcal{H}_{\theta}$ is to be inferred from detections of $x$'s. It is possible, however, to consider the odd situation of detecting $\theta$'s and making inferences about the pointer variables in $\mathcal{H}_x$. Through this computation we find that the resulting joint probability distribution $\rho(x,\theta)$ of $\hat{\rho}_{x,\theta}$ is independent of the order in which $x$'s and $\theta$'s are detected, which is in agreement with results of the delayed choice experiment.

Consider rewriting a similarly entangled pointer variable prior density matrix $\hat{\varphi}_{x,\theta}$ from (\ref{3.39}) as,
\begin{eqnarray}
\hat{\varphi}_{x,\theta}=\sum_{x,x'}\ket{x}_x\bra{x'}\otimes (A_{x}\hat{\varphi}_{\theta}A^{\dag}_{x'})=\sum_{\theta,\theta'}(A_{\theta}\hat{\vartheta}_xA^{\dag}_{\theta'})\otimes\ket{\theta}_{\theta}\bra{\theta'},
\end{eqnarray}
where the $A_{\theta}$'s are a bit messy but obtained from moving and relabeling the components of the unitary matrices. If another projection measurement device is designed to detect $\theta$ states, rather than $x$ states, an analogous argument can be used to decohere the joint prior density matrix in the $\theta$'s,
 \begin{eqnarray}
\hat{\varphi}_{x,\theta}\rightarrow\,\stackrel{\sim}{\vartheta}_{x,\theta}=\sum_{\theta}(A_{\theta}\hat{\vartheta}_xA^{\dag}_{\theta})\otimes\ket{\theta}_{\theta}\bra{\theta},
\end{eqnarray}
rather than the $x$'s.  It should be noted that in general $\stackrel{\sim}{\vartheta}_{x,\theta}\neq\stackrel{\sim}{\varphi}_{x,\theta}$, from the previous section, as they are block diagonal in different Hilbert spaces -- this joint prior density matrix is block diagonal in $\mathcal{H}_{\theta}$. The same analysis from the previous section is made: using $\stackrel{\sim}{\vartheta}_{x,\theta}$ as the prior, maximize the entropy with respect to normalization and the $\theta$ data constraint,
\begin{eqnarray}
\mbox{Tr}(\hat{1}_x\otimes\ket{\theta}\bra{\theta}\cdot\hat{\varrho}_{x,\theta})=\delta_{\theta\theta'},
\end{eqnarray}
and solve for the Lagrange multipliers. This gives
 \begin{eqnarray}
\hat{\varrho}_{x,\theta}=\frac{\stackrel{\sim}{\vartheta}_{x,\theta=\theta'}}{\vartheta(\theta')}\otimes\ket{\theta'}\bra{\theta'},\label{varrhojoint}
\end{eqnarray}
such that the marginal posterior is
\begin{eqnarray}
\hat{\varrho}_{x}\equiv \stackrel{\sim}{\vartheta}_{x|\theta'}=\mbox{Tr}_{\theta}(\hat{\varrho}_{x,\theta})=\frac{\stackrel{\sim}{\vartheta}_{x,\theta=\theta'}}{\vartheta(\theta')}\,,
\end{eqnarray}
which is equivalent to,
\begin{eqnarray}
\hat{\varrho}_{x}=\frac{A_{\theta'}\hat{\vartheta}_{x}A^{\dag}_{\theta'}}{\vartheta(\theta')},
\end{eqnarray}
and is interpreted as the posterior density matrix of the pointer variable after a complementary $\hat{A}_{\theta'}$ measurement operator has been applied and $\theta'$ has been detected. One may make further inferences on $\hat{\varrho}_{x}$, for instance, an inference leading to its ``simple collapse" (\ref{simplecollapse}).

 The posterior density matrices, $\hat{\varrho}_x=\,\stackrel{\sim}{\vartheta}_{x|\theta'}$ and $\hat{\rho}_{\theta}=\, \stackrel{\sim}{\varphi}_{\theta|x'}$, are related in the following way: Notice that although $\stackrel{\sim}{\vartheta}_{x,\theta}\neq\stackrel{\sim}{\varphi}_{x,\theta}$ in general, their components along the diagonal-diagonal are equal, $\bra{x,\theta}\stackrel{\sim}{\vartheta}_{x,\theta}\ket{x,\theta}=\bra{x,\theta}\stackrel{\sim}{\varphi}_{x,\theta}\ket{x,\theta}=\bra{x,\theta}\hat{\varphi}_{x,\theta}\ket{x,\theta}$, because both density matrices are decohered in one way or another from equivalently entangled prior density matrices $\hat{\varphi}_{x,\theta}$. This means
\begin{eqnarray}
\varphi(x,\theta)=\vartheta(x,\theta),
\end{eqnarray}
and because $\varphi(x)=\vartheta(x)$ and $\varphi(\theta)=\vartheta(\theta)$,
\begin{eqnarray}
\varphi(\theta|x)=\frac{\vartheta(x,\theta)}{\varphi(x)}=\frac{\vartheta(x,\theta)}{\vartheta(x)}=\frac{\vartheta(x|\theta)\vartheta(\theta)}{\vartheta(x)}=\vartheta(\theta|x),
\end{eqnarray}
and likewise $\vartheta(x|\theta)=\varphi(x|\theta)$, we see all of the probability relationships hold and may be used interchangeably. It should also be noted that in general: $\stackrel{\sim}{\vartheta}_{x}\neq\stackrel{\sim}{\varphi}_{x}$ and $\stackrel{\sim}{\vartheta}_{\theta}\neq\stackrel{\sim}{\varphi}_{\theta}$ because their off diagonal components may differ. The joint posterior density matrices $\stackrel{\sim}{\vartheta}_{x,\theta}$ and $\stackrel{\sim}{\varphi}_{x,\theta}$, and their posterior marginals potentially differ in how they will evolve in time. Because the use of appropriate measurement devices leads to $\varphi(x,\theta)=\vartheta(x,\theta)$, there is no interpretational issue in the delayed choice experiment because collapse of both $x$ and $\theta$ only occurs after decoherence of both $x$ and $\theta$. The time order of the decoherence becomes irrelevant because the joint probabilities are equal. Essentially what has happened in the delayed choice experiment is that you do not know if you have done a ``which slit" measurement or not, which is like having a ``mixed state of potential measurement outcomes", but, this is precisely what a POVM measurement represents.  

%
%


\paragraph{Quantum Jeffreys Rule via thermal baths:}
Rather than detecting the result of a projective measurement on the pointer variable state, we consider the posterior density matrix one would obtain if the pointer variable is sent into a thermal box -- the result can be naturally generated in the quantum maximum entropy method. We will see inferences of this type allow the experimenter some control over the distribution of the final states of $\theta$ by adjusting the initial temperature of the thermal box that will be used to capture the pointer particle. Here we will let the Hilbert space $\mathcal{H}_x$ of the pointer variable be spanned by $\{\ket{n}\}$, the energy basis eigenstates of the pointer variable in the thermal box having a Hamiltonian $\hat{H}_{n}=\sum_n\epsilon_n\ket{n}\bra{n}$. The joint prior density matrix is prepared similar to how it was prepared above, where $\stackrel{\sim}{\varphi}_{x,\theta}\equiv \stackrel{\sim}{\varphi}_{n,\theta}$, such that,
\begin{eqnarray}
\stackrel{\sim}{\varphi}_{n,\theta}=\sum_{n'}\ket{n'}\bra{n'}\otimes  \stackrel{\sim}{\varphi}_{n=n',\theta}.
\end{eqnarray}
 The following energy expectation value is used to represent the constraint of a pointer variable in a thermal box,
\begin{eqnarray}
\mbox{Tr}(\hat{H}_{n}\otimes\hat{1}_{\theta}\cdot\hat{\rho}_{n,\theta})=\expt{\hat{H}_{n}}.
\end{eqnarray}
Again notice that this information alone is not enough to fully constrain $\hat{\rho}_{n,\theta}$ as there are many $\hat{\rho}_{n,\theta}$ which satisfy this constraint. We therefore employ the quantum maximum entropy method; that is, to maximize the quantum relative entropy subject to normalization, this constraint, and with respect to the prior $\stackrel{\sim}{\varphi}_{n,\theta}$,
\begin{eqnarray}
0=\delta\Big( S-\lambda[\mbox{Tr}(\hat{\rho}_{n,\theta})-1]-\beta[\mbox{Tr}(\hat{H}_{n}\otimes\hat{1}_{\theta}\cdot\hat{\rho}_{n,\theta})-\expt{H_n}]\Big),
\end{eqnarray}
which gives,
 \begin{eqnarray}
\hat{\rho}_{n,\theta}=\frac{1}{Z}\exp\Big(\beta\hat{H}_{n}\otimes\hat{1}_{\theta}+\log(\stackrel{\sim}{\varphi}_{n,\theta})\Big).
\end{eqnarray}
Because the prior density matrix is block diagonal $\log(\stackrel{\sim}{\varphi}_{n,\theta})=\sum_{n}\ket{n}\bra{n}\otimes\log(A_{n}\hat{\varphi}_{\theta}A^{\dag}_{n})=\sum_{n}\ket{n}\bra{n}\otimes\log(\stackrel{\sim}{\varphi}_{n=n,\theta})$ we have that,
\begin{eqnarray}
\hat{\rho}_{n,\theta}=\frac{1}{Z}\sum_{n}\ket{n}\bra{n}\otimes e^{\beta\epsilon_n\hat{1}_{\theta}+\log(\stackrel{\sim}{\varphi}_{n=n,\theta})}=\frac{1}{Z}\sum_{n}e^{\beta \epsilon_n}\ket{n}\bra{n}\otimes\stackrel{\sim}{\varphi}_{n=n,\theta}.
\end{eqnarray}
Imposing normalization gives,
\begin{eqnarray}
Z=\mbox{Tr}\Big(\sum_{n}e^{\beta \epsilon_n}\ket{n}\bra{n}\otimes\stackrel{\sim}{\varphi}_{n=n,\theta}\Big)=\sum_n e^{\beta \epsilon_n}\sum_{\theta}\bra{\theta}\stackrel{\sim}{\varphi}_{n=n,\theta}\ket{\theta}=\sum_{n} e^{\beta \epsilon_n}\stackrel{\sim}{\varphi}(\epsilon_n).
\end{eqnarray}
The expectation value constraint forces,
\begin{eqnarray}
\expt{H_{n}}=\mbox{Tr}(\hat{H}_{n}\otimes\hat{1}_{\theta}\cdot\hat{\rho}_{n,\theta})=\frac{1}{Z}\mbox{Tr}(\hat{H}_{n}\otimes\hat{1}_{\theta}\,\,\sum_{n}e^{\beta \epsilon_n}\ket{n}\bra{n}\otimes\stackrel{\sim}{\varphi}_{n=n,\theta})\nonumber\\
=\frac{1}{Z}\mbox{Tr}(\sum_{n}\epsilon_n e^{\beta \epsilon_n}\ket{n}\bra{n}\otimes\stackrel{\sim}{\varphi}_{n=n,\theta})=\sum_{n}\frac{\epsilon_n e^{\beta \epsilon_n}}{Z}\stackrel{\sim}{\varphi}(\epsilon_n)=\frac{\partial }{\partial \beta}\log (Z),
\end{eqnarray}
meaning one can solve $\beta=\beta(\expt{H_{n}})=-\frac{1}{kT}$ by inverting the above equation after computing $Z$ as is done in Statistical Mechanics. The marginal posterior is a realization of the Quantum Jeffreys Rule using thermalization,
\begin{eqnarray}
\hat{\rho}_{\theta}&=&\mbox{Tr}_{n}(\hat{\rho}_{n,\theta})=\sum_{n}\frac{e^{\beta \epsilon_n}}{Z}\stackrel{\sim}{\varphi}_{n=n,\theta}=\sum_{n}\frac{\varphi(\epsilon_n)e^{\beta \epsilon_n}}{Z}\stackrel{\sim}{\varphi}_{\theta|n}=\sum_{n}\rho(\epsilon_n)\stackrel{\sim}{\varphi}_{\theta|n}\nonumber\\
&=&\sum_{n}\rho(\epsilon_n)\frac{A_{n}\stackrel{\sim}{\varphi}_{\theta}A^{\dag}_{n}}{\varphi(\epsilon_n)},
\end{eqnarray}
in which the state $\hat{\rho}_{\theta}$ of the system may be controlled by forcing the pointer variable into a box with temperature $\beta$ or $\beta'$ as the resulting joint posterior density matrices differ. In particular, if $\beta \ll 0$ this implies $T\approx 0$ and therefore $\hat{\rho}_{\theta}\approx \frac{A_{0}\stackrel{\sim}{\varphi}_{\theta}A^{\dag}_{0}}{\varphi(\epsilon_0)}$ with $n=0$ being the ground state energy of the particles in the thermal box. This is similar in nature to a postselection measurement scheme from \cite{AAV}, which will be reviewed in Chapter \ref{Measurement in Entropic Dynamics}, except here we used a thermal box and the quantum maximum entropy.

\section{Generalizations}
General inferences of $\hat{\rho}$ on the basis of a prior state of knowledge $\hat{\varphi}$ and arbitrary expectation value constraints $\{\expt{\hat{A}_i}\}$ give the following general updating rule,
\begin{eqnarray}
\hat{\rho}=\frac{1}{Z}\exp\Big(\sum_i\alpha_i\hat{A}_i+\ln(\hat{\varphi})\Big),\label{gensolhere}
\end{eqnarray}
from $\hat{\varphi}\rightarrow \hat{\rho}$ in light of new information about $\{\expt{\hat{A}_i}\}$. This is of-course the general solution to the quantum maximum entropy method, but now it is clear it may be interpreted as the solution for general purpose inference in QM when applied correctly.
\subsection{Non-commutativity for Simultaneous and Sequential Updates}
The constraints and priors used in the derivation of the QBR and QJR factored in the exponential because they took the form $\exp\Big(\hat{A}\otimes\hat{1}+\hat{1}\otimes\hat{B}\Big)=\exp(\hat{A})\otimes\exp(\hat{B})$, in part because of commutation of $[\hat{A}\otimes\hat{1},\hat{1}\otimes\hat{B}]$, even when the individual operators $[\hat{A},\hat{B}]\neq0$ do not commute themselves. Thus general inferences involving expectation values of non-commuting operators and prior density matrices generalizes these rules and the solution is found by diagonalizing the exponential and using convex optimization methods (i.e. the general solutions for (\ref{gensolhere})). Perhaps the simplest example of such a situation is the mixed spin state example given in Appendix \ref{spinexampleQRE}.


The above inferences have all been instances of simultaneous updating, so a note should be made on sequential updating. Consider updating $\hat{\varphi}$ with respect to a set of constraints ($*1$) such that $\hat{\varphi}\stackrel{*1}{\rightarrow}\hat{\rho}^1$ via the quantum maximum entropy method. If one receives new information, ($*2$), and we consider $\hat{\rho}^1$ to be the new prior, we may update $\hat{\rho}^1\stackrel{*2}{\rightarrow}\hat{\rho}^{1,2}$. These sequential updates give the following inference chain $\hat{\varphi}\stackrel{*1}{\rightarrow}\hat{\rho}^1\stackrel{*2}{\rightarrow}\hat{\rho}^{1,2}$. If one were instead to apply these constraints in an opposite order, even if the operators in ($*1$) and ($*2$) commute, in general $\hat{\varphi}\stackrel{*2}{\rightarrow}\hat{\rho}^2\stackrel{*1}{\rightarrow}\hat{\rho}^{2,1}\neq\hat{\rho}^{1,2}$. The posterior $\hat{\rho}^{1,2}$ is guaranteed to satisfy 2, but not 1, whereas the posterior $\hat{\rho}^{2,1}$ is guaranteed to satisfy 1 but not 2.  The third case is the simultaneous update  using both pieces of information $(*1)(*2)=(*3)$. This gives $\hat{\varphi}\stackrel{*3}{\rightarrow}\hat{\rho}^3$ in one update, which again is not equal to $\hat{\rho}^{1,2}$ or $\hat{\rho}^{2,1}$ in general as it is guaranteed to satisfy both constraints 1 and 2 simultaneously. This type of non-commutation appears also in the standard maximum entropy method \cite{Giffin,GiffinThesis,book}, so its manifestation in the quantum maximum entropy method is not surprising. The conclusion using the quantum maximum entropy method is the same; sequential and simultaneous updating ultimately refer to different states of knowledge, and therefore one should expect the entropic method to give different results. The fact that the applied order of constraints does not in general commute is therefore a \emph{feature} \cite{Giffin} of the quantum maximum entropy method -- if new information rules out old expectation value constraints, then sequential updating to $\hat{\rho}^{1,2}$ or $\hat{\rho}^{2,1}$ is appropriate, whereas, if both constraints are expected to hold in the posterior, then one should simultaneously update the state to $\hat{\rho}^3$.  

\subsection{Canonically Modified Quantum Bayes Rule}

Finally, to the joint prior density matrix $\stackrel{\sim}{\varphi}_{x,\theta}=\sum_x\ket{x}_x\bra{x}\otimes(A_x\hat{\varphi}_{\theta}A_x^{\dag})$ one may simultaneously apply a data constraint to the pointer variable (that would result in a Quantum Bayes Rule) while simultaneously constraining the $\theta$ space with respect to an expectation value $\expt{f(\hat{\theta})}$. This results in a state that cannot alone be inferred from either a von Neumann maximum entropy procedure or a Quantum Bayes procedure alone. The resulting state is, analogous to its probabilistic form in \cite{Giffin}, a canonically modified Quantum Bayes Rule. Maximizing the quantum relative entropy with respect to these constraints and normalization gives,
\begin{eqnarray}
\hat{\rho}_{x,\theta}=\frac{1}{Z}\sum_x\ket{x}\bra{x}\otimes e^{\alpha_x\hat{1}_{\theta}+\beta f(\hat{\theta})+\log(\stackrel{\sim}{\varphi}_{x=x,\theta})}=\frac{1}{Z}\sum_xe^{\alpha_x}\ket{x}\bra{x}\otimes e^{\beta f(\hat{\theta})+\log(\stackrel{\sim}{\varphi}_{x=x,\theta})}.
\end{eqnarray}
Imposing the data constraint, $\rho_D(x)=\mbox{Tr}_{\theta}(\ket{x}\bra{x}\otimes\hat{1}_{\theta}\cdot\hat{\rho}_{x,\theta})$, gives,
\begin{eqnarray}
\rho_D(x)=\frac{e^{\alpha_x}}{Z}\mbox{Tr}\Big(\exp(\beta f(\hat{\theta})+\log(\stackrel{\sim}{\varphi}_{x=x,\theta}))\Big),
\end{eqnarray}
which means the data Lagrange multiplier is,
\begin{eqnarray}
e^{\alpha_x}=\frac{Z\rho_D(x)}{\mbox{Tr}\Big(\exp(\beta f(\hat{\theta})+\log(\stackrel{\sim}{\varphi}_{x=x,\theta}))\Big)}\equiv\frac{Z\rho_D(x)}{\zeta(x,\beta)}.
\end{eqnarray}
We therefore arrive at a canonically modified Quantum Jeffreys Rule,
\begin{eqnarray}
\hat{\rho}_{\theta}=\mbox{Tr}_{x}(\hat{\rho}_{x,\theta})=\sum_x\frac{\rho_D(x)}{\zeta(x,\beta)}\exp\Big(\beta f(\hat{\theta})+\log(\stackrel{\sim}{\varphi}_{x=x,\theta})\Big).
\end{eqnarray}
The special case, if $\rho_D(x)=\delta_{xx'}$, gives a canonically modified Quantum Bayes Rule,
\begin{eqnarray}
\hat{\rho}_{\theta}=\mbox{Tr}_{x}(\hat{\rho}_{x,\theta})=\frac{ \exp\Big(\beta f(\hat{\theta})+\log(A_{x'}\hat{\varphi}_{\theta}A_{x'}^{\dag})\Big)}{\zeta(x',\beta)}.
\end{eqnarray}
If the expectation value constraint is not imposed, or if $\beta=0$, then the canonical factors drop out and the Quantum Bayes Rule is recovered. Note that the values of $\expt{f(\hat{\theta})}$ are mutually compatible with the data constraint and the prior as to guarantee a logical solution -- this logical restriction is present equally well in the canonically modified Bayes Rule \cite{Giffin}.


\section{Conclusions}
In this chapter we applied the quantum maximum entropy method and derived the L\"{u}ders collapse (and partial collapse) rules, the QBR, and the QJR. Furthermore, it was discussed why the order of sequential updates do not commute in general, even when the operators within the sequentially implemented expectation value constraints do. Having derived a Quantum Bayes Rule from the quantum maximum entropy method, the suggestion of such a possibility in \cite{Carlo} is met. The subject matter reviewed in this chapter therefore unifies topics in Quantum Information and quantum measurement through quantum entropic inference just as much as Caticha and Giffin unified topics in information theory and Bayesian inference through standard  entropic inference \cite{GiffinBayes,Giffin,GiffinThesis}.

The article \cite{QBR} shows how the quantum maximum entropy method can eliminate \emph{ad hoc} collapse postulates in QM, in agreement with \cite{Hellmann,Kostecki}; however decoherence is still required. As is demonstrated by the arguments leading up to the PDMT and as the derivation of the above inference rules have showed, the phrase ``collapse of the wavefunction" should be replaced by ``decoherence and the collapse of the mixed state". This is in agreement with L\"{u}ders' notion that the application of a measurement device is to mix the incoming state $\hat{\rho}\rightarrow \sum_{i}\hat{P}_i\hat{\rho}\hat{P}_i$ if collapse is going to occur. In this sense, a pure state is \textit{already} a collapsed state, so trying to directly collapse this state to a different pure state is illogical (by the PDMT) unless the state first decoheres -- ``uncollapsing" the state.

The main conclusion from \cite{GiffinBayes,Giffin,GiffinThesis} is that the \emph{standard} maximum entropy method is the ``universal method for inference" because it can reproduce Bayesian inference as a special case, while also being able to make inferences not achievable through Bayesian inference or maximum entropy inference (having prior's set to 1 in the relative entropy) -- the method updates $\varphi(x)\stackrel{*}{\rightarrow}\rho(x)$ for arbitrary $(*)$. The main conclusion from this chapter is analogous -- the quantum maximum entropy method is the ``universal method of density matrix inference", the method updates $\hat{\varphi}\stackrel{*}{\rightarrow}\hat{\rho}$ for arbitrary $(*)$. Although pure state prior density matrices fail to update $\hat{\varphi}\stackrel{*}{\rightarrow}\hat{\varphi}$ due to the PDMT, there is no contradiction, because one may still correctly unitarily evolve pure state priors $\hat{\varphi}\rightarrow U\hat{\varphi}U^{\dag}$, and so one may simply accept the non-update as one does for biased probability distributions $\varphi(x)\stackrel{*}{\rightarrow}\varphi(x)$ in the standard maximum entropy procedure.  

Because Entropic Dynamics (reviewed next chapter) is able to derive Quantum Mechanics and a density matrix formulation using the standard maximum entropy method and information geometry, it implies that the quantum maximum entropy method is really a special case of the standard maximum entropy, and so, it retains its title of the universal method of inference. The quantum maximum entropy method thus retains the more specialized title of: ``the universal method of density matrix inference". As the quantum maximum entropy method has utilized nothing but techniques from the standard quantum mechanical formalism, it, and its results, may be appended to the standard formalism.

\newpage
\addcontentsline{toc}{part}{Part II: Entropic Dynamics and the Solution to the Quantum Measurement Problem}
\part*{Part II: Entropic Dynamics and the Solution to the Quantum Measurement Problem}

\chapter{Entropic Dynamics\label{Entropic Dynamics}}

The previous chapters review the design of probability, relative entropy, and quantum relative entropy. By developing the foundations, and designing the tools of inference thereafter, we were able to make progress. In particular, we found that the standard and quantum relative entropies are \emph{designed} for the purpose of inference, each formulated from the same Principle of Minimum Updating and design criteria, and that the Quantum Bayes Rule is a special case of quantum entropic inference. We formulated the  Prior Density Matrix Theorem (PDMT), which states that a quantum system can only be inferentially updated in the regions spanned by its eigenspace, using the quantum maximum entropy method. This stems from the logic that if a pure state, or the set of them in $\hat{\rho}$, is known with certainty, than any information that would lead to a deviation from this fact is moot. Hence, the unitary evolution generated from the Schr\"{o}dinger Equation (SE), which is capable of rotating the eigenvectors of $\hat{\rho}$ out of their original eigenspace, is not in general an update that can be implemented by the quantum maximum entropy method due to the PDMT. We would like to understand unitary evolution from the point of view of inference; however, the quantum maximum entropy method cannot provide the type of understanding we seek.

The material in Chapter 2 and 3 is all predicated on the assumed existence of the standard quantum formalism. While the standard quantum formalism is great in practice, the formalism itself is empty -- there is no agreed upon interpretation of QM as it is difficult to formulate why the oddities of QM are the way they are. This is likely because in most approaches to the foundations of QM, one starts with the formalism and then appends an interpretation to it, almost as an
afterthought. Entropic Dynamics (ED) \cite{EDNew,ED,book,ED2011,ED2010,ED2009}, on the other hand, starts with the interpretation, that is, one
specifies what the ontic elements of the theory are, and only then one
develops a formalism appropriate to describe and predict those ontic elements. Laws of physics are derived as applications of standard entropic inference, and thus, ED differs from most theories in physics. 

This framework is extremely
constraining. For example, there is no room for \textquotedblleft quantum
probabilities\textquotedblright\ in Entropic Dynamics. Probabilities are neither classical nor
quantum, they are tools for reasoning that have universal applicability, as is touched upon in previous chapters.
The wavefunction should therefore
be an \emph{epistemic} object (\emph{i.e.} $|\Psi |^{2}$ is a probability) and its
time evolution -- the updating of $\Psi $ -- should not be arbitrary. Given that probability theory has universal applicability, the
dynamics of probabilities in QM should be dictated by the usual rules of inference. Entropic Dynamics is indeed able to formulate QM as an instance of standard probability updating through entropic inference and information geometry. 

It should be noted that ED is not the ``be all end all" in physics -- rather it generates models for inference that happen to be consistent with physics. Along a similar line of thought, at this point in the development of ED, ``the discoveries" are the inferential constraints and pertinent information required to obtain physics from probability theory, rather than the physics equations themselves \cite{Private}. Past, current, and future research in ED involves: reformulating other fields of physics as inference \cite{Shahid,Ipek1,Ipek3}, refining and strengthening methods in ED \cite{Nick,EDNew}, addressing the classical limit \cite{Demme}, giving a derivation of the exact renormalization group \cite{Pedro}, deriving the Black and Scholes equation \cite{Mohammad}, addressing and differentiating between the QM in ED and its Bohmian limit \cite{Dan1,Dan2}, and using ED to address measurement problems and \emph{no-go} theorems in Quantum Mechanics \cite{Johnson,EDMeasurement,EDContextuality} -- the later being the central focus of subsequent chapters. 


This chapter is a review of the ED formalism presented in \cite{ED,ED2011,Nick,EDNew}. It follows sections of \cite{EDMeasurement,EDContextuality} as well as discussions with A. Caticha \cite{Private}. The newest formulation of ED is \cite{EDNew} and the most primitive is \cite{ED2011}. Novel material appears in Section \ref{MixedED} -- it is a straightforward application of ED that produces mixed quantum states.


\paragraph{Ontological Positions and Epistemic Inferables in Entropic Dynamics:\,\,\,\,\,\,\,\,\,\,\,\,\,\,\,\,\,\,\,\,\,}
An assumption that permeates theories of physics is the existence of particles. Whether or not this assumption is ``true", it is nonetheless useful \cite{Pragmatic}.  As ED is an application of inference toward deriving laws of physics, we need something ``physical" or ``ontic" to actually make inferences about, otherwise ED may be critiqued as ungrounded. We therefore assume the existence of particles and understand them to be the primitive ontological elements we wish to make inferences about -- at least at the level of this effective theory or model. The natural follow-up question is, ``What inferences will we be making?".

 For the results of ED to be verifiable, we must be able to perform experiments, collect data, and make inferences. Given that we live in space, it is impossible to \emph{point} to a detector that is \emph{not} located in some region of space. Therefore on some level, any detection made by a detector gives some amount of positional information. Furthermore, we know a detection has taken place when our detector changes, which may be indicated to the observer in any number of convenient, and usually macroscopic, ways (digital text, needle positioning, flashing lights, signal amplification,...). The \emph{detector} is itself in principle a construction of particles, \emph{located in space}, where changes are enacted upon to reach a newly distinguished ``detector" state. Following this line of thought, and after contemplating the nature of what we might mean by a particle, define: \emph{a particle is a piece of ontology at a position}. In this sense, and like Bohmian mechanics, particle positions will play the role of \emph{beables} \cite{Bell1993} in ED.  
 

Because the observer ultimately holds the final desiderata for inference, they hold the degrees of rational belief, \emph{all verifiable quantities in Physics are inferred} and are therefore categorized as what we call \emph{inferables} \cite{EDMeasurement}. As ED is founded on probability and probability updating, Hermitian observables, as well as the wavefunction, are epistemic and do not have an ontological predisposition in ED. Standard Hermitian ``observables" are more aptly referred to as ``inferables" as their values will be inferred from positional measurements or positional correlations in a probability distribution \cite{Johnson,EDMeasurement}.\footnote{As well as other measurement quantities of interest like the complex valued, and a bit mysterious, Weak Values from \cite{AAV}.}  The thought that perhaps the only type of measurements we do in physics are position measurements, and that other quantities are inferred from position, is not original: Bell \cite{Bell1993,BellKS,Jaeger}, Feynman \cite{Feynman}, Caticha \cite{Johnson,EDMeasurement,CatichaAmplitudes,Private}, and other physicists have expressed this shared philosophy on experiment. As position is the only ontological variable, and the fact that objects other than position may be treated as epistemic \emph{inferables}, the interpretation of QM is drastically different in ED. This ends up providing a solution to the quantum measurement problem \cite{Johnson,EDMeasurement} while also not being ruled out by the Bell-Kochen-Specker \cite{BellKS,KS,Mermin} \emph{no-go} theorem (as well as the Bell \cite{BellEPR} and no $\psi$-epistemic theorem \cite{Pusey}) \cite{EDContextuality}.\footnote{The Bell-Kochen-Specker \emph{no-go} theorem rules out interpretations of QM that assign definite values to operators that simultaneously belong to noncommuting sets of internally complete sets of commuting observables.} A common interpretation of the Bell-Kochen-Specker theorem is that observables in QM are \emph{contextual}, meaning that an observable's ``character", ``aspect", or ``value" depend on the remaining set of observables it is considered along-side-with in a measurement setting.

\section{From Entropic Dynamics to Quantum Mechanics\label{Section 4.1}}  

This remainder of section 4.1 is a review of \cite{EDNew,Nick}. In the context from above, Entropic Dynamics seeks to generate useful and dynamical inferential models toward understanding the things we can infer from nature.  These models eventually take the form of Physics equations, but because they were generated from the foundation of probability and probability updating, we can claim to know ``what we needed to know" in our model to solve the problem.

Here we are interested in the constraints and assumptions required to derive QM from the first principles of inference and probability updating using Entropic Dynamics. The first step is to state the universe of discourse, the set of possible microstates or the subject matter, one would like to infer on the basis of incomplete information -- these are the ontological positions of $N$ particles in a flat Euclidean space $\mathbf{X}$ (metric $\delta _{ab}$). Our knowledge of the positions of particles is characterized by a probability density $\rho(x)$, where $x$ is a coordinate in a $3N$ dimensional configuration space $\mathbf{X}_N=\mathbf{X}\times...\times \mathbf{X}$ of particle coordinates $x=\{x^a_n\}\equiv \{x^A\}$, and where $a=1,2,3$ denotes the $a$th spatial axis of the $n$th particle's position, or for short $A=(n,a)$. From the onset, particles have \emph{definite} yet \emph{unknown} positions and are treated as the ``physical" or ``ontological" quantities of interest -- the proposition $x$ in $\rho(x)$ reads: ``the configuration space coordinate $x$ correctly represents the ontological positions of the particles in 3D space".  Expectation values of over $\rho(x)$, $\int x^n\rho(x)\,dx$, or simply integrations $\int_{a}^b\rho(x)\,dx$, are therefore integrations over the \emph{propositions} $x$ rather than the actual ontological positions of the particles \emph{since a particle can only have one position}.\footnote{Later this implies that the eigenvalues $x$ of the position operator $\hat{x}$ are themselves propositions rather than ontological particle positions.} However for brevity, we will simply call $x$ ``the position of the particles". 

The fact that positions are ``ontological" and probabilities are epistemic, immediately separates the interpretation of QM in ED from other mainstream interpretations of QM, like: the Copenhagen interpretation whose ontological values are \emph{created} by the measurement device, the Bohmian interpretation in which wavefunctions are configuration space ontological fields, and the Many Worlds interpretation that perhaps makes the \emph{largest} assumption in physics possible -- that there exists infinitely many ontological and branching universes. Our assumption is simple, ``Particles have ontological positions" and the designed features of probability updating work.

Now that the microstates have been specified, we are inclined to ask how the position of these particles change. In particular, we wish to know how probable it is for $x\rightarrow x'$, that is, we seek a transition probability of the form $P(x'|x)$ to quantify this uncertainty \emph{while being consistent with the notion that particles have definite yet unknown, ontological positions}. We therefore make the following assumptions: 1) particles move along continuous trajectories, 2) particles have a tendency to be correlated and thus undergo interparticle correlated drift based their configuration, 3) particles have a tendency to undergo uncorrelated individual drifts depending on their location in 3D space. Once the form of the transition probability $P(x'|x)$ is found, it will be used to inferentially update $\rho(x)\stackrel{*}{\rightarrow} \rho'(x')=\int P(x'|x)\rho(x)\,dx$. This crucial step is also the reason why ED naturally avoids the PDMT.


\paragraph{1) Continuous motion:}
The first assumption is implemented by making large $\Delta x^a_n=x^{'a}_n-x^a_n$ improbable. This is done by imposing that each particle have small variances, $\kappa_n$, in particle coordinates,  
\begin{equation}
\int  \Delta x_{n}^{a}\Delta x_{n}^{b} \delta _{ab}P(x'|x)\,dx'=\langle \Delta x_{n}^{a}\Delta x_{n}^{b}\rangle \delta _{ab}=\kappa
_{n},\qquad (n=1,\ldots, N)~~  \label{kappa n}
\end{equation}%
where motion is continuous in the limit $\kappa _{n}\rightarrow 0$. We use $N$ Lagrange multipliers $\alpha_n$ to impose these $N$ constraints.  The Lagrange multipliers $\alpha_n$ eventually turn out to be proportional to the masses of the particles.  

\paragraph{2) Interparticle correlation and drift:}
Interparticle correlation and drift is implemented in the following way. Letting $\phi(x)=\phi(x_1,...,x_N)$ be a scalar function over the configuration space of our $N$ particles, we design $\phi(x)$ such that its configuration space partial derivatives $\frac{\d \phi(x)}{\d x^A}\equiv\frac{\d \phi(x)}{\d x^a_n}\equiv \d_{na}\phi(x)\equiv \d_{A}\phi(x)$ regulate the \emph{expected} drift of the particles, i.e., $\phi(x)$ is a ``drift potential". We impose this regulation be distributed over the $N$ particles, and thus impose it with a single constraint over the set of particles,
\begin{equation}
\expt{\Delta\phi}=\expt{\Delta x^A} \d_A \phi(x) =\kappa ^{\prime },  \label{kappa prime}
\end{equation}%
where $\kappa ^{\prime }$ is a small constant. Designed over configuration space, $\phi(x)$ allows for interparticle correlations and drift, and eventually, entanglement. A point is made in \cite{EDNew} to note that the origins of $\phi(x)$ are unexplained, which is an interesting topic for future research.\footnote{One potential explanation is in terms of the entropy of some other microscopic ($y$) variables \cite{ED2011}.} Not giving the precise (perhaps microscopic) origins of $\phi(x)$, but still using it for the purpose of modeling probabilistic updates in the Entropic Dynamics approach, is analogized to not giving the precise microscopic origins forces, but still finding their use in Newton's law. We will use the Lagrange multiplier $\alpha'$ to impose this constraint.

\paragraph{3) Uncorrelated individual particle drift:} Imposing the first two constraints without this one leads to an interesting evolution, but richer forms of dynamics are found by further imposing this constraint. As uncorrelated individual particle drifts are unconcerned with the drifts of other particles, we introduce a field $\chi(x_n)$, with $x_n\in\mathbf{X}$ in 3D space, to regulate the expected independent drift of the $N$ particles. We further assume that the field $\chi$ can be redefined by different amounts $\gamma(x_n)$ at each location such that what is called the ``0" field value at one location may not be the ``0" field value at another location. This is a local gauge symmetry, and the way to compare field values at different locations is through a \emph{connection field} $A_a(x_n)$ that reveals how the field value at $x$ is related to the field value at $x+\Delta x$ in 3D space. The connection field is constructed such that gauge transformations in $\chi\rightarrow\chi+\gamma$ also shift the connection field by $A_a\rightarrow A_a+\d_a\gamma$, and thus $\d_a\chi-A_a$ remains invariant. The gauge invariant particle drift constraints regulate each particle's individual expected drift, 
\begin{equation}
\expt{\Delta x^A}  [ \partial_{A} \chi(x_n)-A_a(x_n) ]=\kappa ^{\prime\prime }_n,  \label{kappa prime prime}
\end{equation}
and thus we require $N$ Lagrange multipliers $\beta_n$ to impose these $N$ constraints. If $\chi(x_n)=\chi(x_n)+2\pi$ has the topology of an angle, it solves Wallstrom's objection to Nelson's stochastic mechanics \cite{Nelson,Wallstrom1,Wallstrom2,Nick}, and the Lagrange multipliers $\beta_n$ eventually turn out to be proportional to the (quantized) electric changes of the particles. This will be discussed a bit more later when it is more relevant.

%
%
%
\paragraph{Maximum Entropy:}

There are many probability distributions $P(x'|x)$ that satisfy the above expectation value constraints (\ref{kappa n} - \ref{kappa prime prime}). We therefore use the standard maximum entropy method \cite{book,Jaynes1,Jaynes2,Jaynesbook} to rank the candidate distributions. Without any prior knowledge, the prior transition distribution $Q(x'|x)$ is a very broad normalizable Gaussian distribution to encode that, given nothing is known about particle motion (equations (\ref{kappa n} - \ref{kappa prime prime}) are yet to be imposed), particles may jump anywhere with near to equal probability -- there is no reason to believe otherwise. 

Maximizing the relevant relative entropy with respect to $P(x'|x)$,
\begin{eqnarray}
S[P(x'|x),Q(x'|x)]=-\int dx' P(x'|x)\log\frac{P(x'|x)}{Q(x'|x)},\label{EDentropy}
\end{eqnarray}
subject to the expectation value constraints, (\ref{kappa n} - \ref{kappa prime prime}), and normalization, via the Lagrange multiplier method \emph{forces the probability updating scheme to evolve probabilities in a way that is consistent with the notion of particles having definite yet unknown, and ontological, positions}.\footnote{This, and the assumption that particles have ontological positions in ED, is of tantamount importance for the remainder of the discussion in this thesis.} The maximum entropy update gives,
\begin{eqnarray}
P(x'|x)=\frac{1}{Z}\exp\Big[-\frac{1}{2}\sum_n\alpha_n\delta_{ab}(\bigtriangleup x^a_n-\expt{\bigtriangleup x^a_n})(\bigtriangleup x^b_n-\expt{\bigtriangleup x^b_n})\Big],\label{4}
\end{eqnarray}
after completing the square. Because $Q(x'|x)$ is nearly constant over regions of interest, it has been absorbed into the normalization constant $Z$. The expected drift of the particles is,
\begin{eqnarray}
\expt{\Delta x^A}=\frac{1}{\alpha_n}\delta^{ab}[\alpha'\partial_{B}\, \phi(x)+\beta_n\d_{B}\chi(x_n) - A_b(x_n)],
\end{eqnarray}
with $B=(n,b)$. We will absorb $\alpha'$ as a scaling constant into $\phi(x)$ without loss of generality. A generic displacement can be expressed as the expected drift plus a fluctuation,
\begin{eqnarray}
\Delta x^a_n=\expt{\Delta x^a_n}+\Delta w^a_n,
\end{eqnarray}
respectively, where,
\begin{eqnarray}
\expt{\Delta w^a_n}=0\mbox{\tab and\tab} \expt{\Delta w^a_n\Delta w^b_n}=\frac{\delta^{ab}}{\alpha_n}.
\end{eqnarray}
One finds that for large $\alpha_n$ the dynamics is dominated by fluctuations, $\Delta w^a_n\sim \expt{\Delta w^a_n\Delta w^b_n}^{1/2}$, which are of order $O(\alpha_n^{-1/2})$ whereas the expected drifts are on the order of $O(\alpha_n^{-1})$. Large $\alpha_n$ implies small $\kappa_n$, meaning that the large $\alpha_n$ limit is the continuous limit of particle motion, and that in this limit, fluctuations dominate.   
\subsection{Entropic Time\label{Section 4.1.1}}
So far, there has been no explicit mention of time; rather, the only assumptions made are that particles have a tendency to change ontological positions and follow drift gradients. But aren't these changes in position exactly what we as observers refer as a mechanism for keeping track of time classically? We therefore introduce time as a bookkeeping parameter to index change. If we have some initial knowledge of the positions of particles $\rho(x)$, we may consider how our distribution changes once our particles have undergone a fluctuation and drift by considering,
\begin{equation}
\rho(x'|t')=\int P(x^{\prime }|x,\Delta t)\rho(x|t)\,dx.\label{step}
\end{equation}
 We have let $t$ label the distribution $\rho(x)\equiv\rho(x|t)$ and let $t'=t+\Delta t$ label the distribution $\rho(x')\equiv\rho(x'|t')$ ``after" the distribution been updated ``entropically" by $P(x^{\prime }|x,\Delta t)$ (\ref{4}). Equation (\ref{step}), and its preliminaries, are what separate Entropic Dynamics from the standard ``time independent" probability updating one is usually accustomed to seeing in the maximum entropy approach. It is natural to call equation (\ref{step}) the ``entropic dynamics update".

Before continuing, we need to introduce the concept of duration $\Delta t$ in our transition probabilities $P(x'|x,\Delta t)$. Because short steps imply short time periods, and because fluctuations $\expt{\Delta w^a_n\Delta w^b_n}=\frac{1}{\alpha_n}\delta^{ab}$ dominate for short steps ($\alpha_n\rightarrow \infty$), the notion of continuous motion must be implemented at the level of short steps. This implies the form of $\alpha_n=\frac{m_n}{\hbar}\frac{1}{\Delta t}\propto \frac{1}{\Delta t}$, where later it will be revealed that the particle specific constant  $m_n$ is the mass of the $n$th particle and the constant $\hbar$, that fixes units, is Planck's constant. We have that ``equal fluctuations" of a particle are equal measures of times.

The information metric of the transition probability in configuration space coordinates is,
\begin{eqnarray}
\gamma_{AB}=C\int dx'\, P(x'|x)\frac{\d \log P(x'|x)}{\d x^A}\frac{\d \log P(x'|x)}{\d x^B}.
\end{eqnarray}
In the limit of short steps, one finds,
\begin{eqnarray}
\gamma_{AB}=\frac{Cm_n}{\hbar\Delta t}\delta_{nn'}\delta_{ab}=\frac{Cm_n}{\hbar\Delta t}\delta_{AB},
\end{eqnarray}
which diverges as $\Delta t$ goes to zero. This is somewhat expected due to the nature of the information metric being a  measure for statistical distinguishability. In this limit $P(x'|x)$ is sharply peaked and thus $P(x'|x)$ and $P(x'|x+\Delta x)$ overlap less and are thus ``more distinguishable". If we choose the arbitrary scale constant $C$ such that it is proportional to $\Delta t$, then this metric can be recast as a ``mass tensor"
\begin{eqnarray}
m_{AB}=\frac{\hbar\Delta t}{C}\gamma_{AB}=m_n\delta_{AB}.
\end{eqnarray} 
The inverse mass tensor is therefore,
\begin{eqnarray}
m^{AB}=\frac{C}{\hbar\Delta t}\gamma^{AB}=\frac{1}{m_n}\delta^{AB}.
\end{eqnarray} 
These tensors will become particularly relevant later in the derivation. 

Using our notion of time, we may reformulate the previously defined quantities. Recasting $\expt{\Delta x^A}$ as,
\begin{eqnarray}
\expt{\Delta x^A}=b^A(x)\Delta t,
\end{eqnarray}
with 
\begin{eqnarray}
b^A(x)= m^{AB}[\hbar\partial_{B}\, (\phi+\overline{\chi})-\overline{A}_B],
\end{eqnarray}
 allows $b(x)$ to be interpreted as the drift velocity of the particle, where
\begin{eqnarray}
\overline{\chi}(x)\equiv \sum_n \beta_n\chi(x_n)~~\mbox{and}~~ \overline{A}_A(x)\equiv\hbar\beta_nA_a(x_n).
\end{eqnarray}
 The new form of,
\begin{eqnarray}
\quad \expt{\Delta w^A\Delta w^B}=\hbar m^{AB}\Delta t,\label{fluctime}
\end{eqnarray}
makes clear time has been designed such that equal measures of a particle's fluctuation occur over equal durations of time. The transition probability is,
\begin{equation}
P(x^{\prime }|x,\Delta t)=\frac{1}{Z}\exp [-\frac{1}{2}\sum_{n}\frac{m_{n}\delta_{ab}}{\hbar \Delta t}%
( \,\Delta x_{n}^{a}-\langle \Delta x_{n}^{a}\rangle ) (
\,\Delta x_{n}^{b}-\langle \Delta x_{n}^{b}\rangle ) ]~,
\label{transtime}
\end{equation}%
which specifies the entropic dynamics update in (\ref{step}). 

Equation (\ref{step}) is the integral form of the Fokker-Planck (diffusion) equation and may be recast as the differential Fokker-Planck equation (a derivation is reviewed in \cite{book}),
\begin{equation}
\partial _{t}\rho =-\d_A(\rho\, b^A)+\frac{1}{2}\hbar m^{AB}\d_A\d_B\rho=-\partial _{A}\left( \rho v^{A}\right),   \label{FP}
\end{equation}%
where the current velocity $v^{A}=b^A+u^A$ is the current = drift $+$ osmotic velocities of the probability flow in configuration space, respectively. Specifically:
\begin{equation}
v^{A}= m^{AB}(\partial _{B}\Phi-\overline{A}_B) \quad
\mbox{and}%
\quad \Phi =\hbar(\phi +\overline{\chi} - \log \rho ^{1/2}),  \label{curr}
\end{equation}
is a function defined in terms of previously defined variables. In this sense $\Phi$ is something like a ``current potential" for the current velocity $v^A$ that tells us how $\rho$ is going to change in time by (\ref{FP}). At this point, $\Phi$'s only time dependence is through $\rho$, but it is important to evaluate what we have been able to derive using ED so far: 

 ED has managed to show that the the Fokker-Planck equation (\ref{FP}) may be interpreted as a mechanism of entropic \emph{probability updating}. The ``current potential" $\Phi$, as argued above, is thus a mechanism or function that guides probability updating, and in this sense, it is purely epistemic -- it is informative. To derive QM, we need an additional mechanism for updating the constraints (\ref{kappa prime}) and (\ref{kappa prime prime}) in the entropic dynamics update (\ref{step}). We also let $\Phi(x)\rightarrow \Phi(x,t)$ to be labeled by time such that $\Phi$ has further functionality in its ability to update and mediate correlations in $\rho$ -- that is, we let $\Phi$ be dynamically informative.

Note that nothing prevents us from rewriting (\ref{FP}) as a functional derivative $\partial _{t}\rho =\frac{\delta \tilde{H}}{\delta \Phi }$, where,
\begin{eqnarray}
\tilde{H}[\rho ,\Phi ]=\int dx\,\Big[ \frac{1}{2}\rho m^{AB}(\partial_{A}\Phi-\overline{A}_A)( \partial_{B}\Phi-\overline{A}_B)
+F[\rho]\Big],\label{Hgen}
\end{eqnarray}
 has an integration functional constant of $\rho$, named $F[\rho]$. At this point the dynamics of $\phi,\chi$, and consequently $\Phi$, are unknown and we need a natural way to tie down the functional form of the time dependence in $\Phi$.  The form of $\Phi$ should originate from considerations of the system of particles being modeled and how we might expect it to respond to changes in $x$ or $\rho$. The current potential $\Phi$ gives the modeler freedom to model different systems in ED. Previous versions of ED that eventually lead to QM impose that the dynamics of $\Phi$ are determined by changes $\Phi'=\Phi+\delta\Phi$ that keep $d\tilde{H}/dt=0$, that is, $\tilde{H}$ plays the role of a Hamiltonian. We will take the newer approach \cite{EDNew}, as it better adheres to the probabilistic foundations in ED. 

  \subsection{Geometry of e-phase space}

In the search for the constraints on $\Phi$ that lead to QM, we look for inspiration from information geometry, and impose that $\chi(x_n)=\chi(x_n)+2\pi$ has the topology of an angle (i.e. now $\Phi$ has the topology of an angle). Imposing that $\chi$ has the topology of an angle solves Wallstrom's objection\footnote{Wallstrom's objection is that stochastic mechanics leads	to phases and wavefunctions that are either both multi-valued or both single-valued. Both alternatives are unsatisfactory because on one hand QM requires single-valued wavefunctions, while on the other hand single-valued phases exclude states that are physically relevant (e.g., states with non-zero angular momentum). \cite{Nick}} to Nelson's stochastic mechanics \cite{Nelson,Wallstrom1,Wallstrom2,Nick}. Working with spin-$1/2$ particles in which a ``spin frame field" also contributes to the updating of $\rho(x)$, $\chi$ becomes one of the orientation angles of the spin frame field \cite{Private,EDNew} (forthcoming \cite{NickSpin}) and the argument for $\chi$ having the topology of an angle becomes more palatable. To obtain a mechanism for updating $\Phi(x)$ we will extend the information metric from a discrete simplex $S_{\nu-1}$ over $\{\rho^i$\} to the $2\nu$-dimensional ensemble phase space (e-phase space) with extended coordinates $\{\rho^i\}\rightarrow\{\rho^i, \Phi^i\}$, and then take the continuous limit and utilize the simplectic structure of the e-phase space. This derivation is similar in nature to \cite{Reginatto3,Reginatto2}; however, the derivation presented in \cite{EDNew} is motivated for the purpose of doing inference in Entropic Dynamics. First, however, in preparation for this derivation, we will review a relevant derivation of the information metric on a discrete statistical manifold \cite{Wooters1981,EDNew}.

Due to normalization, the statistical manifold over a set of discrete probabilities $\{\rho^1, ..., \rho^{\nu} \}$ is the simplex $\mathcal{S}_{\nu-1}$, that is $1=\sum\rho^i$ is the equation of a simplex. One may consider changing coordinates to $\xi^i=(\rho^i)^{1/2}$ and then the normalization condition takes the form $\sum_i^{\nu}(\xi^i)^2=1$, which suggests the $\xi^i$ coordinates parameterize the surface of a sphere. This suggestion can be taken seriously, and from it, declare that the simplex is a $(\nu-1)$ sphere embedded in a $\nu$-dimensional spherically symmetric space. The generic form of a length invariant in a spherical symmetric space takes the form,
\begin{eqnarray}
d\ell^2=(a(|\rho|)-b(|\rho|))(\sum_i^{\nu}\xi^id\xi^i)^2+|\rho|b(|\rho|)\sum_i^{\nu}(d\xi^i)^2,
\end{eqnarray}
where $a(|\rho|)$ and $b(|\rho|)$ are two arbitrary smooth and positive functions of $|\rho|=\sum_i^{\nu}\rho^i$. Changing back to the original $\rho^i$ coordinates and letting the probabilities be normalized to unity gives the information metric up to an overall scale,
\begin{eqnarray}
d\ell^2=b(1)\sum_i^{\nu}\frac{1}{\rho^i}(d\rho^i)^2.
\end{eqnarray}

In the present derivation, we wish to consider an information metric that is extended from the simplex to the 2$\nu$-dimensional e-phase space $(\rho^i,\Phi^i)$ by imposing the following conditions: (A) that the extended space is compatible with the information metric on the simplex, and (B) that $\Phi$ has the topological structure of an angle. Condition (B) suggests the following polar coordinate representation,
\begin{eqnarray}
\xi^i=(\rho^i)^{1/2}\cos \Phi^i/\hbar\mbox{\tab and\tab} \eta^i=(\rho^i)^{1/2}\sin \Phi^i/\hbar,
\end{eqnarray}
such that the equation of the simplex is maintained,
\begin{eqnarray}
|\rho|=\sum_{i}^{\nu}\rho^i=\sum_{i}^{\nu}[(\xi^i)^2+(\eta^i)^2]=1,
\end{eqnarray}
while also suggesting the space is over the surface of a sphere in 2$\nu$ dimensions. To satisfy (A) we will follow the same algorithm that was used to find the information metric. That is, we take the ``spherical suggestion" seriously and declare spherical symmetry in the space of $(\rho,\Phi)$,
\begin{eqnarray}
d\ell^2=(a(|\rho|)-b(|\rho|))[\sum_i^{\nu}(\xi^id\xi^i+\eta^id\eta^i)]^2+|\rho|b(|\rho|)\sum_i^{\nu}[(d\xi^i)^2+(d\eta^i)^2].
\end{eqnarray}
Transforming back to the coordinates $(\rho^i,\Phi^i)$ and setting $|\rho|=1$, gives, again up to an arbitrary proportionality constant,
\begin{eqnarray}
d\ell^2=b(1)\sum_i^{\nu}\Big[\frac{\hbar}{2\rho^i}(d\rho^i)^2+\frac{2}{\hbar}\rho^i(d\Phi^i)^2\Big].
\end{eqnarray}
Taking the continuous limit gives the desired form of the e-phase space metric,
\begin{eqnarray}
d\ell^2=b(1)\int dx\,\Big[\frac{\hbar}{2\rho}(\delta\rho)^2+\frac{2}{\hbar}\rho(\delta\Phi)^2\Big].\label{metric}
\end{eqnarray}
The square displacement may be interpreted as a measure of distinguishability between $(\rho,\Phi)$ and $(\rho+\delta\rho,\Phi+\delta\Phi)$ in the e-phase space. As is proven and reviewed in \cite{EDNew}, this metric has an extra symmetry: a complex structure and thus has a symplectic 2-form. The symplectic 2-form leads to the canonical Hamilton-Jacobi formalism with $\rho$ and $\Phi$ being the canonically conjugate variables. The explicit derivation is not entirely relevant for the remainder of this thesis so it will be omitted, but it may be found in \cite{EDNew}. 

The availability of complex structure allows for the change of coordinates,
\begin{eqnarray}
\Psi=\rho^{1/2}\exp(i\Phi/\hbar)\mbox{\tab and \tab}\Psi^*=\rho^{1/2}\exp(-i\Phi/\hbar),
\end{eqnarray}
under which the metric takes a simple form,
\begin{eqnarray}
d\ell^2=b(1)\int dx\,\delta\Psi^*\delta\Psi.
\end{eqnarray}
In these coordinates, $i\hbar\Psi^*$ is the canonically conjugate momentum to $\Psi$ as they satisfy the following Poisson brackets:
\begin{eqnarray}
[\Psi(x),\Psi^{*}(x')]=\int dx''\,\Big(\frac{\delta\Psi(x)}{\delta\rho(x'')}\frac{\delta\Psi^{*}(x')}{\delta\Phi(x'')}-\frac{\delta\Psi^{*}(x)}{\delta\rho(x'')}\frac{\delta\Psi^{*}(x')}{\delta\Phi(x'')}\Big)]=\frac{1}{i\hbar}\delta(x-x'),
\end{eqnarray}
and,
\begin{eqnarray}
[\Psi(x),\Psi(x')]=[\Psi^*(x),\Psi^*(x')]=0.
\end{eqnarray}
At this point we may let the arbitrary scale constant $b(1)=1/4$ as is common in spherical embedding derivations of the information metric \cite{book}.
\subsection{Synthesis and the Schr\"{o}dinger Equation}
By finding the flow in e-phase space that preserves the symplectic symmetry, we may find a Hamiltonian and find the appropriate dynamics for $\Phi(x)$, thus, the constraints (\ref{kappa prime}) and (\ref{kappa prime prime}) may be updated at each time step of (\ref{step}). As it stands, the geometry of the e-phase space is independent from ED. We must incorporate the following features for the geometry of e-phase space to be useful and consistent with ED: the notion of ontological particle positions, entropic time, and the entropic dynamics update (\ref{step}) that leads to the Fokker-Planck equation (\ref{FP}). The synthesis of these topics with the e-phase space geometry gives the fully equivalent Schr\"{o}dinger Equation (SE) with the added interpretation that $\rho$, $\Phi$, and $\Psi$ are epistemic, whereas particle positions are ontic. 

As positions are the ontic variables, and because there is no notion of time in the e-phase space as of yet, consider the potential positional displacements $\delta x^A=x'^A-x^A$ of particles and evaluate the changes to $\rho$ and $\Phi$:
\begin{eqnarray}
\delta\rho(x)=\d_A\rho\delta x^A\mbox{\tab and\tab} \delta\Phi(x)=(\d_A\Phi-\overline{A}_A)\delta x^A,
\end{eqnarray}
and therefore,
\begin{eqnarray}
\delta\Psi=(\d_A\Psi-\frac{i}{\hbar}\overline{A}_A\Psi)\delta x^A.
\end{eqnarray}
The form of the metric under these variations is found by substitution,
\begin{eqnarray}
d\ell^2=\frac{1}{\hbar}\int( \tilde{h}_{AB}\delta x^{A}\delta x^{B} )\,dx,
\end{eqnarray}
where,
\begin{eqnarray}
\tilde{h}_{AB}=\rho\Big[\frac{1}{2}(\d_A\Phi-\overline{A}_A)(\d_B\Phi-\overline{A}_B)+\frac{\hbar^2}{8\rho^2}\d_A\rho\d_B\rho\Big],
\end{eqnarray}
or in the complex coordinates,
\begin{eqnarray}
\tilde{h}_{AB}= \Big[\frac{\hbar^2}{2}(\d_A\Psi^*-\frac{i}{\hbar}\overline{A}_A\Psi^*)(\d_B\Psi-\frac{i}{\hbar}\overline{A}_B\Psi)\Big].
\end{eqnarray}
This is a tensor in the e-phase space that measures distinguishability under small displacements. To introduce entropic time, we follow \cite{EDNew} and demand that duration be a measure of fluctuations, because fluctuations (\ref{fluctime}, \ref{transtime}) dominate the dynamics as $\Delta t\rightarrow0$. That is, \emph{in probability} $\delta x^{A}\delta x^{B}$ converges to $\expt{\Delta x^{A}\Delta x^{B}}$ for small $\Delta t$,
\begin{eqnarray}
\delta x^{A}\delta x^{B} = \expt{\Delta x^{A}\Delta x^{B}} +o(\Delta t)=\hbar m^{AB}\Delta t +o(\Delta t),
\end{eqnarray}
(from \cite{Nelson} section 5), so therefore \emph{in probability},
\begin{eqnarray}
d \ell^2=\tilde{H}_{AB}m^{AB}\Delta t\equiv\tilde{H}_0\Delta t,\label{HM}
\end{eqnarray}
where $\int \tilde{h}_{AB}\,dx=\tilde{H}_{AB}$. For small duration, the expected square displacement in the e-phase space is effectively proportioned by the particles' expected fluctuations, or equally well, \emph{the reciprocal of their masses}. 

The final synthesizing step is to demand that $\d_t\rho$ have the form of the Fokker-Planck equation. This simply requires the identification of the e-Hamiltonian $\tilde{H}_0=m^{AB}\tilde{H}_{AB}$ with the general form of $\tilde{H}$'s, from (\ref{Hgen}), that are compatible with $\d_t\rho$. As this construction is built for small times, $\tilde{H}_0$ turns out to be the free particle e-Hamiltonian. To account for additional interactions, a potential term $V(x)$ is added to the free e-Hamiltonian and thus the full e-Hamiltonian is,
\begin{eqnarray}
\tilde{H}=\int m^{AB}\rho\Big[\frac{1}{2}(\d_A\Phi-\overline{A}_A)(\d_B\Phi-\overline{A}_B)+\frac{\hbar^2}{8\rho^2}\d_A\rho\d_B\rho+V\Big]\,dx,
\end{eqnarray}
which is allowed, and consistent with, the form of (\ref{Hgen}). Due to the symplectic structure of the e-phase space and its synthesis into the ED analysis above, the e-Hamilton equations
\begin{eqnarray}
\frac{\d\rho}{\d t}=\frac{\delta \tilde{H}}{\delta \Phi}\mbox{\tab and\tab} \frac{\d\Phi}{\d t}=-\frac{\delta \tilde{H}}{\delta \rho},\label{Hamilton}
\end{eqnarray}
correctly satisfy all of the required ED constraints and mechanisms. In the complex coordinates $(\Psi,\Psi^*)$ we have,
\begin{eqnarray}
\tilde{H}[\Psi,\Psi^*]=\int dx\,\Big[\frac{\hbar^2}{2}m^{AB}(\d_A\Psi^*-\frac{i}{\hbar}\overline{A}_A\Psi^*)(\d_B\Psi-\frac{i}{\hbar}\overline{A}_B\Psi)+\Psi^*\Psi V\Big]\nonumber\\
=\int dx\,\Psi^*\Big[-\frac{\hbar^2}{2}m^{AB}(\d_A-\frac{i}{\hbar}\overline{A}_A)(\d_B-\frac{i}{\hbar}\overline{A}_B)+V\Big]\Psi\label{Hamiltonian}
\end{eqnarray}
(the second equality is reached through integration by parts) and therefore the e-Hamilton equation(s) are,
\begin{eqnarray}
\d_t\Psi(x)=\frac{\delta \tilde{H}}{\delta(i\hbar\Psi^*(x))},
\end{eqnarray}
which is the Schr\"{o}dinger Equation (SE),
\begin{eqnarray}
i\hbar\frac{\d\Psi}{\d t}=-\frac{\hbar^2}{2}m^{AB}(\d_A-\frac{i}{\hbar}\overline{A}_A)(\d_B-\frac{i}{\hbar}\overline{A}_B)\Psi+V\Psi.
\end{eqnarray}
In standard notation this is,
\begin{eqnarray}
i\hbar\frac{\d\Psi}{\d t}=-\sum_n\frac{\hbar^2}{2m_n}\delta^{ab}(\frac{\d}{\d x^a_{n}}-i\beta_nA_a(x_n))(\frac{\d}{\d x^b_{n}}-i\beta_nA_b(x_n))\Psi+V(x)\Psi,
\end{eqnarray}
where one identifies $\hbar$ as Planck's constant, $m_n$ as the particle masses, and $\beta_n=\frac{q_n}{\hbar c}$ as proportional to particle charges. The condition for compatibility of the probabilistic and linear structure of the SE that leads to full equivalence between ED and the QM of charged particles is that charges are quantized \cite{EDNew,Nick}. Entropic Dynamics has managed to derive general unitary evolution of pure states as an application of the standard maximum entropy method and information geometry, as was desired in the introduction.
%

 At this point the standard Hilbert space formalism may be adopted to represent
the epistemic state $\Psi (x)$ as a vector for convenience, 
\begin{equation}
|\Psi \rangle =\int dx\,\Psi (x)|x\rangle \quad \mbox{with\quad }\Psi
(x)=\langle x|\Psi \rangle .
\end{equation}%
The expression of $\ket{\Psi}$ in another basis may be interpreted as a potentially convenient way of expressing position space wavefunctions in ED.  


\section{Mixed States in ED\label{MixedED}}
The derivation of mixed quantum states in ED involves a straightforward application of probability theory. First, consider the derivation of pure state QM up to equation (\ref{step}),
\begin{equation}
\rho(x'|t')=\int P(x^{\prime }|x,\Delta t)\rho(x|t)\,dx.\label{stepagain}
\end{equation}
Recall that the transition probability is,
\begin{equation}
P(x^{\prime }|x,\Delta t)=\frac{1}{Z}\exp [-\sum_{n}\delta_{ab}(\frac{m_{n}}{2\hbar \Delta t}%
( \,\Delta x_{n}^{a}-\langle \Delta x_{n}^{a}\rangle ) (
\,\Delta x_{n}^{b}-\langle \Delta x_{n}^{b}\rangle ) ]~,
\label{transtime2}
\end{equation}
having the expected drifts,
\begin{eqnarray}
\expt{\Delta x^A}=m^{AB}[\hbar\partial_{B}\, (\phi+\overline{\chi})-\overline{A}_B]\Delta t,
\end{eqnarray}
and
\begin{eqnarray}
\overline{\chi}(x)\equiv \sum_n \beta_n\chi(x_n)~~\mbox{and}~~ \overline{A}_A(x)\equiv\hbar\beta_nA_a(x_n).
\end{eqnarray}
The evolution of $\rho$ in (\ref{stepagain}) is generated based on the implicit assumption that \emph{if} the potentials $(\phi,\chi,A_a)$, contributing to the expected drift of the particle $\expt{\Delta x^A}$, are known with certainty -- and $\rho(x|t)$ is given -- that the form of $\rho(x'|t')$ is known with certainty. That is, given that the initial conditions $\{\phi,\chi,A_a,\rho(x|t)\}$ are known, we may find $\rho(x'|t')$.

 Let's suppose, that in addition to our uncertainty in the definite yet unknown particle positions, that there exists an uncertainty in the set of initial conditions under which the system has been prepared. Let $p(k)$ represent the probability the particles evolve according to the $k$th set of initial conditions $\{\phi,\chi,A_a,\rho(x|t)\}_k\equiv\{\phi_k,\chi_k,A_{ak},\rho(x|k,t)\}$. The probability distribution of interest is the joint probability $\rho(x,k)=p(k)\rho(x|k)$, so we seek updates of the form $\rho(x,k|t)\rightarrow\rho(x',k|t')$. This can be accomplished by considering
\begin{equation}
\rho(x',k|t')=\int P(x^{\prime }|x,k,\Delta t)\rho(x,k|t)\,dx,\label{step a}
\end{equation}
however, the situational probabilities $p(k)$ factor out,
\begin{equation}
p(k)\rho(x'|k,t')=p(k)\int P(x^{\prime }|x,k,\Delta t)\rho(x|k,t)\,dx,\label{step b}
\end{equation}  
which means that for all $k$,
\begin{equation}
\rho(x'|k,t')=\int P(x^{\prime }|x,k,\Delta t)\rho(x|k,t)\,dx\label{step c}.
\end{equation}
We therefore may evaluate the entropic dynamics update of $N$ particles in the $k$th preparation $\rho(x|k)$ independently from the other preparations. Because the original entropic dynamics that leads to pure state QM is indeed a special case of this entropic dynamics when the initial conditions $k$ are known with certainty, we may follow the exact derivation of pure state QM from (\ref{step}) onwards and arrive at the SE for the $k$th preparation,
\begin{eqnarray}
i\hbar\frac{\d\Psi_k}{\d t}=-\sum_n\frac{\hbar^2}{2m_{n}}\delta^{ab}(\frac{\d}{\d x^a_{n}}-i\beta_{n}A_{ak}(x_n))(\frac{\d}{\d x^b_{n}}-i\beta_{n}A_{bk}(x_n))\Psi_k+V_k(x)\Psi_k.\label{SEk}
\end{eqnarray}
Because each of the $k$ preparations are over exactly the same $N$ particles, intrinsic particle qualities having to do with $n$, such as masses $m_n$ and the quantized charges $\beta_n=\frac{q_n}{\hbar c}$, are independent of the initial conditions; i.e, we may not know $k$ but we do know that the masses and charges are independent of $k$. The index $k$ is left on the external vector potential and the scalar potential $V_k$ to include instances in which the observer in question does not know, as part of $k$, the potential used in the preparation procedure of $k$. Because the inference of $k$ is over the same $N$ particles, the scalar potential is expected to factor $V_k(x)=V_{in}(x)+v_k(x)$ into an internal potential $V_{in}$ (like particle-particle Coulomb interactions between the $N$ particles), which are independent of $k$, and the externally prescribed potential $v_k(x)$ that does depend on $k$. If the observer in question is the one applying the external potential(s), then the $k$ indices on the potentials drop out, but $\Psi_k$ retains its index through $\rho(x|k)$ and $\Phi_k$. 

At this point the standard Hilbert space formalism may be adopted to represent
the epistemic state $\Psi_k (x)$ as a vector, 
\begin{equation}
|\Psi_k \rangle =\int dx\,\Psi_k (x)|x\rangle \quad \mbox{with\quad }\Psi_k
(x)=\langle x|\Psi_k \rangle .
\end{equation}%
If the observer in question is interested in calculating the probability that the set of $N$ particles are located at the point $x$ in configuration space, it is
\begin{eqnarray}
\rho(x)=\sum_kp(k)\rho(x|k)=\sum_kp(k)|\braket{x}{\Psi_k}|^2=\sum_kp(k)\braket{x}{\Psi_k}\braket{\Psi_k}{x}=\bra{x}\hat{\rho}\ket{x},
\end{eqnarray}
and we may extract the density matrix $\hat{\rho}=\sum_kp(k)\ket{\Psi_k}\bra{\Psi_k}$ as a matter of convenience. Expectation values are treated in the usual way by introducing the trace,
\begin{eqnarray}
\expt{f(x)}=\sum_k\int f(x)\rho(x,k)dx=\sum_kp(k)\expt{f(x)}_k
=\mbox{Tr}(f(\hat{x})\hat{\rho}).
\end{eqnarray}
Probabilities are expectation values of projectors $p(a)=\mbox{Tr}(\ket{a}\bra{a}\hat{\rho})$. One should note that in general,
\begin{eqnarray}
p(\Psi_{k})=\mbox{Tr}(\ket{\Psi_{k}}\bra{\Psi_{k}}\hat{\rho})=\sum_{k'}p({k'})|\braket{\Psi_{k}}{\Psi_{k'}}|^2\neq p(k),
\end{eqnarray}
meaning the probability $p(k)$ that the particles were prepared in $k$, is not equal in general to the probability that the state could end up being inferred as $\ket{\Psi_k}$, due to the fact that $\braket{\Psi_{k}}{\Psi_{k'}}\neq \delta_{kk'}$ in general. 

The time derivative of this density matrix is the quantum Liouville equation $\frac{d\hat{\rho}}{dt}=\frac{1}{i\hbar}[\hat{H},\hat{\rho}]$, thus, using ED, it can be interpreted as an application of inference. This completes the derivation of density matrices in ED with static probabilities $p(k)$, which is sufficient for this thesis. A future topic of research is to recast the general dynamics of density matrices as an application of inference using ED.

Now that density matrices are specified in ED, one may go ahead and use the quantum relative entropy,
\begin{eqnarray}
S(\hat{\rho},\hat{\varphi} )=-\mbox{Tr}(\hat{\rho} \log \hat{\rho} -\hat{\rho}\log \hat{\varphi}),
\end{eqnarray}
for the purpose of updating density matrices as we did in earlier Chapters. It is now clear that because QM was derived from standard probability theory, that the quantum maximum entropy method cannot be a generalization of ``probability updating" and that density matrices cannot be ``generalizations of probability". Rather than density matrices being generalizations, they are ``particularizations", as their function is specifically designed for describing quantum systems \cite{Private}. All probability updating in this instance has been accomplished by the use of the standard relative entropy and information geometry in ED. This will be discussed further in the next chapter on Quantum Measurement in ED. 

\chapter{Solution to the Quantum Measurement Problem in Entropic Dynamics\label{Measurement in Entropic Dynamics}}

This chapter follows \cite{Johnson,EDMeasurement} and expands upon a few sections in \cite{EDMeasurement}.\footnote{Reference \cite{EDMeasurement} extends the treatment to von Neumann, weak measurements, and fully specifies the solution to the preferred basis problem.} In the previous chapter, Quantum Mechanics (QM) was derived as an application of probability updating and information geometry in Entropic Dynamics (ED). ED itself is an inference framework that is general enough to process arbitrary information in the form of dynamical constraints on the probability distribution in question. In standard QM, wavefunctions follow one of two modes of dynamical evolution that are usually considered to be detached from one another: the Schr\"{o}dinger Equation (SE), which evolves states unitarily from one pure state to another, and its discontinuous collapse once a detection has been made \cite{vonNeumann}. In ED, these modes of evolution are both described as instances of entropic probability updating and are therefore two sides of the same coin \cite{Johnson}. In the sense of \cite{GiffinBayes,Giffin,GiffinThesis}, there is only one universal probability updating mechanism (entropic updating), and therefore in ED there is no reason to privilege one instance of probability updating (SE or collapse) over another. 

As is mentioned in the preamble of the previous chapter, in an inference framework such as ED, the common reference to ``observables" is misguided. ``Observables" should be replaced by Bell's term ``beables" \cite{Bell1993} for ontic elements such as particle positions, and ``inferables" for those epistemic elements associated
to probability distributions \cite{EDMeasurement}.\footnote{%
	Although, beables are inferables too.} This distinction between ontological and epistemic variables is essential toward the future development of this thesis, and indeed using more rigorous verbiage leads to a clearer interpretation of QM \cite{Bell1993,Jaeger}. Ultimately, it is these notions that prevent ED from being ruled out by any of the aforementioned QM \emph{no-go} theorems \cite{EDContextuality}, as well as providing clear interpretations of the quantum measurement process in ED \cite{Johnson,EDMeasurement}.


In ED it is possible to infer \textquotedblleft
observables\textquotedblright\ other than position, e.g. momentum,
energy, and spin \cite{Johnson,EDMeasurement}. While positions are the only ontic elements, other ``observables" are purely epistemic; they are properties of a probability distribution (or equally well an epistemic wavefunction), not of the particle \cite{Johnson}. These ideas can be pushed to an extreme when discussing the notion of a Weak Value \cite{AAV,duck,Dressel} of an operator. The Weak Value of an operator $\hat{A}$ in
which the system is prepared in an initial state $\ket{\Psi}$ and ``post-selected" in state $|\Psi ^{\prime }\rangle $ is defined to be,
\begin{equation}
A_{w}=\frac{\langle \Psi ^{\prime }|\hat{A}|\Psi \rangle }{\langle \Psi ^{\prime
	}|\Psi \rangle }~.
\end{equation}%
The Weak Value of an operator may take values far outside the acceptable range of eigenvalues while also being complex in general. They are therefore not Hermitian observables, and yet, they can
still be \emph{inferred}. Interpreting Weak Values as being part of the ontology leads to paradoxes, which are usually advertised in the titles of various Weak Value articles: ``How the result of a measurement of a component of the spin of a spin-1/2 particle can turn out to be 100" \cite{AAV},  ``The quantum pigeonhole principle and the nature of quantum correlations"\footnote{From \cite{CaiPigeon}, ``The classical pigeonhole principle states that if $N$ objects are placed in $M < N$ separate boxes,
	then at least one box must contain more than one object. In \cite{AharonovPigeon} it is argued that quantum systems do not obey this principle. "} \cite{AharonovPigeon,CaiPigeon}, as well as the ``Quantum Cheshire Cats" \cite{Cats} in which properties of a particle (its spin in this case) seemingly travels along one arm of an interferometer while the particle itself travels along the other. Although stimulating, many of these paradoxes can be resolved by treating a Weak Value as nothing more than a potentially interesting epistemic inferable, which we will do later in this chapter. It should be noted that weak measurement and Weak Values have been used as a practical amplification technique given a large number of measurements are made \cite{AAV,Dressel}. At this point we will take a step back from Entropic Dynamics and review the quantum measurement problem(s).
\section{The Quantum Measurement Problems}
There are two measurement problems outlined in \cite{Schlosshauer}: the first is the problem of \emph{definite outcomes}, and the second is the problem of \emph{preferred basis} (or degenerate basis). We will introduce them here:

 The problem of \emph{definite outcomes} stems from the difference between how Quantum Mechanics (QM) describes the world, and how the world is described in everyday experience. In QM, particles evolve from one pure state to another, and in some sense never ``settle down" to a definite final state, in all but the most trivial cases. This is in stark contrast to our everyday experience and the detected results of quantum mechanical experiments. Although the experimental results match the predictions of QM in probability, they fail to match in formalism. ``Wavefunction collapse" is usually tacked onto the formalism \emph{ad hocly} to cover the blemish of QM's lack of definite outcomes. The quantum formalism fails to predict when its unitary evolution will halt\footnote{``Halt", precisely in the sense of computability theory.} and collapse the state, as was given in the 2012 paper ``Quantum measurement occurrence is undecidable" \cite{halt}. 

The second measurement problem, the problem of \emph{preferred basis} or \emph{degenerate basis}, is a bit more technical in nature. In the von Neumann measurement scheme, the system we would like to measure, $\ket{\Psi}=\sum \alpha _{n}\ket{a_{n}}$, is entangled with a ``pointer variable" that indicates the state of a measurement device. The pointer variable is treated quantum mechanically and correlated with the system of interest in such a way that by detecting its state, we may infer the state $\ket{a_n}$ with certainty. Starting the pointer variable in a ``ready state" $\ket{r}$, it is entangled with the system of interest via a unitary time evolution, 
\begin{eqnarray}
\Big(\sum_n \alpha_n\ket{a_n}\Big)\ket{r}\stackrel{t}{\longrightarrow}\sum_{n,m}\alpha_{n}\delta_{nm}\ket{a_n}\ket{b_m}\label{vN a}.
\end{eqnarray}
This is the von Neumann measurement procedure. A detection that finds the pointer variable in state 
$|b_{n}\rangle $ seemingly allows the observer to infer that the system of interest has become the state $%
|a_{n}\rangle $; however, the catch is that the system and measurement device evolve into a special entangled state called a \textquotedblleft
biorthogonal\textquotedblright\ state \cite{Schlosshauer}, which has the form of a Schmidt decomposition.  The problem is, without a proper specification of the ontology, the right hand side of (\ref{vN a}) can be expanded in other bases, and due to possible degeneracies in the probabilities of the measurement outcomes and the unspecified ontology of the pointer variable, it can be unclear what state was actually detected, and therefore, what state was inferred  -- this specifies the problem of \emph{preferred basis} \cite{Schlosshauer}.  These \emph{degenerate basis} raise the question of which basis should be the \emph{preferred basis}, i.e. the basis which is actually present in the ontology of the quantum experiment. A pedagogical example is that an entangled spin state $\ket{\Psi}=\frac{1}{\sqrt{2}}(\ket{+-}-\ket{-+})$ in $\sigma_z$ may be expanded into the eigenbasis of $\sigma_x$, $\ket{\Psi}=\frac{1}{\sqrt{2}}(\ket{x_+x_-}-\ket{x_-x_+})$, which both have an identical probability spectrum. There is no ontological matter of fact about which outcome has been obtained in the measurement process ($\ket{\pm\mp}$ or $\ket{x_{\pm}x_{\mp}}$) because both instances are probabilistically indistinguishable.  There is no reason to believe that $\sigma_x$ pointer states should be more ontological than $\sigma_z$ pointer states, or the reverse, and take the privileged role of the \emph{preferred basis} in general measurement setups. Thus, because the standard quantum formalism fails to specify the \emph{preferred basis}, quantum measurement procedures are insufficient in general.

\section{Position and Inference: The Solution to the Quantum Measurement Problem in Entropic Dynamics}


The two major quantum measurement problems outlined in \cite{Schlosshauer}, and introduced above, are the problem of \emph{definite outcomes} (collapse) and the problem of \emph{preferred basis} or \emph{degenerate basis}. Solutions to the measurement problem are often called ``interpretations of QM", the idea being that all such ``interpretations" agree on the formalism and thus the experimental predictions \cite{Wallace}. Other solutions involve making modifications to Quantum Mechanics \cite{Wallace}. Entropic Dynamics presents a third type of solution to the quantum measurement problem(s). By showing that QM is a subset of the available inference applications in entropic inference, representing collapse through inferential Bayesian updates is self-contained within ED's theoretical framework \cite{Johnson,EDMeasurement}.\footnote{Reference \cite{EDMeasurement} extends the treatment to von Neumann, weak measurements, and fully specifies the solution to the preferred basis problem.} Furthermore, stating that the positions are the ontological variables of interest, and updating probabilities in a way that is consistent with this notion,\footnote{Specifically equations (\ref{kappa n}) - (\ref{4}), and (\ref{step}).} solves the problem of preferred basis. This matches the conclusion in \cite{Hitoshi} that the preferred basis must be supplied as an additional postulate outside of quantum mechanical law, and indeed, this is how the preferred basis is prescribed in ED \cite{Johnson,EDMeasurement}. Thus, Entropic Dynamics solves the quantum measurement problems in a way that afflicts other interpretations of QM (and their solutions to the quantum measurement problems).

How it is that ED is not ruled out by the Bell-Kochen-Specker Theorem\footnote{The Bell-Kochen-Specker Theorem is a \emph{no-go} theorem that rules out interpretations of QM that assign ontological status to the eigenvalues of operators that simultaneously belong to multiple (noncommuting) sets of internally complete sets of commuting observables. The interpretation of the Bell-Kochen-Specker theorem is that operators in QM are \emph{contextual}, meaning that their character or value depend on the remaining set of commuting observables in a measurement setting.} and other \emph{no-go} theorems is discussed in the next chapter. The remaining measurement problems in ED are to describe collapse and the inference of \emph{inferables} other than position. Due to the nature of ``inferables" in ED, we are able to interpret Weak Values in the weak measurement scheme as interesting epistemic inferables, which will be discussed later.


\subsection{Detection as an Entropic Update: The ``Collapse" of the Wavefunction\label{collapse section}}

The ``problem" of wavefunction collapse is never truly encountered in ED for the same reason that it is not encountered in epistemic ``classical" probability theory. No one asks how the probability distribution of a die role collapses during measurement; this just follows from the inductive logic expressed by Bayes Rule, which again, is a special case of entropic probability updating \cite{GiffinBayes,Giffin,GiffinThesis}. If a particle with wavefunction $\Psi$ is detected at $x_D$ with certainty, the prior probability $\varphi(x) = |\braket{x}{\Psi}|^2$ is updated to $\rho(x) =\delta (x-x_D)$ -- the entropic dynamics updating scheme comes to a halt and addresses the new data via the standard maximum entropy method reviewed in previous chapters. What should be emphasized is that, in Entropic Dynamics, \emph{both the unitary evolution of the wavefunction and its collapse are probability updates compatible with entropic inference} -- probability is always updated with respect to the available information. A question of interest in Entropic Dynamics is ``What is the inference procedure for detecting particles?". This requires the specification of measurement devices in ED. 

As is discussed in the preamble of the previous chapter, detectors are themselves made of particles with definite positions. Let the internal state of a detector be represented by the positional configuration of its constituting particles $\{\mathbf{d}\}$ (the bold face is to differentiate this $\mathbf{d}$ from the differential $d$). As the particles of interest in ED also have ontological positions $x$, the detector's particles with ontological positions $\mathbf{d}$ may be described equally well in ED, and positional correlations between the system and measurement device $\varphi(x,\mathbf{d})=\varphi(x)\varphi(\mathbf{d}|x)\neq \varphi(x)\varphi(\mathbf{d})$ may be generated and capitalized upon by the observer for the purpose of inference. In principle, we have access to the joint probability $\varphi(x,\mathbf{d})$ as it can be generated from the SE. Correlations of this type are possible in ED because detectors ultimately consist of particles, and thus, their inner workings may be described purely from that basis. For instance, one may describe a voltmeter as measuring voltage in terms of the displacements of the voltmeter's internal particle configurations (a current that generates an amplified signal in the sense below). The inclusion of the positional detector states $\{\mathbf{d}\}$ into the analysis further actualizes the notion of a detector given in \cite{Johnson,EDMeasurement}; however, the final results are the same.


The internal mechanisms of a detector that lead to a \emph{detection} can be expressed ``classically" as an amplification \cite{Johnson}, and so, we may let the signal amplification of a detector be represented (for the purpose of inference) by an ``amplification" likelihood function $q(D|\mathbf{d})$ that gives the probability of a (usually macroscopic) detection signal $D$ when its particles have transitioned from a ready configuration $\mathbf{d}_r$ to a final configuration $\mathbf{d}$. The likelihood functions $q(D|\mathbf{d})$ are capable of representing a large class of detectors from CCD cameras, to bubble chambers, or even our own eyes, each having various likelihood functions specific to the effectiveness of the amplification process of the detector in question. The full measurement inference includes amplified detection signals $D$, detector states $\mathbf{d}$, and the system of interest $x$. This is incorporated into the inference procedure by considering a larger space of variables $\varphi(x,\mathbf{d})\rightarrow\varphi(x,\mathbf{d},D)$. 

Here is a quick summary of the discussion so far: the particles of interest with configuration space coordinate $x$ are correlated with the positions $\mathbf{d}$ of particles that are considered to be part of the ``detector". These correlations result in the joint probability $\varphi(x,\mathbf{d})$, that may be generated using QM. Furthermore, for the convenience of the observer, the final result of $\mathbf{d}$ is amplified through the detector, which, by design, gives the signal $D$. The effectiveness of conveying the result of $\mathbf{d}$ through $D$ is represented by the amplification likelihood function $q(D|\mathbf{d})$. The entire inference process may be stated by the joint probability, $\varphi(x,\mathbf{d},D)=\varphi(x|\mathbf{d},D)q(D|\mathbf{d})\varphi(d)$. The conditional probability $\varphi(x|\mathbf{d},D)=\varphi(x|\mathbf{d})$ is independent of $D$ in general because, by construction, $D$ is some probabilistic function of $\mathbf{d}$, and thus $D$ cannot give any more information about $x$ than can $\mathbf{d}$ through the conditional probability $\varphi(x|\mathbf{d})$ that was generated using QM. Thus the joint probability of interest takes a slightly simplified form $\varphi(x,\mathbf{d},D)=\varphi(x|\mathbf{d})q(D|\mathbf{d})\varphi(\mathbf{d})=\varphi(x,\mathbf{d})q(D|\mathbf{d})$.

An ideal device has amplified signals $D_{\mathbf{d}}$ that are in a one to one correspondence with $\mathbf{d}$ such that $q(D_{\mathbf{d}}|\mathbf{d}')= \delta_{\mathbf{d},\mathbf{d}'}=\frac{q(D_{\mathbf{d}})}{\varphi(\mathbf{d}')}q(\mathbf{d}'|D_{\mathbf{d}})=q(\mathbf{d}'|D_{\mathbf{d}})$, and $q(D_{\mathbf{d}})=\varphi(\mathbf{d})$. Ideal amplification of this type is impossible to implement in practice when the number of relevant detector particles, having the configuration space coordinate $\mathbf{d}$, is macroscopic in number. However, if the relevant number of detector particles is manageable, for instance when detector particles are initialized in an unstable equilibrium that when perturbed results in a chain reaction that amplifies its signal, the number of relevant macroscopic (amplified) detector signals $\{D_{\mathbf{d}}\}$ are manageable. We will assume these probability relationships to hold for now as it leads to the description of an ideal measurement device.

The purpose of $D$ is to provide the observer with a convenient macroscopic interface for detection so the observer doesn't need to concern themselves with the potentially complicated internal detector states $\{\mathbf{d}\}$. These states may be marginalized over and indeed give $\varphi(x,D)=\varphi(x|D)\varphi(x)$, where $\varphi(x|D)$ is equal in value to the correlations originally presented by $\varphi(x|\mathbf{d})$, if the efficiency of the detector amplification $q(D_{\mathbf{d}}|\mathbf{d}')=\delta_{D_{\mathbf{d}},\mathbf{d}'}=q(\mathbf{d}'|D_{\mathbf{d}})$ is ideal. The proof can be seen by writing the marginalization in the following form, 
\begin{eqnarray}
\varphi(x,D_{\mathbf{d}})&=&\int d\mathbf{d}'\,\varphi(x,\mathbf{d}',D_{\mathbf{d}})=\int d\mathbf{d}'\,\varphi(x,\mathbf{d})q(D|\mathbf{d})\nonumber=\int d\mathbf{d}'\,\varphi(x|\mathbf{d}')\varphi(\mathbf{d}')\delta_{D_{\mathbf{d}},\mathbf{d}'}\nonumber\\
&=&\varphi(x|D_{\mathbf{d}})\varphi(D_{\mathbf{d}}).\label{margd}
\end{eqnarray}
  Here we see the likelihood function for the detector particle (configuration) is $\varphi(\mathbf{d}|x)=\varphi(D_{\mathbf{d}}|x)\equiv \varphi(D|x)$, which is equal to that of its amplified signal, and in some sense, the correlations between $(x,\mathbf{d})$ have propagated to  $(x,D)$ through the amplification process. Given an idealized amplification process, the probability that the particle was at $x$ given the amplified signal $D$ is,
\begin{eqnarray}
\rho(x)=\varphi(x|D)=\frac{\varphi(x)\varphi(D|x)}{\varphi(D)},\label{detect}
\end{eqnarray}
by Bayes rule, which may be generated using entropic methods \cite{GiffinBayes,Giffin,GiffinThesis} (reviewed in the introduction of Chapter \ref{Entropy Applications}). An \emph{idealized} inference process allows one to detect the presence of a particle in an infinitesimally small region centered around $x_D$, that is, (\ref{detect}) is $\rho(x)=\delta(x-x_{D})$. This requires \emph{both} the amplification likelihood function to be ideal $q(D_{\mathbf{d}}|\mathbf{d}')=\delta_{D_{\mathbf{d}},\mathbf{d}'}$ as well as an ideal detector particle configuration conditional probability $\varphi(x|\mathbf{d})=\delta (x-x_{\mathbf{d}})$. Correlations such as $\varphi(x|\mathbf{d})=\delta (x-x_{\mathbf{d}})$ are sought after by the observer and this is indeed why the von Neumann measurement process is valuable.

In most physical situations there is a lack of efficiency in the amplification process such that $q(D_{\mathbf{d}}|\mathbf{d}')\rightarrow q(D|\mathbf{d}')\neq\delta_{D,\mathbf{d}'}$. There are many such examples of non-idealized amplification likelihood functions $q(D|\mathbf{d})$: a Gaussian (or other) distribution for $D$ having an average value $f(\mathbf{d})$, a CCD camera that clicks with certainty if $\mathbf{d}$ is within a certain interval\footnote{Unlike the maximally efficient detector whose likelihood function is a delta, the likelihood function for a CCD camera $q(D|\mathbf{d})$ is an indicator function.}, or any such detector that has signals $D=D(\mathbf{d})$ that are a surjective function of the detector particle configurations (in practice $D$ maps $\mathbf{d}$ from its high dimensional space to $D$'s lower dimensional space). In any of these cases, the Bayes update becomes an instance of Jeffreys Rule,
\begin{eqnarray}
\rho_J(x)=\varphi_J(x|D)= \frac{\int\varphi(x,\mathbf{d},D)\, d\mathbf{d}}{q(D)}= \int\varphi(x|\mathbf{d})\varphi(\mathbf{d})\frac{q(D|\mathbf{d})}{q(D)}d\mathbf{d}= \int\varphi(x|\mathbf{d})q(\mathbf{d}|D)\, d\mathbf{d},\nonumber\\\label{Jeff}
\end{eqnarray}
such that a detection at $D$ specifies $x$ with probability $\varphi_J(x|D)$. Even if $x$ and $\mathbf{d}$ are equally correlated $\varphi(x,\mathbf{d})$, here and in the ideal case, the lack of efficiency in signal amplification on part of the detector $q(D|\mathbf{d}')\neq\delta_{D,\mathbf{d}'}$ leads to further uncertainty in the final position of $x$ (\ref{Jeff}).  It should be noted that the type of uncertainty presented in signal amplification $q(D|\mathbf{d}')\neq\delta_{D,\mathbf{d}'}$ is not necessarily quantum mechanical in origin. Jauch was able to show this experimentally, his measurements of position had less uncertainty than did the particle's quantum mechanical statistical uncertainty \cite{Jauch}, i.e., you can detect the position of a particle with a finer precision than the original quantum statistical uncertainty permits. This statement is as simple as differentiating between the statistical uncertainty of a coin flip and our ability to detect that it indeed landed on heads or tails with certainty. Detection and collapse in ED are therefore characterized by (\ref{detect}) and (\ref{Jeff}).

\paragraph{Compatibility with the the quantum maximum entropy method:}
The explanation presented in \cite{EDMeasurement} for collapse, and reiterated above, is completely consistent with the quantum maximum entropy method \cite{QBR} from Chapter \ref{Entropy Applications}. This is because the data update of the density matrix $\stackrel{\sim}{\varphi}\rightarrow\hat{\rho}$ using the quantum maximum entropy method is simply a probabilistic update of its components $\varphi(x)\rightarrow\rho(x)$. In the language of density matrices, the prior pure state $\hat{\varphi}$, with probability $\varphi(x) = |\braket{x}{\Psi}|^2$, needs to decohere $\hat{\varphi}\rightarrow\stackrel{\sim}{\varphi}$ with the measurement device before it could be collapsed using the quantum maximum entropy method to avoid the PDMT. This step of decoherence may be expressed using the standard QM tools from Chapter \ref{Entropy Applications} in the ED of mixed states formalism, or one can simply skip to a more probabilistic explanation as was given in the above discussion as the positional detector states $\{\mathbf{d}\}$ are marginalized over. The states $\{\mathbf{d}\}$ are in principle the same detector states from Chapter \ref{Entropy Applications}.  If however, the prior probability between $x$ and the macroscopic detector states $D$ are already known $\varphi(x,D)$, one could simply use the standard maximum entropy method with the appropriate data constraint $\rho(D)$, to derive the appropriate Bayes or Jeffreys rule \cite{Johnson,GiffinBayes,Giffin,GiffinThesis} -- all of which are available tools within the ED framework. Depending on the type of detector, the final inferred state(s) of the particle(s) in question may continue to evolve under the same or a new Hamiltonian, by some other inferential means, or perhaps be thrown away entirely in preparation for the next experiment. 

\subsection{Unitary Measurement Devices\label{Unitary measurement devices}}
 A question of interest answered in \cite{Johnson,Caticha2000} is, ``How can we measure (infer) observables (or in the present case ``inferables") other than position (given that position is the preferred basis) in ED?". The subtext has been added for clarity and flow in this thesis, and the surprisingly simple arguments of \cite{Johnson,Caticha2000} are reviewed here. 

Consider the wavefunction of a single particle living on a discrete lattice, 
\begin{equation}
|\Psi \rangle =\sum\nolimits_{i}c_{i}|x_{i}\rangle \quad \mbox{with}\quad 
\mbox{Prob}(x_{i})=p_{i}=|\langle x_{i}|\Psi \rangle |^{2}=|c_{i}|^{2}~.
\end{equation}%
As the particles have ontological positions in ED, $p_i$ gives the probability the particle is at location $x_i$. In a more \textquotedblleft complicated\textquotedblright\ measurement, the
particle is subject to additional interactions right before reaching the
position detector. Let such a setup $\mathcal{A}$ be described by a
particular \emph{unitary} time evolution $\hat{U}_{A}$ that is designed to take particles from an initial state $|a_{i}\rangle $  to the position $|x_{i}\rangle $ on the discrete lattice with certainty
 -- that is, $\hat{U}_{A}|a_{i}\rangle
=|x_{i}\rangle $. The unitary measurement device is the analog of a light prism; it takes well defined momentum states of particles and deflects them onto a screen for position detection. Since the set $\{|x_{i}\rangle \}$ is orthonormal and
complete, the set $\{|a_{i}\rangle \}$ is also orthonormal and complete. To
figure out the effect of $\mathcal{A}$ on some generic initial state vector $%
|\Psi \rangle $, expand the state of interest \emph{in the basis of the inferables of interest} $|\Psi \rangle
=\sum\nolimits_{i}c_{i}|a_{i}\rangle$, where$\ c_{i}=\langle a_{i}|\Psi
\rangle $. Then the state $|\Psi \rangle $ evolves according to $\hat{U}_{A}$
into a new state at a later time $t'$,
\begin{equation}
\hat{U}_{A}|\Psi \rangle =\sum\nolimits_{i}c_{i}\hat{U}_{A}|a_{i},t\rangle
=\sum\nolimits_{i}c_{i}|x_{i},t'\rangle,\label{unitary}
\end{equation}%
which, invoking the Born rule for \emph{position} measurements, implies that
the probability of finding the particle at the position $x_{i}$ at a later time is $%
p_{i}=|c_{i}|^{2}=|\langle x_{i}|\Psi(t') \rangle |^{2}=|\langle a_{i}|\Psi(t) \rangle |^{2}$, which is equal to the probability of $a_i$ at an earlier time. These positional probabilities then may evolve with $\{\mathbf{d}\}$ on the screen in the above section and one may collapse the system and infer an original positional state $\ket{a_i}=\sum_{j} c_{ij}\ket{x_j}$. That is, directly from \cite{EDMeasurement} (and inspired by \cite{Johnson,Caticha2000}), ``From a physics perspective there is nothing more to say but we can adopt 
\emph{a different language}: we can say that the particle has been
\textquotedblleft measured\textquotedblright\ \emph{as if} it had earlier been in
the state $|a_{i}\rangle $" -- although, in-fact, it was not, because the full state at that time was a superposition of $\ket{a_i}$'s. Thus, the setup $\mathcal{A}$ is a device that in principle
\textquotedblleft measures\textquotedblright\ all operators of the form $%
\hat{A}=\sum\nolimits_{i}\lambda _{i}|a_{i}\rangle \langle a_{i}|$ where
the eigenvalues $\lambda _{i}$ are \emph{arbitrary} scalars. Note that there
is no implication that the particle previously had or now currently has the
value $\lambda _{i}$. In the context of this thesis and in ED, the states $\{\ket{a_i}\}$ and operator $\hat{A}$ are inferables of the theory that are inferred from detections of the preferred ontological position basis. 

If one wants to infer a continuous variable from a state like $\ket{\Psi}=\int da\,\,\psi(a)\ket{a}$ one uses a unitary device $\hat{U}_{A}$ with the property $\hat{U}_{A}\ket{a,t}=\ket{x,t'}$. A change of variable is required $a=a(x)\rightarrow x$, where $a(x)$ is the appropriate monotonic correspondence function of the unitary device (for simplicity consider $a(x)=cx$, where $c$ converts and scales what positions on a screen correspond to what values of $\ket{a}$ at an earlier time). This amounts to,
\begin{eqnarray}
\hat{U}_A\ket{\Psi}&=&\int dx\,\Big(\frac{d a(x)}{dx}\Big)^{1/2}\ket{x,t'}\braket{a(x),t}{\Psi}=\int dx\,\Big(\frac{d a(x)}{dx}\Big)^{1/2}\ket{x,t'}\bra{a(x),t}\int da\, \psi(a) \ket{a}\nonumber\\
&=&\int dx\,\Big(\frac{d a(x)}{dx}\Big)^{1/2}\psi(a(x))\ket{x,t'},
\end{eqnarray}
such that $\rho(x)=|\psi(a(x))|^2|da/dx|$ and therefore $p(a)=|\psi(a,t)|^2da=|\psi(x,t')|^2dx=p(x)$ at the later time $t'$. Again Bayes Rule may be used as an instance of collapse as is specified above. Observables of the form $\hat{A}$ are thus inferred from position.

\subsection{Von Neumann Measurements in Entropic Dynamics}
This subsection discusses the method for generating correlations between the detector states $\mathbf{d}$ and the inferables of interest using a von Neumann measurement procedure in ED. As well, this discussion provides an example that shows how both the measurement problems of \emph{definite outcomes} and \emph{preferred basis} are resolved within the ED framework. 

Given a state of interest $\ket{\Psi_i}=\sum_n\alpha_n\ket{a_n}$, which has been expanded \emph{in the basis of the inferables of interest} $\{a_n\}$, a von Neumann measurement is one in which $\ket{\Psi_i}$ evolves with an auxiliary ``pointer variable" (or detector) state, $\ket{\Phi_i}=\ket{\mathbf{d}_0}\rightarrow \{\ket{\mathbf{d}_m}\}$, of which both become entangled, such that $a_n$ may be inferred from detections of $\mathbf{d}_n$. In ED, the natural pointer variables are the positions of (detector) particles as position is the preferred basis. These ``pointer variables" $\mathbf{d}_m$ may be interpreted as the internal positional states of a detector or as the position of a single particle being fed into a detector, from Section \ref{collapse section}. The resulting state from a von Neumann measurement is the entangled state,
\begin{eqnarray}
\Big(\sum_n \alpha_n\ket{a_n}\Big)\ket{\mathbf{d}_0}\stackrel{\Delta t}{\longrightarrow}\sum_{n,m}\alpha_{n}\delta_{nm}\ket{a_n}\ket{\mathbf{d}_m},
\end{eqnarray}
that's unitary evolution will be discussed now.

Initially, the state of interest is $\ket{\Psi_i}=\sum_n\alpha_n\ket{a_n}$ and the pointer variable is in its ``ready" state $\ket{\Phi_i}=\ket{\mathbf{d}_0}$. The initial joint system is represented by the tensor product of the initial states $\ket{\Psi_i}\otimes\ket{\Phi_i}\equiv\ket{\Psi_i}\ket{\Phi_i}$. The pure state evolves with the detector states $\{\mathbf{d}\}$ via a unitary evolution to a final entangled state $\ket{\Psi_f,\Phi_f}$,
\begin{eqnarray}
\ket{\Psi_i}\ket{\Phi_i}=\Big(\sum_n \alpha_n\ket{a_n}\Big)\ket{\mathbf{d}_0}\longrightarrow \ket{\Psi_f,\Phi_f}=\hat{U}\ket{\Psi_i}\ket{\Phi_i}.
\end{eqnarray}
The general form of this unitary operator is,
\begin{eqnarray}
\hat{U}=\sum_{n,n',m,m'}u_{n,n',m,m'}\ket{a_n}\ket{\mathbf{d}_m}\bra{a_{n'}}\bra{\mathbf{d}_{m'}}=\sum_{m,m'}B_{\mathbf{d}_m\mathbf{d}_{m'}}\otimes \ket{\mathbf{d}_m}\bra{\mathbf{d}_{m'}},
\end{eqnarray}
with the sub-block matrices,
\begin{eqnarray}
B_{\mathbf{d}_m\mathbf{d}_{m'}}=\sum_{n,n'}u_{n,n',m,m'}\ket{a_n}\bra{a_{n'}}.
\end{eqnarray}
We define a good detector as one in which the $a_n$th state only entangles with the $\mathbf{d}_n$th positional state of the detector, which is an argument for the sub-block matrix to take a simple form,
\begin{eqnarray}
B_{\mathbf{d}_m\mathbf{d}_0}=\sum_{n,n'}\delta_{m,n}\delta_{m',n'}\ket{a_n}\bra{a_{n'}}=\ket{a_m}\bra{a_m}.
\end{eqnarray}
This then gives a fully entangled (von Neumann measurement) state at a later time,
\begin{eqnarray}
\ket{\Psi_f,\Phi_f}=\sum_{n,m}\alpha_{n}\delta_{nm}\ket{a_n}\ket{\mathbf{d}_m}.\label{vnED}
\end{eqnarray}
Because position plays the role of the \emph{preferred basis} in ED, the state does not have \emph{degenerate bases} and the measurement problem of preferred basis is solved -- particle or particles are ontologically present at one of the locations $\mathbf{d_n}$. This solution is equally valid for the \emph{beable} particle positions in Bohmian Mechanics. Although ED does \emph{assume} that particles have definite yet unknown (and ontological) positions from the start, this assumption is carried throughout, and in-fact, helps guide the derivation of QM from ED. The probability updating scheme evolves probabilities $\rho(x|t)$ in a way that is consistent with notion that particles having definite yet unknown, and ontological, positions as can be seen in the discussion before, around, and through equations (\ref{kappa n}) - (\ref{4}), and (\ref{step}).

Given the von Neumann measurement above in ED, the joint probability is, 
\begin{eqnarray}
P(a_n,\mathbf{d}_m)=|\bra{a_n}\bra{\mathbf{d}_m}\sum_{n',m'}\alpha_{n'}\delta_{n'm'}\ket{a_{n'}}\ket{\mathbf{d}_{m'}}|^2=|\alpha_{n}\delta_{nm}|^2=|\alpha_{n}|^2\delta_{nm},
\end{eqnarray}
which is normalized $\sum_{n,m}|\alpha_{n}|^2\delta_{nm}=\sum_{n}|\alpha_{n}|^2=1$. Given ideal amplification $\mathbf{d}\rightarrow D$ from Section \ref{collapse section}, and that the value of the data is $D$, the (standard) entropic updating procedure (from the introduction in Chapter \ref{Entropy Applications} or \cite{GiffinBayes,Giffin,GiffinThesis}) gives Bayes Theorem,
\begin{eqnarray}
P(a_n|D)=\frac{P(a_n,D)}{P(D)}=\frac{|\alpha_{n}\delta_{nD}|^2}{\sum_n|\alpha_{n}\delta_{nD}|^2}=\delta_{nD},
\end{eqnarray}
which is collapse in Entropic Dynamics \cite{Johnson,EDMeasurement}. For a von Neumann measurement, a detection of $D$ allows a precise inference of the inferable in question. The resulting state is accurately \emph{describable} by a particular eigenvector $\ket{a_D}$, which is understood to be a convenient representation of its position space wavefunction $\ket{a_D}=\int \Psi_{a_D}(x)\ket{x}\,dx$. 

Again we emphasize that in Entropic Dynamics, both the unitary evolution of the wavefunction \emph{and} its collapse are \emph{both} probability updates compatible with entropic inference. The state evolves unitarily when the information available is with respect to the continuous positional \emph{expected} drift of the particles (i.e. it undergoes the entropic dynamics update with updating drift potentials), and if there is new information in the form of data, the entropic dynamics update halts, addresses the data, and updates the current probability distribution accordingly. Thus, the measurement problem of \emph{definite outcomes} is solved using inference methods that are completely contained within the inference framework that is Entropic Dynamics. Both of the quantum measurement problems outlined in \cite{Schlosshauer} are therefore solved in Entropic Dynamics.

These notions can easily be extended to ``weak measurements", which we will discuss now. 


\section{Weak Measurements and Weak Values in Entropic Dynamics}
This section discusses the method for generating correlations between the detector states $\mathbf{d}$ and inferables of interest, in ED.
\subsection{Weak Measurements in Entropic Dynamics}
To avoid clutter we will set $\hbar=1$ and save the normalization factors until the end of the calculation. An ideal pointer variable, or detector state, in ED has a definite initial state in position space,
\begin{eqnarray}
\ket{\Phi_{ideal}}=\int e^{-\frac{(\mathbf{d}-\mathbf{d}_i)^2}{4\epsilon^2}}\ket{\mathbf{d}}\,d\mathbf{d},
\end{eqnarray}
i.e., $\epsilon \ll1$ and $\ket{\Phi_{ideal}}$ approaches a delta function.  We consider a more general case in which the initial state is,
\begin{eqnarray}
\ket{\Phi_{i}}=\int e^{-\frac{\mathbf{d}^2}{4\bigtriangleup^2}}\ket{\mathbf{d}}\,d\mathbf{d},\label{weak pointer}
\end{eqnarray}
where $\bigtriangleup$ is some finite volume, which reproduces the ideal measurement device when $\bigtriangleup\rightarrow \epsilon$ after normalization.  Using the completeness relation $1=\int dp \ket{p}\bra{p}$, the state of the pointer particle can be represented in momentum space as,
\begin{eqnarray}
\ket{\Phi_{i}}=\int e^{-\frac{\mathbf{d}^2}{4\bigtriangleup^2}}\int  \ket{p}\braket{p}{\mathbf{d}}\,d\mathbf{d}dp=\int e^{-\frac{\mathbf{d}^2}{4\bigtriangleup^2}}\int  e^{-ip\cdot \mathbf{d}}\,d\mathbf{d}\ket{p}dp=\int e^{-\bigtriangleup^2p^2} \ket{p}\,dp.
\end{eqnarray}
In this section we will consider $\mathbf{d}$ to be the position of a single particle in 3D space. The quantum system to be entangled with the measurement device is a preselected or prepared superposition state, that when expanded \emph{in the basis of the inferable of interest} $\hat{A}$, is
\begin{eqnarray}
\ket{\Psi_{i}}=\sum_{n}\alpha_n\ket{A=a_n}.
\end{eqnarray}
A particularly telling situation is one in which the measuring device and system of interest are coupled by a coupling or interaction Hamiltonian,
\begin{eqnarray}
\hat{H}=-g(t)\hat{p}\hat{A},\label{H}
\end{eqnarray}
where $\hat{p}$ is the canonical conjugate of the pointer variable, which thereby generates translations in the pointer variable $\hat{\mathbf{d}}$, and $g(t)$ is a coupling constant with compact support near the time of measurement that integrates to a finite number \cite{AAV}. We can assume that the coupling Hamiltonian will dominate over the full Hamiltonian for the, assumed small, period of measurement. The time evolution of our entangled system is,
\begin{eqnarray}
\ket{\Psi_{f},\Phi_{f}}=U\ket{\Psi_{i}}\ket{\Phi_{i}}=e^{-i\int \hat{H} dt}\ket{\Psi_{i}}\ket{\Phi_{i}}\nonumber
\end{eqnarray}
\begin{eqnarray}
=\sum_{n}\alpha_n\int e^{-\bigtriangleup^2p^2}e^{i\hat{p}\hat{A}}\ket{A=a_n}\ket{p}\,dp
=\sum_{n}\alpha_n \int e^{-\bigtriangleup^2p^2}e^{ipa_n} \ket{A=a_n}\ket{p}\,dp\label{ent},
\end{eqnarray}
which, in the position space representation, is 
\begin{eqnarray}
=\sum_{n}\alpha_n\int e^{-\bigtriangleup^2p^2}e^{-ip(\mathbf{d}-a_n)}\ket{A=a_n}\ket{\mathbf{d}}\,dp\,d\mathbf{d}=\sum_{n}\alpha_n\int e^{-\frac{(\mathbf{d}-a_n)^2}{4\bigtriangleup^2}} \ket{A=a_n}\ket{\mathbf{d}}\,d\mathbf{d}.\label{24}
\end{eqnarray}
This is a superposition of potentially overlapping Gaussian distributions that have peaks at the eigenvalues of $\hat{A}$. When the Gaussian wavefunctions overlap we have a so-called ``weak measurement" \cite{AAV}; when the Gaussian distributions are neatly resolved we have a ``strong'' or von Neumann measurement ($\Delta\rightarrow \epsilon$) \cite{AAV}. The joint probability of the system is,
\begin{eqnarray}
P_f(a_n,\mathbf{d})=|\braket{a_n,\mathbf{d}}{\Psi_{f},\Phi_{f}}|^2=\frac{|\alpha_n|^2e^{-\frac{(\mathbf{d}-a_n)^2}{2\bigtriangleup^2}}}{Z}  .
\end{eqnarray}
In the present case, we may infer the probability the system of interest is accurately \emph{describable} by a particular eigenvector $\ket{a_n}$ by making detections of the position of the pointer particle and entropically collapse the system following Section \ref{collapse section}. The result is given by Bayes Rule,
\begin{eqnarray}
P(a_n)=P_f(a_n|D)=\frac{P_f(a_n,D)}{P_f(D)}=\frac{|\alpha_n|^2  e^{-\frac{(D-a_n)^2}{2\bigtriangleup^2}}}{\sum_{m}|\alpha_m|^2  e^{-\frac{(D-a_m)^2}{2\bigtriangleup^2}}}.\label{22a}
\end{eqnarray}
Thus a detection of $D$ (generally macroscopic) allows us to infer the probability that the system is accurately describable by $\ket{a_n}$. Bayes Rule leads to a ``weak" collapse of the state of knowledge of $\{\ket{a_n}\}$ because the initial state of the pointer variable $\mathbf{d}$ is itself uncertain. As is pointed out in \cite{AAV}, by making repeated measurements on identically prepared weak measurement states, one can infer $|\alpha_n|^2$ with certainty. In the limit $\bigtriangleup\rightarrow 0$ and a detection $D$, we can infer the eigenstate $\ket{a_n}$ with probability 1 using (\ref{22a}) -- this results in a full state collapse in ED. 

So far we have taken into account the uncertainty in the preparation of the pointer variable but we have assumed that the pointer variable $\mathbf{d}$ has been amplified with 100\% efficiency. If the detector is not completely efficient such that $q(D|\mathbf{d}')\neq \delta_{D_{\mathbf{d}},\mathbf{d}'}$ then there is a second source of uncertainty that contributes in addition to the initial uncertainty of the pointer variable in (\ref{weak pointer}).  The probability the state is accurately described by $\ket{a_n}$, if the detection is noisy, is given by Jeffreys Rule,
\begin{eqnarray}
P_J(a_n)=\varphi_J(a_n|D)=\frac{\int P(a_n,\mathbf{d}',D)\, d\mathbf{d}'}{P(D)}=\int P(a_n|\mathbf{d}')q(\mathbf{d}'|D)\, d\mathbf{d}',\label{38}
\end{eqnarray}
which we see only gives the same state of knowledge as (\ref{22a}) if $q(D|\mathbf{d}')= \delta_{D_{\mathbf{d}},\mathbf{d}'}=q(\mathbf{d}'|D)$ is an ideal amplification, otherwise the detector introduces extra uncertainty (as is discussed in Section \ref{collapse section}). For instance, a CCD camera's likelihood function might be uniform over the width of the corresponding pixel, which changes the probability of $a_n$ after a detection $D$ of that pixel. 

It should be noted a more robust set of operators $\hat{B}$ can be measured in ED by compounding a unitary device $U_v$, from (\ref{unitary}), with the weak measurement. This is a combination of a unitary measurement device with the weak measurement scheme, and doing this maps a non-position pointer variable $\{\ket{v}\}$ to a position pointer variable $\{\ket{\mathbf{d}}\}$ for inference. Consider the coupling Hamiltonian 
\begin{eqnarray}
H_c=g(t)\hat{B}\hat{v}_{conj},\label{HB}
\end{eqnarray}
 where $\hat{v}_{conj}$ is the Fourier conjugate to $\hat{v}$ such that  $\braket{v}{v_{conj}}=e^{-iv_{conj}v}$ (for instance $\hat{v}=k\hat{p}$ where $k$ has arbitrary units compatible with $\hat{B}$). Evolution under this Hamiltonian entangles the states,
\begin{eqnarray}
\ket{\Psi_{f},\Phi_{f}}=U\ket{\Psi_{i}}\ket{\Phi_{i}}=\sum_{n}\alpha_n \int e^{-\bigtriangleup^2v_{conj}^2}e^{iv_{conj}b_n} \ket{B=b_n}\ket{v_{conj}}\,dv_{conj},
\end{eqnarray}
which, in the $v$ space representation,
\begin{eqnarray}
=\sum_{n}\alpha_n\int e^{-\bigtriangleup^2v_{conj}^2}e^{-iv_{conj}(v-b_n)}\ket{B=b_n}\ket{v}\,dv_{conj}\,dv=\sum_{n}\alpha_n\int e^{-\frac{(v-b_n)^2}{4\bigtriangleup^2}} \ket{B=b_n}\ket{v}\,dv.\nonumber\\
\end{eqnarray}
Sending the pointer variable state through a unitary measurement device $U_v$ with the property $U_v\ket{v,t}=\ket{\mathbf{d},t'}$ and a simple device correspondence function $v=g(\mathbf{d})=c\mathbf{d}$ \cite{Johnson}, gives,
\begin{eqnarray}
\ket{\Psi_{f},\Phi_{f}}'=I\otimes U_v\ket{\Psi_{f},\Phi_{f}}=c\sum_{n}\alpha_n\int e^{-\frac{c^2(\mathbf{d}-b_n/c)^2}{4\bigtriangleup^2}} \ket{B=b_n}\ket{\mathbf{d}}\,d\mathbf{d}.
\end{eqnarray}
This is again a superposition of potentially overlapping Gaussian distributions having peaks at the eigenvalues of $\hat{B}/c$. Detections of $D$ at $t'$ allow us to infer the most likely $\ket{b_n}$ at $t$,
\begin{eqnarray}
P(b_n|D)=\frac{P(b_n,D)}{P_f(D)}=\frac{|\alpha_n|^2  e^{-\frac{c^2(D-b_n/c)^2}{2\bigtriangleup^2}}}{\sum_{m}|\alpha_m|^2  e^{-\frac{c^2(D-b_m/c)^2}{2\bigtriangleup^2}}}.
\end{eqnarray}
Measurement device uncertainty can be included using the same arguments from (\ref{38}). The mechanics for the Stern-Gerlach experiment is well known \cite{Holland,AAV,duck}; however a brief discussion of the inference of spin from position detections (including weak measurements) is given in Appendix \ref{spinexampleED}.

\subsection{Exotic Inferables and Weak Values in Entropic Dynamics}
The Weak Value $A_w\equiv \bra{\Psi'}\hat{A}\ket{\Psi}/\braket{\Psi'}{\Psi}$ of a Hermitian operator $\hat{A}$ was first introduced by Aharonov, Albert, and Vaidman \cite{AAV} (AAV) in 1989 as an interesting application of a weak measurement that gives nonintuitive results (recent review \cite{Dressel}). What is particularly significant in AAV's paper \cite{AAV} is not that they defined an odd quantity associated to $\hat{A}$, but rather they found, after a series of approximations (see \cite{duck}) a way in which $A_w$ could be ``measured.'' A Weak Value is a complex number which may lie outside the set of eigenvalues of $\hat{A}$ when $\braket{\Psi'}{\Psi}$ is sufficiently small. Due to this, the interpretation of Weak Values has had a ``colorful history" \cite{Dressel}, much of which can be summarized by the question, ``Are Weak Values ontic properties of a particle?". The answer given by ED is negative in that respect -- Weak Values are inferables of the theory. 
\paragraph{Postselection:}
The method for inferring a Weak Value involves two steps: first the system must undergo a weak measurement to couple $\ket{\Psi}\ket{\Phi}\rightarrow\ket{\Psi_f,\Phi_f}$ as in (\ref{24}) and then postselect the final state of the system of interest. A postselection is in principle no different than a preselection (in the sense of \cite{Ballentine}), which is the appropriate filtering and selecting of an initial quantum state except it happens after, rather than before, the system is correlated with the pointer variable \cite{AAV}. Because the system of interest's final state is known $\ket{\Psi}\rightarrow\ket{\Psi'}$ when successfully postselected to $\ket{\Psi'}$, the probability of the pointer variable taking a value $\mathbf{d}$ is,
\begin{eqnarray}
P(\mathbf{d}|\Psi')=\frac{P(\mathbf{d},\Psi')}{P(\Psi')}=\frac{|\braket{\Psi',\mathbf{d}}{\Psi_f,\Phi_f}|^2}{\int d\mathbf{d}\, |\braket{\Psi',\mathbf{d}}{\Psi_f,\Phi_f}|^2},
\end{eqnarray}
in agreement with the analysis in \cite{Ban} that concludes that postselections may be represented simply through conditional probability. 


\paragraph{Weak Values:}
In ED, variables other than position either: are inferred from position, are useful parameters in the position probability distribution of the particle, or may be a particularly convenient basis for representing the position space probability distribution $\rho(x)=|\Psi(x)|^2$. From this perspective it is clear that inferring $A_w$ does not indicate that the particle is ontically expressing $A_w$, in the same way as inferring the momentum from position detections does not imply it is ontic. Consider the final state of an entangled joint system from (\ref{ent}) that has been postselected into a peculiar state $\ket{\Psi'}=\sum \alpha_n'\ket{A=a_n}$, such that,
\begin{eqnarray}
\ket{\Phi_f}=\bra{\Psi'}U\ket{\Psi}\ket{\Phi_i}=\bra{\Psi'}e^{-i\int \hat{H} dt}\ket{\Psi}\ket{\Phi_{i}}
\approx (\braket{\Psi'}{\Psi}+i\hat{p}\bra{\Psi'}\hat{A}\ket{\Psi}+...)\ket{\Phi_i}\nonumber\\
=\braket{\Psi'}{\Psi}(1+i\hat{p}A_w+...)\ket{\Phi_i} \approx \braket{\Psi'}{\Psi}\int dp e^{ipA_w}e^{-\bigtriangleup^2p^2}\ket{p},
\end{eqnarray}
which in the position space representation is,
\begin{eqnarray}
\ket{\Phi_f}\approx \braket{\Psi'}{\Psi}\int d\mathbf{d}\,\exp\Big(-\frac{(\mathbf{d}-A_w)^2}{4\bigtriangleup^2}\Big)\ket{\mathbf{d}},
\end{eqnarray}
given the postselection is in the desired range of validity \cite{duck,Dressel}. In the ED framework we are interested in the probability distribution of $\mathbf{d}$ postselected into $\Psi'$,
\begin{eqnarray}
P(\mathbf{d}|\Psi')\approx\frac{1}{Z}\exp\Big(-\frac{(\mathbf{d}-\mbox{Re}[A_w])^2}{2\bigtriangleup^2}\Big),
\end{eqnarray}
and note the real part of the weak value $\mbox{Re}[A_w]$ is taken as a feature of the probability distribution of $\mathbf{d}$.  Because $A_w$ appears as a parameter in the probability distribution of $\mathbf{d}$, we may consider $P(\mathbf{d})=P(\mathbf{d}|\mbox{Re}[A_w])$ and invert the problem to ask, ``what is the probability the parameter $\mbox{Re}[A_w]$ has a certain value given a detection of the pointer particle at $x'$" -- that is we may use $P(\mbox{Re}[A_w]|\{\mathbf{d}\})\propto\exp\Big(-(\mbox{Re}[A_w]-\overline{\mathbf{d}})^2\slash (2\bigtriangleup^2/N)\Big)$ and the parameter estimation scheme in \cite{book} to find $\mbox{Re}[A_w]$ in agreement with \cite{AAV}. Equally well we can find the imaginary part of the Weak Value by using a unitary measurement device (\ref{unitary}). Consider using $U_p\ket{p,t}=\ket{\mathbf{d},t'}$ with a simple correspondence function $p=g(\mathbf{d})=c\mathbf{d}$,
\begin{eqnarray}
U_p\ket{\Phi_f}\approx c\braket{\Psi'}{\Psi}\int d\mathbf{d} e^{ic\mathbf{d}A_w}e^{-(c\bigtriangleup \mathbf{d})^2}\ket{\mathbf{d}},
\end{eqnarray}
and after completing the square and normalizing one finds,
\begin{eqnarray}
P(\mathbf{d}|\mbox{Im}[A_w])=\frac{1}{Z}\exp\Big(-2(c\bigtriangleup)^2(\mathbf{d}+\mbox{Im}[A_w]/2c\bigtriangleup^2)^2\Big),
\end{eqnarray}
where $\mbox{Im}[A_w]$ is the imaginary part of $A_w$ and again after many detections, $\mbox{Im}[A_w]$ and thus $A_w=\mbox{Re}[A_w]+i\,\mbox{Im}[A_w]$ may be inferred in full.

There are several potentially interesting Weak Values, but in particular consider that the operator $\hat{A}=\ket{x}\bra{x}$ postselected in the zero momentum state,
\begin{eqnarray}
A_{w}=\frac{\braket{p}{x}\braket{x}{\Psi}}{\braket{p}{\Psi}}\stackrel{p=0}{\rightarrow}  k\Psi(x),
\end{eqnarray}
is proportional to the full wavefunction, where $k$ is a constant that will be removed after normalization \cite{Lundeen1}. Lundeen \emph{et al} show that the real and imaginary parts of $\Psi(x)$ are proportional to the position and momentum shifts of the pointer variable and claim they are ``directly measuring the wavefunction". From the ED perspective it is more appropriate to say that the value of the wavefunction at each $x$ is being ``directly" inferred. 
If the complex valued $\Psi(x)$ is inferred with certainty then it is possible to also solve for the phase of the wavefunction $\Phi$ (up to an additive constant $2\pi n$), and thus, $\Phi$ is also an inferable. This provides a link to Wiseman's use of Weak Values to measure the probability current in Bohmian Mechanics \cite{Wiseman}. 

Especially in cases as exotic as the above, ED takes the standpoint that Weak Values and quantities other than position (energy, momentum, spin, etc.) are best considered as epistemic ``inferables" rather than ontic beables or observables.\\\\

\section{POVM Measurements in Entropic Dynamics}
A POVM measurement can be looked at as a generalization of a von Neumann measurement, where detections result in the inference of a density matrix rather than a pure state. As POVM measurement can be described using the designed quantum maximum entropy method \cite{QRE,QBR}, it can naturally be incorporated into Entropic Dynamics. In particular, the Section on the Quantum Bayes Rule (QBR) from Chapter \ref{Entropy Applications} can be added into the ED framework wholesale with the requirement that the $x$'s and $\mathbf{d}$'s which appear in the QBR are in-fact particle positions. The resulting density matrix in the Quantum Bayes Rule (\ref{QBayes3}), $\hat{\rho}_{\theta}=\,\stackrel{\sim}{\varphi}_{\theta|x}$, is the inferable that results from the position detection $x$ after a POVM measurement. This is analogous to the inferable $\ket{a_n}$ that results from the detection of $\mathbf{d}_n$ in the von Neumann measurement scheme (\ref{vnED}).

\section{Conclusion}
Both of the quantum measurement problems outlined in \cite{Schlosshauer} are completely solved within the Entropic Dynamics framework \cite{Johnson,EDMeasurement}. Both the unitary evolution of the wavefunction \emph{and} its collapse are probability updates compatible with entropic inference. The state evolves unitarily when the information available corresponds to the continuous positional motion of the particles (i.e. it undergoes the entropic dynamics update (\ref{step}) with updating drift potentials (\ref{Hamilton})), and if there is new information in the form of data, the entropic dynamics update halts, addresses the data, and updates the current probability distribution accordingly. Entropic updates with respect to data yield Bayesian probability updates \cite{GiffinBayes,Giffin}. Thus, the measurement problem of \emph{definite outcomes} is solved using inference techniques that are completely contained within the inference framework that is Entropic Dynamics. 

The measurement problem of \emph{preferred basis} is solved within ED due to the assumption, and ontological treatment throughout the derivation and measurement process, of definite yet unknown particle positions. ``Observables" other than positions are more aptly called ``inferables" as their values are inferred on the basis of position detections. The title of ``observables" is downgraded to the title of ``inferables", which plays a central role into why ED is not ruled out by the Bell-Kochen-Specker theorem next chapter.

\chapter{Entropic Dynamics and the sense in which \emph{no-go} theorems can \emph{go} again\label{No-go}}
This chapter follows \cite{EDContextuality} and continues to get back to the original quote by Jaynes (A) in the introduction. Now that we have derived precise notions of probability, probability updating, density matrix updating, Quantum Mechanics (QM) and quantum measurement in Entropic Dynamics (ED), we can address age-old \emph{no-go} theorems and tactfully avoid pitfalls that some interpretations of QM fail to do. In particular, we will discuss the following relevant \emph{no-go} theorems: Bell's theorem, the Bell-Kochen-Specker (BKS) theorem, and the more recent $\psi$-epistemic \emph{no-go} theorem by Pusey-Barret-Rudolph (PBR), in the context of ED. 

 On one hand, Quantum Mechanics is hugely successful in its ability to predict the set of eigenvalues, expectation values, and operators for a particle system of interest.  On the other hand, the states of interest hold intrinsic unpredictability, quantified by a probability distribution, except for a few trivial cases. This unpredictable nature, coupled with a desire to solve the quantum measurement problem(s) which are left open by QM's standard formalism, leaves a space for the many interpretations of QM to coexist inharmoniously within the community -- a community, no doubt, easily bothered by disharmony of any type. 

The community reduces and organizes this disharmony by ruling out interpretations and foundational theories of QM that disagree with the predictable findings of QM. This is done by first making a few seemingly reasonable assumptions a theory of QM may obey, and then by showing these assumptions lead to contradictions in the formalism, construct a \emph{no-go} theorem. This is the basis of Bell's theorem \cite{BellEPR}, the BKS theorem   \cite{BellKS,KS,Mermin}, as well as the findings of Pusey-Barret-Rudolph (PBR) \cite{Pusey} (reviewed in   \cite{Leifer}) on the epistemic interpretation of the wavefunction. 

Bell's theorem and the BKS theorem are intimately connected -- the failure of one sometimes implying the failure of the other, i.e., they are both capable of ruling out local hidden variable theories.\footnote{A review of many joint proofs of the BKS and Bell theorems, together with references to the original papers, can be found in \cite{Aravind}.} The interpretation of the BKS theorem is that operators in QM are \emph{contextual}, meaning that their character (or value) depend on the remaining set of commuting observables in a measurement setting. After considering the final results of these \emph{no-go} theorems, theories and interpretations of QM are sometimes classified using tables, for instance the 2 by 2 table: 
\begin{center}
\begin{tabular}{c|c|c| }

   & $\psi$-ontic & $\psi$-epistemic \\ 
\hline
 contextual & A & B \\  
\hline
 noncontextual & C& D\\
\hline
\end{tabular}
,
\end{center}
might be followed by statements like, ``theories of type ``C" or ``D" which have noncontextual operators are ruled out by the BKS theorem and ``B" is ruled out by PBR". A reader may be inclined to conclude that QM must be a theory of type ``A" (potential interpretation of Bohmian mechanics or Many Worlds).  The 2 by 2 table is by its nature an over simplification; it fails to span the entire set of plausible theories, and consequently interpretations, of QM. This is due the fact that $\emph{no-go}$ theorems are proofs by contradiction, and only theories which strictly adhere to their assumptions are ruled out.

In particular, this chapter shows that Entropic Dynamics (ED) is a theory of QM that lies on the line between theories ``B" and ``D", while not being ruled out by any of the aforementioned  \emph{no-go} theorems. We classify ED as a hybrid-contextual theory of QM because the positions of particles are treated noncontextually, as they are the preferred basis \cite{Johnson,EDMeasurement}, while it is shown that all other observables are treated contextually -- the main result of \cite{EDContextuality}. Although being ``hybrid-contextual", ED is not ruled out by the BKS theorem. Concepts in ED are naturally communicated in the language of probability, and for this reason, the operator language used in contextuality proofs do not naturally coincide with the language in ED -- this will be touched upon more later. We will discuss the $\psi$-epistemic \emph{no-go} theorem (PBR), Bell's Theorem, and finally the BKS theorem in the context of ED.

\section{\label{sec:level1}$\psi$-epistemic?}

In the previous chapters we claimed that $\psi$ is an epistemic object that helps represent our current knowledge of the system in question. This seemingly runs into conflict with the $\psi$-epistemic \emph{no-go} theorem from \cite{Pusey}; however, there is no issue. An excellent review of the $\psi$-epistemic/ontic dichotomy is presented in   \cite{Leifer}; however, the $\psi$-epistemic classifications there (and in \cite{Pusey}) do not meet the exact $\psi$-epistemic classification that ED formulates from the first principles of probability and probability updating. 

The first assumption in   \cite{Pusey} is 1) that ``a $\psi$-epistemic system has `physical states' upon which inferences may be made" (paraphrased). The ``physical states" of a system are denoted by $\lambda$. ED agrees with this assumption, and the variables which are ``physical" in ED are alone the definite yet unknown positions of particles.  The second assumption 2) is that ``systems which are prepared independently have independent physical states" (PIP). The PBR theorem finds a contradiction between assumption 2) and QM for systems which are prepared independently and then measured in an entangled basis, i.e., the physical states have dependencies in the entangled basis that are seemingly not present in their preparation. Thus, throwing away assumption 2) seemingly implies that $\psi$ must be ontic by their definition. 

The second assumption (called the PIP) is considered to have a weak point \cite{Leifer}, ``In my view, the weakest part of the PIP is the CPA, i.e., the idea that there should be no global properties of a system that are not reducible to properties of its subsystems when it is prepared in a product state". It is further stated ``... the only time global properties would
necessarily have to play a role is when a joint measurement is made, e.g.  a measurement in an entangled basis" (on which the PBR theorem depends) and, `` It would still be possible to work with separate systems completely independently
of  each  another,  in  blissful  ignorance  of  the  global  properties,  until  we  decide  to  do  an  experiment  that necessarily involves bringing the systems together.". The description of bringing the systems together in space is not present in the mathematical formalism of the PBR proof.  The measurement process in the PBR is treated like a ``black box", having inputs and outputs where nothing is discussed about what happens in the middle. Caticha \cite{Private} finds this to be a weak point in the PBR, it fails to take Bell's advice about measurement -- we should be careful and describe the full inference procedure. Rather than discussing the second assumption and the entire PBR proof, we will discuss how the notions of ``physicality" and $\psi$-epistemology in \cite{Pusey} and ED differ, which is reason enough for our version of $\psi$-epistemic states to not be ruled out.

The conclusion of \cite{Pusey} is that wavefunctions are ``physical properties" of quantum systems because the second assumption that ``systems which are prepared independently have independent physical states" fails to hold up in their definition of $\psi$-epistemic states. By their definition, a ``property" $L=L(\lambda)$ is a function of the ``physical states" $\lambda$ of the probability distribution $\mu(\lambda)$ in question. The set of properties $\{L,L',...\}$ are considered to be ``physical properties" iff $\mu_L(\lambda)$ and $\mu_{L'}(\lambda)$ do not overlap in $\lambda$. This guarantees that a measurement of $\lambda$ allows $L$ or $L'$ to be inferred uniquely. The following classical particle example is given in the PBR paper \cite{Pusey}:
\begin{quotation}
\noindent
``... if an experimenter knows only that
the system has energy $E$, and is otherwise completely
uncertain, the experimenter's knowledge corresponds to
a distribution $\mu_E(x, p)$ uniform over all points in phase
space with $H(x, p) = E$. Seeing as the energy is a physical property of the system, different values of the energy
$E$ and $E'$ correspond to disjoint regions of phase space,
hence the distributions $\mu_E(x, p)$ and $\mu_{E'}(x, p)$ have disjoint supports. On the other hand, if two probability
distributions $\mu_L(x, p)$ and $\mu_{L'}(x, p)$ have overlapping supports, i.e. there is some region $\Delta$ of phase space where
both distributions are non-zero, then the labels $L$ and $L'$ cannot refer to a physical property of the system."
\end{quotation}
From this classical example we see that the physicality of a ``property" $L$ is determined on the basis of whether or not any physical states $\lambda$ are shared between $\mu_L$ and $\mu_{L'}$. These definitions are considered in QM where one would like to know if the wavefunction is a ``physical property" of the system ($\psi$-ontic) or if it is not a physical property ($\psi$-epistemic). From \cite{Pusey}:
\begin{quotation}
\noindent
``Suppose that, for any pair of distinct quantum states $\ket{\psi_0}$
and $\ket{\psi_1}$, the distributions $\mu_0(\lambda)$ and $\mu_1(\lambda)$ do not overlap: then, the quantum state $\ket{\psi}$ can be inferred uniquely
from the physical state of the system and hence satisfies
the above definition of a physical property. Informally,
every detail of the quantum state is “written into” the
real physical state of affairs. But if $\mu_0(\lambda)$ and $\mu_1(\lambda)$ overlap for at least one pair of quantum states, then $\ket{\psi}$ can
justifiably be regarded as “mere” information."
\end{quotation}
In ED, the only ``physical states" and ``physical properties" are the positions of particles $x$ -- all other ``properties" $I$ are ``inferables". Independent of whether two probability  distributions are overlapping in $x$, such that $I$ or $I'$ can (or cannot) be uniquely inferred from position detections, the ``properties" $I$ and $I'$ are epistemic inferables (positions are ontic however). In ED, the ``physicality" of a property is not determined by one's ability to make an inference with certainty. The previous chapter shows this.\footnote{It also shows that the wavefunction itself is an inferable through position detections using the weak measurement and Weak Value scheme.} Where \cite{Pusey} finds contradiction between the notion that prepared quantum states cannot in general be uniquely inferred from measurements of $\lambda$, ED finds no contradiction -- probability theory is designed for the purpose of addressing one's lack of complete information, and thus, indeterminacy is commonplace in the ED framework. 

In \cite{Leifer} the following relevant comment is made about Bohmian Mechanics when addressed in the context of the PBR theorem:
\begin{quotation}
\noindent
``The  particle  positions  are  supposed  to  be  the  things  in  the
theory that provide a direct picture of what reality looks like to us, e.g.  when we observe the pointer of a
measurement device pointing to a specific value then it is the positions of the particles that make up the
pointer that determine this.  Nevertheless, the wavefunction is still needed as part of the ontology because it
determines how the particles move via the guidance equation.  The response of a measurement device to an
interaction with a system it is measuring depends on the wavefunction of the system as well as the particle
positions, so the wavefunction is still part of the ontic state, even if it is in some sense less primitive than
the particle positions."
\end{quotation}

Consider the following: In ED, ``interactions" occur \emph{between} the positions of the particles in the system of interest and positions of the particles in the detector. These interactions cause changes in the detector particle's positions, and then we make inductive inferences. The realization that these interacting fields are \emph{intermediary}, that they are predicated in experiment between particle-based-apparatuses and the particles of interest themselves, speaks to an interpretation that their function may simply be a convenient representation, or set of mechanisms, for \emph{describing} peculiar positional correlations between particles \cite{Private}. The particles themselves are ``doing whatever they are doing" and our model does its best to make inferences on the basis of the available information. The phase $\Phi$ is guided by the e-Hamilton equations (\ref{Hamilton}) and Hamiltonian, which informs us about the \emph{expected} drift of the particles through the expectation value constraints (\ref{kappa prime}, \ref{kappa prime prime}), rather than dictating the motion of the particles directly -- $(\rho,\Phi)$ is a location in e-phase space. The ontology of any intermediating fields can only be specified as far as the epistemic correlations (conditional dependencies) they build between ontic particle positions in $\rho(x)$. The conclusion from this assessment is not that ``fields are not real", but rather, that there is space for the inquisition ``must we demand in our model that these fields are real?" - the answer in ED is ``no, particles with peculiar probabilistic correlations is enough". In the absence of QED in ED, this is just a conjecture, but regardless, the conjecture works well enough for the model of QM in ED \cite{Private}.

As the leading assumptions of what entails a $\psi$-epistemic state differ, the $\psi$-epistemic \emph{no-go} theorem does not apply, which is admitted as a possible exemption to their \emph{no-go} theorem in the conclusion of \cite{Pusey}. We are therefore justified in treating $\psi$ epistemically by our own definition -- that $\psi=\sqrt{\rho}e^{i\Phi}$ is a convenient representation for epistemic probability distributions $\rho=|\psi|^2$ and how $\rho$ is inferentially updated, that has nothing to do with probability distribution overlap or preparation. Although the PBR theorem may be valuable for ``ontological models" \cite{Leifer}, it is not particularly valuable here.

\section{\label{sec:level1}Hidden Variables, Realism, and Non-locality}
The subject of hidden variables, realism, and non-locality in ED has been touched upon in   \cite{book,Johnson} and it will be further explored here. In Bell's landmark paper   \cite{BellEPR}, he found a contradiction between QM and hidden variable theories which claimed local realism. It was accomplished by considering a hidden variable $\lambda$, which if known, would give the outcome of an experiment (an eigenvalue of an operator) with certainty $a_0=A(\lambda=\lambda_0)$. By integrating over the probability of a hidden variable,
\begin{eqnarray}
\expt{A(\lambda)B(\lambda)}=\int p(\lambda) A(\lambda)B(\lambda)\, d\lambda,\label{Bellexpt}
\end{eqnarray}
he showed that such expectation values do not always agree with the expectations values of QM, for general $p(\lambda)$. 

In ED there is no such hidden variable. The particle dynamics is non-deterministic as can be seen by the Brownian like paths particles take due to the form of the transition probability $P(x'|x)$ in (\ref{transtime}), or after ``energy conservation", that the particles are undergoing a non-dissipative diffusion. The process that is deterministic in ED is the evolution of the probability distribution as it follows the e-Hamilton equations from (\ref{Hamilton}) given the appropriate constraints, boundary, and initial conditions are known. The phase of the wavefunction $\Phi(x)$ updates the probability distribution of particle locations rather than guiding each particle at every point. In the same fashion as above, ``The particles themselves are ``doing whatever they are doing" and our model does its best to make inferences on the bases of the available information. The phase $\Phi$ is guided by the e-Hamilton equations (\ref{Hamilton}), and Hamiltonian, which informs us about the \emph{expected} drift of the particles ... ".  The nonlocal nature of probability as a means for quantifying knowledge (of the future, past, or present) accounts for the ``quantum mechanical" nonlocal correlations between particles in ED.

As spin states are regularly used in Bell-type experiments, before ED can give a full account of the Bell experiment, the ED of spin must be developed fully. At this point it seems like spin states are represented by ``spin frames" \cite{Private,NickSpin} (Forthcoming by Caticha, Cararra) that provide additional information about the positions of particles in $\rho(x)$ much in the same fashion that $\Phi(x)$ does, and therefore, spin should be an epistemic inferable. 
Note that by detecting positions in a von Neumann or weak measurement setting, the spin may be inferred from position detections as is reviewed in Appendix \ref{spinexampleED}.
 
Using a multiple observer probability analysis, Bell's theorem was investigated in epistemic frameworks of QM \cite{MOPA}. As any collapse is an epistemic change in the system, each viable observer is obligated to assign distributions that coincide with their current state of knowledge of the system. If Alice and Bob are stationed at space-like separated measurement devices, they have access to different information throughout the experiment due to the observed order of events being different. I find that, at best, the Bell and the related CHSH inequality \cite{CHSH,Cavalcanti} can only be ``nonlocally violated counterfactually" as the CHSH is generated from the posing of an if-then question due to the asymmetry of each observer's local information.\footnote{\,\,i.e. \emph{if} Alice's measurement setting and outcome is $\theta_A$ and $+$, then Bob would expect the measurement outcomes to be ... ; however Bob does not actually know Alice's measurement setting and outcome until after she communicates it.} As the CHSH and Bell inequalities are expectation values, they are themselves epistemic inferables. The final result of \cite{MOPA} is that probabilities in QM over nonlocal measurement settings must have counterfactual (if-then) dependencies on their measurement settings as it cannot be verified otherwise by any local observer. This provides support for epistemic interpretations of the wavefunction and their use as a tool for inference.

 \section{\label{sec:level1}BKS type Theorems\label{section BKS}}

The BKS theorem shed light on the incompatibility of hidden variable theories and Quantum Mechanics   \cite{BellKS,KS}. Years later Mermin \cite{Mermin} demonstrated what is considered to be the simplest expression of what is usually an algebra and geometry intensive BKS theorem using observables. BKS proofs have been generalized to the $N$-qubit Pauli group   \cite{Cai}, and   \cite{Cabellocont} gives a BKS proof using continuous position and momentum observables. In   \cite{Cai}, they give a simple algorithm to convert observable based BKS proofs to a large number of projector based BKS proofs, so here we will focus on the simpler observable based proofs.

The class of hidden variable theories excluded by the BKS theorem satisfy the following seemingly reasonable conditions. The value of an operator is \emph{definite} yet \emph{unknown} such that we may assign it a preexisting value (its eigenvalue) by applying what is called a valuation \cite{Ballentine,Mermin,Cai}. The reason for introducing valuations is to make a connection to hidden variable theories (\ref{Bellexpt}) in which, given the hidden variable $\lambda$ is known, $A(\lambda)$ is known too. The alleged power of the proof below is that it is independent of the state $\ket{\psi}$. This immediately conflicts with the inferential ideology of QM in ED, however, we will continue to introduce the proof. The valuation of an operator $\hat{A}$ at any time is one of its eigenvalues,
\begin{eqnarray}
v(\hat{A})=\bra{a}\hat{A}\ket{a}=a.\label{val}
\end{eqnarray} 
 It is also assumed that functional relationships between the operators $f(\hat{A},\hat{B},\hat{C},...)$ should hold throughout the valuation process,
\begin{eqnarray}
v(f(\hat{A},\hat{B},\hat{C},...))=f(v(\hat{A}),v(\hat{B}),v(\hat{C}),...),\label{func}
\end{eqnarray}
as the values of the operators are supposed to take definite values. Thus it is found that operators must commute $v(\hat{A}\hat{B})=v(\hat{A})v(\hat{B})=v(\hat{B}\hat{A})$ when taking valuations for (\ref{func}) to hold. Mermin demonstrates the contradiction of equations (\ref{val}) and (\ref{func}) with Quantum Mechanics by considering what is now know as the Peres-Mermin Square: 
\begin{center}
\begin{tabular}{ |c|c|c| }
\hline
 ZI & IX & ZX \\ 
\hline
 IZ & XI & XZ \\  
\hline
 ZZ & XX & YY\\
\hline
\end{tabular}
.
\end{center}
Each table entry is an observable from the 2-qubit Pauli group consisting of a joint eigenbasis consisting of 4 eigenvectors. As a notational convenience we will omit tensor products when there is no room for confusion and let $Z=\sigma_z$ such that an arbitrary table entry $XI$ represents $\sigma_x^{(1)}\otimes I^{(2)}$, following the notational structure in   \cite{Cai}.  The standard matrix product of the operators along a given row or column is the rank 4 identity $I(4)=II$ (in this notation) with the exception of the last row, which is  $-II$. Consider the valuation of the standard matrix product of the elements of the first row,
\begin{eqnarray}
v(ZI \cdot IX \cdot ZX)=v(II)=1.
\end{eqnarray}
Supposing (\ref{func}) is true then
\begin{eqnarray}
v(ZI \cdot IX \cdot ZX)=v(ZI)v(IX)v(ZX)=1.\label{6.5}
\end{eqnarray}
The valuation of the $ij$th element $A^{ij}$ in the table is $v(A^{ij})=\pm 1$, and therefore (\ref{func}) imposes a constraint on the individual valuations $v(ZI)v(IX)v(ZX)=1$, which is only satisfied if either 0 or 2 of the valuations are $-1$. This cuts the number of possibilities from $2^3=8$ to $4$. Let $A^{i\odot}$ be the product of the operators in the $i$th row and $A^{\odot j}$  the product of the operators in the $j$th column such that above $A^{1\odot}=$$ZI\,\cdot\,$$IX\,\cdot\,$$ZX$ is the standard matrix product between the listed operators. Mermin showed his square indeed leads to a contradiction when considering the product of the row and column valuations,
\begin{eqnarray}
\prod_{i}v(A^{i\odot})v(A^{\odot i})=v(II)^5v(-II)=-1,\label{agree}
\end{eqnarray}
whereas applying (\ref{func}) to each row and column, $v(A^{i\odot})=\prod_jv(A^{ij})$, gives,
\begin{eqnarray}
\prod_{i}v(A^{i\odot})v(A^{\odot i})\rightarrow \prod_i\prod_jv(A^{ij})^2=1\label{contradiction},
\end{eqnarray}
which is a contradiction. This is due to the fact that not all of the elements in Mermin's square commute and therefore all of the observables cannot simultaneously be assigned definite eigenvalues. Quantum mechanical formalism and experiment agree with (\ref{agree}) and not with (\ref{contradiction}) and thus (\ref{func}) is ruled out. Bell makes a point that it may be overconstraining for the valuation to produce identical values when different sets of commuting observables are being considered, just to refute it by noting that a space-like separated observer could change which set of commuting observables he/she wishes to measure mid-flight. A hidden variable theory would then have to explain this nonlocal change in the valuation meaning that the BKS theory only refutes local hidden variables theories \cite{BellKS}.

\subsection{\label{sec:level1}Interpreting the Contradiction: Contextuality}

The standard interpretation of the contradiction by Bell, Kochen, Specker, Mermin and others is that quantum mechanical observables are \emph{contextual}, meaning that the operator's ``aspect", ``character", or ``value" depend on the remaining set of commuting observables under which it is considered, that is, which observables it would be measured along side with.  Any observable that does not depend on the remaining set of commuting observables in this way is called \emph{noncontextual}, which, for example, are the individual observables $v(A^{ij})$ from the Mermin square and (\ref{contradiction}). 

In more recent years the interpretation of the BKS theorem, which in principle would rule out all local hidden variable theories obeying (\ref{val}) and (\ref{func}), has been under scrutiny, in essence, for having a more restricted interpretation than the theorem claims. The work by   \cite{Clifton,Kent,Meyer} opens a loophole due to the impracticality of infinite measurement precision, and thus the BKS theorem is ``nullified" in their language. Appleby (and others) find the ``nullified" critique to be too harsh of a criticism   \cite{Appleby}. De Ronde   \cite{deRonde} points out that epistemic and ontic contextuality are consistently being scrambled into a omelet when perhaps the yoke and egg whites should be cooked separately.  He defines ``ontic contextuality" as the formal algebraic inconsistency of the operator and valuation formalism of Quantum Mechanics within the BKS theorem -- having nothing to do with measurement. Its epistemic counterpart is more aligned with the principles of Bohr in that Quantum Mechanics involves an interaction between the system and measurement apparatus whose outcomes are inevitably communicated in classical terms -- the context is given by the measurement device. The difference is subtle but, as noted, ``ontic contextuality" is defined to be independent of the differing interpretations of quantum mechanics whereas epistemic contextuality need not be. Our treatment of contextuality does separate in this fashion; however, de Ronde's usage of the word ``ontic" refers to the quantum formalism, whereas our usage only refers to ontic particle positions in ED.

\subsection{Critiques on representing onticity with valuations in QM\label{crit}}
As shown, the assumptions (\ref{val}) and (\ref{func}) lead to contradictions. The main critique we present is, ``how do we know that the valuation of an observable $v(\hat{A})$ accurately represents the notion of  \emph{definite, preexisting values} of an operator, that would be obtained if a measurement is carried out?".  The alleged strength of the BKS theorem is that the analysis has been done independent of the particular state $\ket{\Psi}$ and thus it should hold for all $\ket{\Psi}$ in general. This is troubling for a number of reasons, the first being that a particular $\ket{\Psi}$ may not have components along every eigenvector of an operator $\hat{A}$, in which case a zero probability event could be assigned a definite existence, and one would never know because $\ket{\Psi}$, which all of the observables in question pertain to, has not been specified. This issue here is an interplay between the ontic and epistemic contextuality given by de Ronde, because only sensible valuations may be given if the state of the system is known -- in general the density matrix $\hat{\rho}$. 

If the valuation process is to be applicable to arbitrary ``observables" independent of the state at hand, then one runs into another logical inconsistency when attempting to apply valuations to a density matrix, $\hat{\rho}$, because it represents the probabilistic state of a system. It makes little sense to have different sets of commuting observables $\{\hat{\rho}_{1},\hat{\rho}_{2},\hat{\rho}_{3}...\}$ which are required to span the same Hilbert space as the state in question $\ket{\Psi}$ (or $\hat{\rho}$). Furthermore, the valuation of a density matrix $\hat{\rho}=\sum_ip_i\ket{i}\bra{i}$ gives one of its eigenvalues, $p_i$, which are probabilities themselves and are never \emph{directly} observed, but are usually inferred from the frequency of a large number of independent trials. One cannot possibly claim that a system is ontically expressing a definite preexisting probability value $p_i$. Probability by its nature is a measure of the uncertainty of a state $\ket{i}$ rather than a value (physically) carried by the state $\ket{i}$  -- which is as epistemic as it gets! If Alice knowingly prepares one system and Bob does not know which system Alice has prepared, then it is clear that $p_i$'s cannot have a definite existence because both Alice and Bob disagree about said values over the same single ``ontic" system of interest. Furthermore, when a measurement is made to determine the state, the probability value updates (the eigenvalue changes) and in this sense the assignment of an eigenvalue $\hat{\rho}$ through valuation represents nothing physical about the state of the system's \emph{definite, preexisting values} that would in principle be obtained if a measurement was carried out. If this one $\hat{\rho}$ valuation counter example can be found, it is unclear how many other observables would also be counter examples. In general the eigenvalues of operators do not represent  \emph{definite, preexisting} (noncontextual) \emph{values} of an operator that would be obtained if a measurement was carried out.

Due to these critiques, and that in ED one may infer eigenvalues from position detections, it is difficult to know what precisely a valuation procedure represents meta-physically, besides the simple choice of a matrix element. As discussed, the valuation of an operator may not always represent an ontic value of an observable, and therefore we suggest relaxing this notion and replacing it by the more general statement, ``The valuation of an operator (or set of operators) represents a quantity that in principle may be inferred", or in the language of   \cite{EDMeasurement}, ``The valuation of an observable is an \emph{inferable} of the theory". In this sense, the problem presented by the BKS theorem never truly arises in the ED framework -- operators, simply put, are epistemic inferables.

\subsection{\label{sec:level1}ED: A hybrid-contextual theory\label{section HCT}}

It should be noted that in Entropic Dynamics, the idea of valuation is very unnatural. An inference based theory allows us to state, quantify, and represent how much we do \emph{not} know about the state of a system through a probability distribution, upon which we use the rules of inference and probability updating to determine what we do. The resulting interpretation of the BKS theorem (contextuality) when performed in the ED framework ``is that operators and their eigenvalues do not in general pertain to the ontology of particles" -- they are, in general, inferables. Although a precisely prepared state $\ket{a}$ can be measured (inferred) with certainty using a unitary measurement device from Section \ref{Unitary measurement devices}, it is important to recall the note, ``... there
is no implication that the particle previously had or now currently has the value $a$" \cite{Johnson,EDMeasurement}. 

Strictly speaking, the BKS theorem discards realist theories in which all of the considered operators are treated ontically through their valuation. This leaves open the possibility for a hybrid-contextual theory in which only a subset of commuting observables are \emph{definite yet unknown}, or noncontextual, while other variables (or sets of commuting observables) are contextual inferables. 

The only operators that are required to undergo valuation in ED are the $3N$-particle position coordinates with their corresponding $3N$ operators $\hat{X}^{(n)}$ as they are the preferred basis. In the language of valuations, we have,
\begin{eqnarray}
v_x(\hat{x}^{(n)}_i)\equiv \bra{x_i^{(n)}}\hat{x}^{(n)}_i\ket{x_i^{(n)}}=x^{(n)}_i,
\end{eqnarray}
for a particular coordinate $x_i^{(n)}$.
Position operators trivially obey (\ref{func}),
\begin{eqnarray}
v_x(f(\hat{x}^{(n)}_i,\hat{x}^{(m)}_j,...))=f(v(\hat{x}^{(n)}_i),v(\hat{x}^{(m)}_j),...),
\end{eqnarray}
for any function $f$, because all position operators mutually commute.  No parity contradiction in the sense of   \cite{Mermin,Cai} can be reached. At the end of the day, the BKS proofs are proofs by contradiction which means that a set of counter examples has been found which rule out the general applicability of assigning definite yet unknown values to all operators all the time. However, as seen above, there are instances in which there is no contraction and the assignment of definite yet unknown values in this instance is consequently \emph{not} ruled out.

Operators other than position, $A^{ij}$, need not be noncontextual in ED as they are considered to be epistemic in nature. In this case, one should not claim $A^{ij}$, one of its eigenvalues $a_n^{ij}$, or a state $\ket{a^{ij}}=\int dx \,\psi_{a^{ij}}(x)\ket{x}$, to have a definite existence outside of characterizing our knowledge of the definite yet unknown positions of particles $x$. That being said, the operators $A^{ij}$ \emph{can in general be expanded and interpreted in the position basis}. When applying positional valuations to $A^{ij}$, we find that they are naturally contextual in the sense that equation (\ref{func}) fails to hold in general. Although the following argument does hold true mathematically, the valuation process is unwarranted in ED as ED never claims the operator $A^{ij}$ any of its eigenvalues, $a^{ij}_n$, or the results of their valuations, to be ontic -- it is meaningless to talk about the noncontextuality of non-position variables. We will proceed anyhow for completeness. 

\paragraph{Proof that position valuations of non-position operators are contextual:}
If one were to perform the valuation of an arbitrary (non-position) operator $A^{ij}$ in ED before measurement, because the positions of particles are the preferred basis, the only valuation function worth considering, $v_x$, is the one that considers definite particle positions.\footnote{The Hermitian $A^{ij}$ operators are arbitrary: they could be tensor products of 1D projectors, standard QM observables of interest like $\hat{H}$ or $\hat{p}$, or subsets of them.} Thus, we should consider valuation of the diagonal matrix elements of $A^{ij}$ in the position basis (here let $\ket{x}\equiv\ket{x_1}\otimes\ket{x_2}...\otimes\ket{x_N}$ for the $N$ particles and $N$ operators that are tensor multiplied in $A^{ij}$), 
\begin{eqnarray}
v_{x_0}(A^{ij})&\equiv& \bra{x_0}A^{ij}\ket{x_0}=\bra{x_0}\sum_n\ket{a_n^{ij}}a_n^{ij}\braket{a_n^{ij}}{x_0}\nonumber\\
&=&\sum_na_n^{ij}|\braket{x_0}{a_n^{ij}}|^2\neq a_n^{ij},
\end{eqnarray}
where in this case it is supposed that the definite yet unknown value of $x$ is $x_0$. The positional valuation $v_x(A^{ij})$ is not one of the eigenvalues or ``observables" of $A^{ij}$, but in ED, $A^{ij}$ is an inferable and so is $v_x(A^{ij})$ (through for instance the use of a Weak Value \cite{CaiPigeon}). The positional valuation $v_x(A^{ij})$ is some real number that in principle may be assigned to the $x$th position coordinate, and potentially has nothing to do with the state of the system $\ket{\Psi}$ as is remarked in the critiques in Section \ref{crit}. The position valuation $v_x(A^{ij})$ may be interpreted as an expected value of the operator $A^{ij}$ if the position of the particle(s) were known to be exactly at the value $x_0$ in configuration space, that is, $\int v_x(A^{ij})\delta(x-x_0)dx=\int\bra{x}A^{ij}\ket{x}\delta(x-x_0)dx= \expt{A^{ij}}_{x_0}=v_{x_0}(A^{ij})$. The actual location of the particle is not known in general and should be weighted by the appropriate $\rho(x)$ if one is considering expectation values of $A^{ij}$. Parity type proofs of the BKS theorem require  $A^{ij}$ to be simultaneously part of an even number of sets of commuting observables   \cite{Cai}. This means an operator $A^{ij}$ is simultaneously diagonalized in (at-least two) different basis,
\begin{eqnarray}
A^{ij}=\sum_n\ket{a_n^{i\odot}}a_n^{ij}\bra{a_n^{i\odot}}=\sum_n\ket{a_n^{\odot j}}a_n^{ij}\bra{a_n^{\odot j}},
\end{eqnarray}
where, for example in the Peres-Mermin square, $\ket{a_n^{i\odot}}$ refers to the eigenvectors of the commuting set of variables from the $i$th row and $\ket{a_n^{\odot j}}$ refers to the eigenvectors of the commuting set of variables from the $j$th column; however, the argument is valid for all (non-spin) KS-sets. The largest number of distinct sets of eigenvectors is equal to the number of sets of commuting observables in the BKS proof (the number of rows and columns of the Peres-Mermin square). Using this notation we may denote the product of the operators in a commuting set by,
\begin{eqnarray}
A^{i\odot}=\sum_n\ket{a_n^{i\odot}}a_n^{i1}a_n^{i2}...a_n^{iN}\bra{a_n^{i\odot}}=\sum_n\ket{a_n^{i\odot}}a_n^{i\odot}\bra{a_n^{i\odot}},\nonumber\\
\end{eqnarray}
where $N$ is the number of commuting observables in the set of commuting observables $\{A^{ij}\}_i$. In general, the application of (\ref{func}) to the position valuations of $\{A^{ij}\}_i$ will not hold,
\begin{eqnarray}
v_x(A^{i\odot})\rightarrow \prod_jv_x(A^{ij}),
\end{eqnarray}
because it would require,
\begin{eqnarray}
\sum_n|\braket{x_0}{a_n^{i\odot}}|^2a_n^{i\odot}= \prod_j\Big(\sum_na_n^{ij}|\braket{x_0}{a_n^{i\odot}}|^2\Big)_j,\label{notequal}
\end{eqnarray}
which holds in a few trivial cases (identity, null, orthogonal), but not in general. Note that this equation is the analog of (\ref{6.5}), expect with position valuations, and here, equation (\ref{func}) already fails to hold in general. This poses no issue in ED because $A^{i\odot}$ or the individual $A^{ij}$ need only exist epistemically, so their valuations (matrix elements) need not agree - the product of matrix elements need not be the matrix element of the product so imposing equality is nonsensical. Equations like (\ref{func}) do not hold true in general because, if valuations are interpreted as \emph{inferables} (Section \ref{crit}), then expecting something like $v_x(A^2)=v_x(A)^2$ to hold true is analogous to expecting expectation values like $\expt{A^2}_x=\expt{A}_x^2$ to hold true, which of-course is not true in general.  

Furthermore, ``ontic contextuality" in the sense of de Ronde (contextuality due to the operator formalism of QM) is preserved among non-position observables (for noncontextual position). Furthermore, if (\ref{func}) is applied to the product of all of the commuting sets of observables, 
\begin{eqnarray}
\prod_{i}v_x(A^{i\odot})v_x(A^{\odot i})\rightarrow \prod_i\prod_jv_x(A^{ij})^2\geq0,
\end{eqnarray}
for situations when the LHS is less than zero or it is simply not equal to the RHS (compounding from (\ref{notequal})). This calculation shows that definite (noncontextual) positions before measurement do not imply definite (noncontextual) $A^{ij}$, and therefore, we are further justified in treating the operators $A^{ij}$ contextually - which means we should not apply valuations to them, or if we do, we should not expect (\ref{func}) to hold. The current form of ED would potentially be ruled out if $A^{ij}$ were required to be ontological - but this is not the case.

Because position operators always mutually commute with one another, and are therefore all simultaneously diagonalizable in the same set of position eigenvectors (i.e. $\ket{x}=\ket{x^{i\odot}}=\ket{x^{\odot j}}=\ket{x^{\odot\odot}}$), they may be treated noncontextually together. If an operator is a (tensor or dot) product of contextual and noncontextual operators, it remains contextual as it is epistemic (or equally well due to the lack of equality in (\ref{notequal})). This can be seen by applying position space valuations to the continuous operators defined in \cite{Cabellocont}. Spin in ED is epistemic (forthcoming \cite{Private,NickSpin}) so their noncontextual valuations are not required as well and thus spin will be contextual in general. As noted in the critiques, the valuation of an operator may not always express the definite yet unknown values of an observable --  it may be best to relax this notion such that the valuation of an operator represents a quantity that in principle may be inferred, an inferable, in general.
\paragraph{Inferring contextual operators:} In the previous section, the positional valuations of the operators are a bit obscure, partially because the measurement process in ED was not included (it is as if positions measurements were made before applying a unitary measurement device, so the eigenvalues of $A$'s were not actually being inferred). A question of interest is, how, if everything is to be measured or inferred using a (non-contextual) position basis (Chapter \ref{Measurement in Entropic Dynamics} and   \cite{Johnson,EDMeasurement}), is the contextual nature of a set of contextual operators $\{\hat{A}^{ij}\}$ non-contradictory? This question is especially tricky because it mixes the epistemic and ontic notions of contextuality in the sense of   \cite{deRonde}, who, quote Mermin , ``the whole point of an experimental test of BKS [theorem] misses the point.".  ED perhaps sheds some light onto Mermin's statement about the lack of an experimental test of the BKS. 

Suppose Alice prepares a two particle system and sends it to Bob who has a compound unitary measurement device (\ref{unitary}) for each set of commuting observables (each row and column) of the Mermin square (for simplicity), but really this is applicable to any construction of sets of commuting observables. Because Bob can only measure one row or column for a given pair of particles sent from Alice, him choosing the $i$th row or column means he has chosen and applied the unitary measurement device $\hat{U}_{A^{i\odot}}$ to the incoming state and mapped it to position coordinates for detection and inference. This measurement device is designed to infer one set of commuting observables at a time from position detections. That is, the physical application of $\hat{U}_{A^{i\odot}}$ picks, $P^i$, the $i$th set of commuting observables,\footnote{It should be noted that the square of $A$'s is depicted for the purpose of illustrating that measurement devices pick a set of commuting observables. This and the previous sections refer to arbitrary (non-spin) KS-sets, not just those that can be arranged in squares.} 

\begin{center}
$P^i \,*\,$
\begin{tabular}{ |c|c|c| }
\hline
 $A^{11}$ & $A^{12}$ & $A^{13}$ \\ 
\hline
 $A^{21}$& $A^{22}$ & $A^{23}$\\  
\hline
 $A^{31}$& $A^{32}$ & $A^{33}$\\
\hline
\end{tabular}
$\longrightarrow \hat{U}_{A^{i\odot}}$ \begin{tabular}{ |c|c|c| }
\hline
 $A^{i1}$ & $A^{i2}$ & $A^{i3}$ \\ 
\hline
 \end{tabular}
\end{center}
\begin{center}
$\longrightarrow$
\begin{tabular}{ |c|c|c| }
\hline
 $x^{i1}$ & $x^{i2}$ & $x^{i3}$ \\ 
\hline
 \end{tabular}
,
\end{center}
and at a later time one may apply valuation(s) to the associated position operators if one wishes because the operators are diagonal in the position basis (at that later time) $A^{ij}\rightarrow A^{ij}(t)=\sum_n\ket{x^{ij}_n}a^{ij}_n\bra{x^{ij}_n}$. The positions may be detected and the associated commuting set of $A^{ij}$ may be \emph{inferred}. The notion of detectors picking sets of commuting observables is mentioned in \cite{Mermin}, but here the process is specified to show how contextuality is preserved. Observables $A^{ij}$ not in row $i$ (after $\hat{U}_{A^{i\odot}}$ is applied) are not in general diagonal in $x$ as they did not originally commute with all of the observables in $i$ by construction, and naturally, their eigenvalues are not being inferred. It is as if picking a set of commuting observables, by the application of a unitary measurement device, rotates the entire KS-set (the square in this illustration) about the picked ``axis" (the picked set of commuting observables), the picked set of commuting observables now being diagonal in position. Picking other sets of commuting observables requires using a different unitary measurement device $\hat{U}_{A^{i'\odot}}$, which rotates the whole KS-set differently (about a different ``axis" for the sake of the analogy). Thus, mixtures of noncontexutal position operators with epistemic contextual operators are contextual as a unitary measurement device will be needed for the inference of the contextual operators, which ``rotates" the KS-set. 

The operators $A^{ij}$ are treated contextually, or hybrid contextually in the preferred basis.  As only one set of commuting observables may be picked at a time by Bob, the quantum mechanical expectation values match that as inferred by Bob (and are therefore in the form of (\ref{agree})). Alice, being in the dark, does not know which row Bob will pick and is free to assign a probability Bob picks the $i$th row or column, and after learning the chosen row or column may she update her probability accordingly \cite{MOPA}. 

\section{Discussion}
The most natural inferential tool in ED is probability. The critiques given in Section \ref{crit} are further motivation for the use of probability to make rational inferences, while the interpretation of valuation functions as assigning ontic values, which inevitably lead to paradox in the BKS theorem, is not generally applicable.  There, reason was given for the need of a more general interpretation of the valuation of an operator, which was stated, ``The valuation of an operator (or set of operators) represents a quantity that in principle may be inferred". Because the probability of a state is only defined in terms of its set of commuting observables, and because there is no way to generate a unique joint probability distribution among non-commuting observables   \cite{Cohen}, a rational discussion on the potential simultaneous onticity of non-commuting observables is not possible. As ED formulates QM as an application of entropic inference while assuming the position of particles to be the only ontic variables, the positions form the preferred basis for inference, and therefore ED is a ``hybrid-contextual" theory of QM that does not violate any of the relevant aforementioned $\emph{no-go}$ theorems. Although it is unverifiable if reality truly consists of particles with ontic position, the concept and model is demonstratively useful for making predictions and for analyzing the results of experiment. In a pragmatic approach \cite{Pragmatic,Private}, when a model turns out to be trustworthy and reliable, we recognize its success and say the model or theory is ``true" and the ontic elements are ``real".

\chapter{Conclusions}

Presented in this thesis is the synthesis of probability and entropic inference with Quantum Mechanics and quantum measurement. The standard and quantum relative entropies were shown to be designed for the purpose of updating probability distributions and density matrices, respectively, from the same inferential origins. The Quantum Bayes Rule and collapse were derived as an application of the quantum relative entropy maximization, which unifies topics in quantum measurement and Quantum Information through entropic inference -- similar to the unification of Bayesian probability updating and the standard maximum entropy method in \cite{GiffinBayes,Giffin,GiffinThesis}. Because the quantum maximum entropy method is able to simultaneously process inferences that neither a Quantum Bayes procedure nor a maximum von Neumann entropy procedure can process alone, the designed quantum maximum entropy method may be considered ``a universal method of density matrix inference". Because the quantum maximum entropy method only utilizes the standard quantum mechanical formalism, it, and its results, may be appended to the standard quantum mechanical formalism. In this sense, the quantum measurement problem of collapse is solved within the quantum mechanical formalism, but the quantum measurement problem of preferred basis is not.

The derived interpretations of probability and its entropic updating allows one to formulate Quantum Mechanics as an application of entropic inference called Entropic Dynamics. In Entropic Dynamics, particles have definite yet unknown positions and probabilities are purely epistemic. This separates Entropic Dynamics from other theories and interpretations of Quantum Mechanics because one may address the quantum measurement problem and quantum \emph{no-go} theorems in a new light. Pivotal to both of these discussions is the concept that \emph{observables} in Quantum Mechanics may better be stated as \emph{inferables}, that is, quantities one may wish to infer. Entropic Dynamics suggests this change of language, \emph{observables} $\rightarrow$ \emph{inferables}, when it derives Quantum Mechanics as an application of inference; and although on the surface this change of language seems purely semantical, it has deep implications for the interpretation of Quantum Mechanics. The combination in Entropic Dynamics of epistemic inferables that lack an ontological predisposition, the positions of particles that are ontological, and use of epistemic probability theory, end up resolving the measurement problems of preferred basis and definite outcomes, as well as the \emph{no-go} theorems of $\psi$-epistemic, Bell's inequality, and the BKS theorem, in effect, without ever being a problem in Entropic Dynamics in the first place \cite{Johnson,EDMeasurement,EDContextuality}.  Because Entropic Dynamics derives Quantum Mechanics from the standard maximum entropy method, as well as the beginnings of a density matrix formalism added here, the quantum maximum entropy method can be derived from Entropic Dynamics; and therefore, the standard maximum entropy method retains its title as the ``universal method of inference" \cite{GiffinBayes,Giffin,GiffinThesis}.

In the spirit of E.T. Jaynes' quote (A) presented in the introduction to this thesis, through the precise design of probability theory, the maximum entropy method, and the use of Entropic Dynamics, we have managed to push the theory of Quantum Mechanics past some of the derailing ambiguities of its standard formalism. The culmination of the articles reviewed in this thesis has led to a better understanding of Quantum Mechanics and quantum measurement through inference and Entropic Dynamics.

\bibliographystyle{unsrtnat}
\bibliography{bibthesis}


\newpage
\appendix
\addcontentsline{toc}{chapter}{Appendices:}
\chapter{Functional Equations and Entropy\label{appendix1}}
The Appendix loosely follows the relevant sections in \cite{Aczel}, and then uses the methods reviewed to solve the relevant functional equations for $\phi$, following \cite{QRE}.
\section{Simple Functional Equations}


If Cauchy's functional equation
\begin{eqnarray}
f(x+y)=f(x)+f(y)\label{Cauchy}
\end{eqnarray}
is satisfied for all real $x$, $y$, and if the function $f(x)$ is (a) continuous at one point, (b) nonegative for small positive $x$'s, or (c) bounded in an interval, then,
\begin{eqnarray}
f(x)=cx\label{linear}
\end{eqnarray}
is the solution to (\ref{Cauchy}) for all real $x$. If (\ref{Cauchy}) is assumed only over all positive $x$, $y$, then under the same conditions,~(\ref{linear}) holds for all positive $x$.

The most natural assumption for our purposes is that $f(x)$ is continuous at a point (which later extends to continuity all points as given by Darboux \cite{Darboux}). Cauchy solved the functional equation by induction. In particular, Equation (\ref{Cauchy}) implies,
\begin{eqnarray}
f(\sum_{i}x_i)=\sum_if(x_i),
\end{eqnarray}
and if we let each $x_i=x$ as a special case to determine $f$, we find 
\begin{eqnarray}
f(nx)=nf(x).
\end{eqnarray}
We may let $nx=mt$ such that
\begin{eqnarray}
f(x)=f(\frac{m}{n}t)=\frac{m}{n}f(t).
\end{eqnarray}
Letting $\lim_{t\rightarrow 1}f(t)=f(1)=c$ gives
\begin{eqnarray}
f(\frac{m}{n})=\frac{m}{n}f(1)=\frac{m}{n}c,
\end{eqnarray}
and, because for $t=1$,  $x=\frac{m}{n}$ above, we have
\begin{eqnarray}
f(x)=cx,
\end{eqnarray}
which is the general solution of the linear functional equation. In principle, $c$ can be complex. The~importance of Cauchy's solution is that it can be used to give general solutions to the following Cauchy equations:
\begin{eqnarray}
f(x+y)&=&f(x)f(y),\\
f(xy)&=&f(x)+f(y),\\
f(xy)&=&f(x)f(y),
\end{eqnarray}
by performing consistent substitution until they are the same form as (\ref{Cauchy}), as given by Cauchy. We will briefly discuss the first two.

The general solution of $f(x+y)=f(x)f(y)$ is $f(x)=e^{cx}$ for all real or for all positive $x,y$ that are continuous at one point and, in addition to the exponential solution, the solution $f(0)=1$ and $f(x)=0$ for ($x>0$) are in these classes of functions.

The first functional $f(x+y)=f(x)f(y)$ is solved by first noting that it is strictly positive for real $x$, $y$, $f(x)$, which can be shown by considering $x=y$,
\begin{eqnarray}
f(2x)=f(x)^2> 0.
\end{eqnarray}
If there exists $f(x_0)=0$, then it follows that $f(x)=f((x-x_0)+x_0)=0$, a trivial solution, hence the reason why the possibility of being equal to zero is excluded above. Given $f(x)$ is nowhere zero, we are justified in taking the natural logarithm $\ln(x)$, due to its positivity $f(x)>0$. This gives,
\begin{eqnarray}
\ln(f(x+y))=\ln(f(x))+\ln(f(y)),
\end{eqnarray}
and letting $g(x)=\ln(f(x))$ gives,
\begin{eqnarray}
g(x+y)=g(x)+g(y),
\end{eqnarray}
which is Cauchy's linear equation, and thus has the solution $g(x)=cx$. Because  $g(x)=\ln(f(x))$, one finds in general that $f(x)=e^{cx}$.

If the functional equation $f(xy)=f(x)+f(y)$ is valid for all positive $x,y$ then its general solution is $f(x)=c\ln(x)$ given it is continuous at a point. If $x=0$ (or $y=0$) are valid, then the general solution is $f(x)=0$. If all real $x,y$ are valid except $0$, then the general solution is $f(x)=c\ln(|x|)$. 

 In particular, we are interested in the functional equation $f(xy)=f(x)+f(y)$ when $x,y$ are positive. In~this case, we can again follow Cauchy and substitute $x=e^u$ and $y=e^v$ to get,
\begin{eqnarray}
f(e^ue^v)=f(e^u)+f(e^v),\label{87}
\end{eqnarray}
and letting $g(u)=f(e^u)$ gives $g(u+v)=g(u)+g(v)$. Again, the solution is $g(u)=cu$ and, therefore, the general solution is $f(x)=c\ln(x)$ when we substitute for $u$. If $x$ could equal $0$, then $f(0)=f(x)+f(0)$, which has the trivial solution $f(x)=0$. The general solution for $x\neq0$, $y\neq0$ and $x,y$ positive is therefore $f(x)=c\ln(x)$. 

\section{Functional Equations with Multiple Arguments}

From \cite{Aczel} pages 213--217. Consider the functional equation,
\begin{eqnarray}
F(x_1+y_1,x_2+y_2,...,x_n+y_n)=F(x_1,x_2,...,x_n)+F(y_1,y_2,...,y_n),\label{GenCauchy}
\end{eqnarray}
which is a generalization of Cauchy's linear functional Equation (\ref{Cauchy}) to several arguments. Letting $x_2=x_3=...=x_n=y_2=y_3=...=y_n=0$ gives
\begin{eqnarray}
F(x_1+y_1,0,...,0)=F(x_1,0,...,0)+F(y_1,0,...,0),
\end{eqnarray}
which is the Cauchy linear functional equation having solution $F(x_1,0,...,0)=c_1x_1$, where $F(x_1,0,...,0)$ is assumed to be continuous or at least measurable majorant. Similarly,
\begin{eqnarray}
F(0,...,0,x_k,0,...,0)=c_kx_k,
\end{eqnarray}
and if you consider
\begin{eqnarray}
F(x_1+0,0+y_2,0,...,0)=F(x_1,0,...,0)+F(0,y_2,0,...,0)=c_1x_1+c_2y_2,
\end{eqnarray}
and, as $y_2$ is arbitrary, we could have let $y_2=x_2$ such that in general
\begin{eqnarray}
F(x_1,x_2,...,x_n)=\sum_i c_ix_i,\label{genlinearsol}
\end{eqnarray}
formulating the general solution.

\section{Relative Entropy\label{appendix1}}
We are interested in the following functional equation:
\begin{eqnarray}
\phi(\rho_1\rho_2,\varphi_1\varphi_2)=\phi(\rho_1,\varphi_1)+\phi(\rho_2,\varphi_2).
\end{eqnarray}
This is an equation of the form,
\begin{eqnarray}
F(x_1y_1,x_2y_2)=F(x_1,x_2)+F(y_1,y_2),
\end{eqnarray}
where $x_1=\rho_1$, $y_1=\rho_2$, $x_2=\varphi_1$, and $y_2=\varphi_2$. First, assume all $\rho$ and $\varphi$ are greater than zero. Then, substitute: $x_i=e^{x_i'}$ and $y_i=e^{y_i'}$ and let $F'(x_1',x_2')=F(e^{x_1'},e^{x_2'})$ and so on such that
\begin{eqnarray}
F'(x_1'+y_1',x_2'+y_2')=F'(x_1',x_2')+F'(y_1',y_2'),
\end{eqnarray}
which is of the form of (\ref{GenCauchy}). The general solution for $F$ is therefore
\begin{eqnarray}
F'(x_1'+y_1',x_2'+y_2')&=&a_1(x_1'+y_1')+a_2(x_2'+y_2')=a_1\ln(x_1y_1)+a_2\ln(x_2y_2)\nonumber\\&=&F(x_1y_1,x_2y_2),
\end{eqnarray}
which means the general solution for $\phi$ is
\begin{eqnarray}
\phi(\rho_1,\varphi_1)&=&a_1\ln(\rho_1)+a_2\ln(\varphi_1).
\end{eqnarray}
In such a case, when $\varphi(x_0)=0$ for some value $x_0\in \mathcal{X},$ we may let $\varphi(x_0)=\epsilon$, where $\epsilon$ is as close to zero as we could possibly want---the trivial general solution $\phi=0$ is saturated by the special case when $\rho=\varphi$ from DC1'. Here, we return to the text.

\section{Matrix Functional Equations}
(This derivation is implied in \cite{Aczel} pages 347--349). First, consider a Cauchy matrix functional equation,
\begin{eqnarray}
f(\hat{X}+\hat{Y})=f(\hat{X})+f(\hat{Y}),\label{102}
\end{eqnarray}
where $\hat{X}$ and $\hat{Y}$ are $n\times n$ square matrices. Rewriting the matrix functional equation in terms of its elements gives
\begin{eqnarray}
f_{ij}(x_{11}+y_{11},x_{12}+y_{12},...,x_{nn}+y_{nn})=f_{ij}(x_{11},x_{12},...,x_{nn})+f_{ij}(y_{11},y_{12},...,y_{nn})
\end{eqnarray}
and is now in the form of (\ref{GenCauchy}), and, therefore, the solution is
\begin{eqnarray}
f_{ij}(x_{11},x_{12},...,x_{nn})=\sum_{\ell,k=0}^n c_{ij\ell k}x_{\ell k}
\end{eqnarray}
for $i,j=1,...,n$. We find it convenient to introduce super indices, $A=(i,j)$ and $B=(\ell, k)$ such that the component equation becomes
\begin{eqnarray}
f_{A}=\sum_{B} c_{AB}x_{B},\label{cab}
\end{eqnarray}
and resembles the solution for the linear transformation of a vector from \cite{Aczel}. In general, we will be discussing matrices $\hat{X}=\hat{X}_1\otimes\hat{X}_2\otimes...\otimes\hat{X}_N$ which stem from tensor products of density matrices. In this situation, $\hat{X}$ can be thought of as $2N$ index tensor or a $z\times z$ matrix where $z=\prod_{i}^N n_i$ is the product of the ranks of the matrices in the tensor product or even as a vector of length $z^2$. In such a case, we may abuse the super index notation where $A$ and $B$ lump together the appropriate number of indices such that (\ref{cab}) is the form of the solution for the components in general. The matrix form of the general solution is
\begin{eqnarray}
f(\hat{X})=\widetilde{C}\hat{X},
\end{eqnarray}
where $\widetilde{C}$ is a constant super-operator having components $c_{AB}$.
\section{Quantum Relative Entropy\label{appendix2}} The functional equation of interest is
\begin{eqnarray}
\phi\Big(\hat{\rho}_{1}\otimes\hat{\rho}_2,\hat{\varphi}_1\otimes\hat{\varphi}_2\Big)=\phi\Big(\hat{\rho}_{1}\otimes\hat{1}_2,\hat{\varphi}_1\otimes\hat{1}_2\Big)+\phi\Big(\hat{1}_1\otimes\hat{\rho}_2,\hat{1}_1\otimes\hat{\varphi}_2\Big).
\end{eqnarray}
These density matrices are Hermitian, positive semi-definite, have positive eigenvalues, and are not equal to $\hat{0}$. Because every invertible matrix can be expressed as the exponential of another matrix, we can substitute $\hat{\rho}_{1}=e^{\hat{\rho}_{1}'}$, and so on for all four density matrices giving,
\begin{eqnarray}
\phi\Big( e^{\hat{\rho}_{1}'}\otimes e^{\hat{\rho}_2'},e^{\hat{\varphi}_1'}\otimes e^{\hat{\varphi}_2'}\Big)=\phi\Big(e^{\hat{\rho}_{1}'}\otimes\hat{1}_2,e^{\hat{\varphi}_1'}\otimes\hat{1}_2\Big)+\phi\Big(\hat{1}_1\otimes e^{\hat{\rho}_2'},\hat{1}_1\otimes e^{\hat{\varphi}_2'}\Big).
\end{eqnarray}
Now, we use the following identities for Hermitian matrices: 
\begin{eqnarray}
 e^{\hat{\rho}_{1}'}\otimes e^{\hat{\rho}_2'}=e^{\hat{\rho}_{1}'\otimes\hat{1}_2+ \hat{1}_1\otimes\hat{\rho}_2'}
\end{eqnarray}
and 
\begin{eqnarray}
 e^{\hat{\rho}_{1}'}\otimes \hat{1_2}=e^{\hat{\rho}_{1}'\otimes\hat{1}_2},
\end{eqnarray}
to recast the functional equation as,
\begin{eqnarray}
\phi\Big(e^{\hat{\rho}_{1}'\otimes\hat{1}_2+ \hat{1}_1\otimes\hat{\rho}_2'},e^{\hat{\varphi}_{1}'\otimes\hat{1}_2+ \hat{1}_1\otimes\hat{\varphi}_2'}\Big)=\phi\Big(e^{\hat{\rho}_{1}'\otimes\hat{1}_2},e^{\hat{\varphi}_{1}'\otimes\hat{1}_2}\Big)+\phi\Big(e^{\hat{1}_1\otimes \hat{\rho}_2'},e^{\hat{1}_1\otimes \hat{\varphi}_2'}\Big).
\end{eqnarray}
Letting $G(\hat{\rho}_{1}'\otimes\hat{1}_2,\hat{\varphi}_{1}'\otimes\hat{1}_2)=\phi\Big(e^{\hat{\rho}_{1}'\otimes\hat{1}_2},e^{\hat{\varphi}_{1}'\otimes\hat{1}_2}\Big)$, and so on, gives
\begin{eqnarray}
G(\hat{\rho}_{1}'\otimes\hat{1}_2+ \hat{1}_1\otimes\hat{\rho}_2',\hat{\varphi}_{1}'\otimes\hat{1}_2+ \hat{1}_1\otimes\hat{\varphi}_2')
=G(\hat{\rho}_{1}'\otimes\hat{1}_2,\hat{\varphi}_{1}'\otimes\hat{1}_2)+G(\hat{1}_1\otimes\hat{\rho}_2',\hat{1}_1\otimes\hat{\varphi}_2').\nonumber\\
\end{eqnarray}
This functional equation is of the form
\begin{eqnarray}
G(\hat{X}_1'+\hat{Y}_1',\hat{X}_2'+\hat{Y}_2')=G(\hat{X}_1',\hat{X}_2')+G(\hat{Y}_1',\hat{Y}_2'),
\end{eqnarray}
which has the general solution
\begin{eqnarray}
G(\hat{X}',\hat{Y}')=\stackrel{\,\,\sim}{A}\hat{X}'+\stackrel{\,\,\sim}{B}\hat{Y}',
\end{eqnarray}
analogous to (\ref{genlinearsol}), and finally, in general,
\begin{eqnarray}
\phi(\hat{\rho},\hat{\varphi})=\stackrel{\,\,\sim}{A}\ln(\hat{\rho})+\stackrel{\,\,\sim}{B}\ln(\hat{\varphi}),
\end{eqnarray}
where $\stackrel{\,\,\sim}{A},\stackrel{\,\,\sim}{B}$ are super-operators having constant coefficients. Here, we return to the text.

\chapter{Hermitian variations of $\hat{\rho}$ and maximum entropy \label{deltarhos}}

Here we fill in the steps between equation (\ref{2.39}) and (\ref{52}). Letting $\delta\hat{\rho}$ be Hermitian is imposing $\delta\rho_{ij}=\delta\rho_{ji}^*$. The variation of interest is equation (\ref{2.39}),
\begin{eqnarray}
\mbox{Tr}\Big(\Big(\frac{\delta S(\hat{\rho},\hat{\varphi})}{\delta \hat{\rho}^T}-\lambda\hat{1}-\alpha\hat{A}\Big)\delta\hat{\rho}\Big)=0,
\end{eqnarray}
which we write as,
\begin{eqnarray}
\mbox{Tr}\Big(\hat{B}\delta\hat{\rho}\Big) =0,
\end{eqnarray}
with $\hat{B}\equiv\frac{\delta S(\hat{\rho},\hat{\varphi})}{\delta \hat{\rho}^T}-\lambda\hat{1}-\alpha\hat{A}$, being a Hermitian matrix. Rewritten in component form, the trace is
 \begin{eqnarray}
\sum_{ij}B_{ji}\delta\rho_{ij}=0,
\end{eqnarray}
which when partitioned by the diagonal, upper triangle, and lower triangle components of $\delta\rho_{ij}$, gives
\begin{eqnarray}
\sum_{i}B_{ii}\delta\rho_{ii}+\sum_{i<j}B_{ji}\delta\rho_{ij}+\sum_{i<j}B_{ij}\delta\rho_{ji}=0.
\end{eqnarray}
Because $\delta\rho_{ii}$ is real, for arbitrary variations of  $\delta\rho_{ii}$, $B_{ii}=0$, which reduces the sum to,
\begin{eqnarray}
\sum_{i<j}B_{ji}\delta\rho_{ij}+\sum_{i<j}B_{ij}\delta\rho_{ji}=0.
\end{eqnarray}
Because $\hat{B}$ and $\delta\hat{\rho}$ are Hermitian, $\delta\rho_{ij}=\delta\rho_{ji}^*$ and $B_{ij}=B_{ji}^*$. Substitution above gives, 
\begin{eqnarray}
\sum_{i<j}(B_{ji}\delta\rho_{ij}+B_{ji}^*\delta\rho_{ij}^*)=0.
\end{eqnarray}
Breaking the matrices' components into real and imaginary parts gives,
\begin{eqnarray}
\sum_{i<j}\Big( \Big( \mathbb{Re}(B_{ji})+i~\mathbb{I}(B_{ji})\Big)\Big( \mathbb{Re}(\delta\rho_{ij})+i~\mathbb{I}(\delta\rho_{ij})\Big)\nonumber\\+\Big( \mathbb{Re}(B_{ji})-i~\mathbb{I}(B_{ji})\Big)\Big( \mathbb{Re}(\delta\rho_{ij})-i~\mathbb{I}(\delta\rho_{ij})\Big)\Big)=0,
\end{eqnarray}
which reduces to
\begin{eqnarray}
2\sum_{i<j}\mathbb{Re}(B_{ji})\mathbb{Re}(\delta\rho_{ij})-2\sum_{i<j}\mathbb{I}(B_{ji})\mathbb{I}(\delta\rho_{ij})=0.
\end{eqnarray}
Therefore, arbitrary Hermitian variations of $\delta\hat{\rho}$ require $\mathbb{Re}(B_{ji})=\mathbb{I}(B_{ji})=B_{ji}=0$, and therefore $\hat{B}=0$. Therefore, at the maximum, the variational derivative of $S$ must satisfy (\ref{52}),
\begin{eqnarray}
\frac{\delta S(\hat{\rho},\hat{\varphi})}{\delta \hat{\rho}^T}=\lambda\hat{1}+\alpha\hat{A},
\end{eqnarray}
for arbitrary or Hermitian variations of $\hat{\rho}$.
\chapter{Spin Example - Quantum Maximum Entropy Method\label{spinexampleQRE}}
This follows \cite{QRE}. Consider an arbitrarily mixed prior (in the $\pm$ spin-$z$ basis for convenience) with $a,b\neq0$,
\begin{eqnarray}
\hat{\varphi}=a\ket{+}\bra{+}+b\ket{-}\bra{-}
\end{eqnarray}
and a general Hermitian matrix in the spin-$1/2$ Hilbert space, 
\begin{eqnarray}
c_{\mu}\hat{\sigma}^{\mu}=c_1\hat{1}+c_x\hat{\sigma}_x+c_y\hat{\sigma}_x+c_z\hat{\sigma}_z
\end{eqnarray}
\begin{eqnarray}
=(c_1+c_z)\ket{+}\bra{+}+(c_x-ic_y)\ket{+}\bra{-}+(c_x+ic_y)\ket{-}\bra{+}+(c_1-c_z)\ket{-}\bra{-},
\end{eqnarray}
having a known expectation value,
\begin{eqnarray}
\mbox{Tr}(\hat{\rho}c_{\mu}\hat{\sigma}^{\mu})=c.
\end{eqnarray}
Maximizing the entropy with respect to this general expectation value and normalization is:
\begin{eqnarray}
0=\Big(\delta S-\lambda[\mbox{Tr}(\hat{\rho})-1]-\alpha(\mbox{Tr}(\hat{\rho}c_{\mu}\hat{\sigma}^{\mu})-c)\Big),
\end{eqnarray}
which after varying gives the solution,
\begin{eqnarray}
\hat{\rho}=\frac{1}{Z}\exp(\alpha c_{\mu}\hat{\sigma}^{\mu}+\log(\hat{\varphi}))\label{spinrhoexample}.
\end{eqnarray}
Letting
\begin{eqnarray}
\hat{C}=\alpha c_{\mu}\hat{\sigma}^{\mu}+\log(\hat{\varphi})
\end{eqnarray}
gives
\begin{eqnarray}
\hat{\rho}=\frac{1}{Z}e^{\hat{C}}=Ue^{U^{-1}\hat{C}U}U^{-1}=\frac{1}{Z}Ue^{\hat{\lambda}}U^{-1}\nonumber\\
=\frac{e^{\lambda_+}}{Z}U\ket{\lambda_+}\bra{\lambda_+}U^{-1}+\frac{e^{\lambda_-}}{Z}U\ket{\lambda_-}\bra{\lambda_-}U^{-1},
\end{eqnarray}
where $\hat{\lambda}$ is the diagonalized matrix of $\hat{C}$ having real eigenvalues. They are
\begin{eqnarray}
\lambda_{\pm}=\lambda\pm\delta\lambda,
\end{eqnarray}
due to the quadratic formula, where explicitly:
\begin{eqnarray}
\lambda=\alpha c_1+\frac{1}{2}\log(ab),
\end{eqnarray}
and
\begin{eqnarray}
\delta\lambda=\frac{1}{2}\sqrt{\Big(2\alpha c_z+\log(\frac{a}{b})\Big)^2+4\alpha^2(c_x^2+c_y^2)}.
\end{eqnarray}
Because $\lambda_{\pm}$ and $a,b,c_1,c_x,c_y,c_z$ are real, $\delta\lambda$ is real and $\geq 0$.
The normalization constraint specifies the Lagrange multiplier $Z$,
\begin{eqnarray}
1=\mbox{Tr}(\hat{\rho})=\frac{e^{\lambda_+}+e^{\lambda_-}}{Z},
\end{eqnarray}
so $Z=e^{\lambda_+}+e^{\lambda_-}=2e^{\lambda}\cosh(\delta\lambda)$. The expectation value constraint specifies the Lagrange \mbox{multiplier $\alpha$},
\begin{eqnarray}
c=\mbox{Tr}(\hat{\rho}c_{\mu}\sigma^{\mu})=\frac{\d}{\d\alpha}\log(Z)=c_1+\tanh(\delta\lambda)\frac{\d }{\d\alpha}\delta\lambda,
\end{eqnarray}
which becomes
\begin{eqnarray}
c=c_1+\frac{\tanh(\delta\lambda)}{2\delta\lambda}\Big(2\alpha(c_x^2+c_y^2+c_z^2)+c_z\log(\frac{a}{b})\Big),\nonumber
\end{eqnarray}
or
\begin{eqnarray}
c=c_1+\tanh\Big(\frac{1}{2}\sqrt{\Big(2\alpha c_z+\log(\frac{a}{b})\Big)^2+4\alpha^2(c_x^2+c_y^2)}\Big)\frac{2\alpha(c_x^2+c_y^2+c_z^2)+c_z\log(\frac{a}{b})}{\sqrt{\Big(2\alpha c_z+\log(\frac{a}{b})\Big)^2+4\alpha^2(c_x^2+c_y^2)}}.\nonumber\\ \label{spinsol}
\end{eqnarray}
This equation is monotonic in $\alpha$ and therefore it is uniquely specified by the value of $c$. Ultimately, this is a consequence from the concavity of the entropy. The specific proof of (\ref{spinsol})'s monotonicity is below:

For $\hat{\rho}$ to be Hermitian, $\hat{C}$ is Hermitian and $\delta\lambda=\frac{1}{2}\sqrt{f(\alpha)}$ is real---furthermore, because $\delta\lambda$ is real $f(\alpha)\geq0$ and thus $\delta\lambda\geq0$. Because $f(\alpha)$ is quadratic in $\alpha$ and positive, it may be written in vertex~form,
\begin{eqnarray}
f(\alpha)=a(\alpha-h)^2+k,
\end{eqnarray}
where $a>0$, $k\geq 0$, and $(h,k)$ are the $(x,y)$ coordinates of the minimum of $f(\alpha)$. Notice that the form of (\ref{spinsol}) is
\begin{eqnarray}
F(\alpha)=\frac{\tanh(\frac{1}{2}\sqrt{f(\alpha)})}{\sqrt{f(\alpha)}}\times\frac{\d f(\alpha)}{\d\alpha}.
\end{eqnarray}
Making the change of variables $\alpha'=\alpha-h$ centers the function such that $f(\alpha')=f(-\alpha')$ is symmetric about $\alpha'=0$. We can then write
\begin{eqnarray}
F(\alpha')=\frac{\tanh(\frac{1}{2}\sqrt{f(\alpha')})}{\sqrt{f(\alpha')}}\times 2a\alpha',
\end{eqnarray}
where the derivative has been computed. Because $f(\alpha')$ is a positive, symmetric, and monotonically increasing on the (symmetric) half-plane (for $\alpha'$ greater than or less that zero), $S(\alpha')\equiv\frac{\tanh(\frac{1}{2}\sqrt{f(\alpha')})}{\sqrt{f(\alpha')}}$ is also positive and symmetric, but it is unclear whether $S(\alpha)$ is strictly monotonic in the half-plane or not. We may restate
\begin{eqnarray}
F(\alpha')=S(\alpha')\times 2a\alpha'.
\end{eqnarray}
We are now in a convenient position to perform the first derivate test for monotonic functions:
\begin{eqnarray}
\frac{\d}{\d\alpha'}F(\alpha')&=&2aS(\alpha')+2a\alpha'\frac{\d}{\d\alpha'} S(\alpha')\\
&=& 2aS(\alpha')\Big(1-\frac{a\alpha'^2}{a\alpha'^2+k}\Big)+a\frac{a\alpha'^2}{a\alpha'^2+k}\Bigg(1-\tanh^2(\frac{1}{2}\sqrt{a\alpha'^2+k})\Bigg)\\
&\geq& 2aS(\alpha')\Big(1-\frac{a(\alpha')^2}{a\alpha'^2+k}\Big)\geq0\\
\end{eqnarray}
because $a,k,S(\alpha')$, and therefore $\frac{a\alpha'^2}{a\alpha'^2+k}$ are all $>0$.  The function of interest $F(\alpha')$ is therefore monotonic for all $\alpha'$, and therefore it is monotonic for all $\alpha$, completing the proof that  there exists a unique real Lagrange multiplier $\alpha$ in (\ref{spinsol}).

Although (\ref{spinsol}) is monotonic in $\alpha$, it is seemingly a transcendental equation. This can be solved graphically for the given values $c,c_1,c_x,c_y,c_z$, i.e., given the Hermitian matrix and its expectation value are specified. Equation (\ref{spinsol}) and the eigenvalues take a simpler form when $a=b=\frac{1}{2}$ because, in this instance, $\hat{\varphi}\propto\hat{1}$ and commutes universally so it may be factored out of the exponential in (\ref{spinrhoexample}).

A specific example might be carry some insight. Consider a prior density matrix $\hat{\varphi}$ diagonal in spin-$z$ and having the components $a=.75$ and $b=.25$. If we then gain new information that the density matrix in question actually has an expectation value of $\expt{\sigma_x}=0.9$, we may impose this expectation value and normalization via the quantum maximum entropy method. In this instance one finds the Lagrange multiplier $\alpha\approx 1.7$, and a posterior density matrix of,
\begin{eqnarray}\hat{\rho}\approx\left( \begin{array}{cc}0.65 & 0.45 \\
0.45 &0.35 
\end{array} \right).
\end{eqnarray}  
This mixed state ($\lambda_{1,2}>0$) is selected over other states that reproduce the expectation value constraints, for instance,  the pure spin $x$-state $\ket{\Psi}=\sqrt{0.95}\ket{+x}+\sqrt{0.05}\ket{-x}$. 

\chapter{Inference of spin from position measurements\label{spinexampleED}}
The inference of spin from position may be done by following the Stern-Gerlach mechanics in \cite{Holland}, or by more-or-less following the mechanics in \cite{AAV,duck}, while demanding ontic particle positions and their detections.
The relevant coupling Hamiltonian from a Stern-Gerlach device is of the form (\ref{HB}),
\begin{eqnarray}
H_s=-\lambda g(t)\hat{\sigma}_z\hat{z},
\end{eqnarray}
where $\lambda\propto \delta B_z/\delta z$ and the particles magnetic moment $\mu$, and $\hat{v}_{conj}=\hat{z}$. Following the weak measurement prescription, that is having $\ket{\Psi_i}$ be Gaussian in position space and letting $\ket{\Phi_i}=\alpha_+\ket{+}+\alpha_-\ket{-}$ be the initial uniform spin vector, the two vectors are entangled via the coupling Hamiltonian, and
\begin{eqnarray}
\ket{\Phi_{f},\Psi_{f}}=U_s\ket{\Phi_{i}}\ket{\Psi_{i}}=\int\sum_{\pm} \alpha_{\pm}e^{\mp iz\lambda}e^{-\bigtriangleup^2z^2}\ket{\pm}\ket{z}\,dz=\int\sum_{\pm} \alpha_{\pm} e^{-\frac{(p\mp\lambda)^2}{4\bigtriangleup^2}} \ket{\pm}\ket{p}\,dp.
\end{eqnarray}
Rather than using a unitary measurement device (\ref{unitary}) to evolve momentum states to position, the particle is usually allowed to travel in free space for an additional time $t$ before hitting a screen, 
\begin{eqnarray}
\ket{\Phi_{f},\Psi_{f}}=\int\sum_{\pm} \alpha_{\pm} e^{-\frac{(p\mp\lambda)^2}{4\bigtriangleup^2}-i\frac{p^2}{2m}t} \ket{\pm}\ket{p}\,dp\propto \int\sum_{\pm} \alpha_{\pm} \exp\Big(-\Delta^2\frac{(z \mp i\frac{\lambda}{2(1+2i\Delta^2t/m)})^2}{(1+2i\Delta^2t/m)^2}\Big) \ket{\pm}\ket{z}\,dz.\nonumber\\\label{spin36}
\end{eqnarray}
One arrives at joint probabilities like,
\begin{eqnarray}
P(\pm,z)\propto\frac{|\alpha_{\pm}|^2}{\sqrt{1+4\Delta^4t^2/m^2}}\exp\Big(-\frac{2\Delta^2}{1+4\Delta^4t^2/m^2}(z\mp \frac{\lambda\Delta t}{m})^2\Big),
\end{eqnarray}
and the distributions are more-or-less resolvable when the average displacement is larger than a few standard deviations of either Gaussian distribution $P(z|\pm)$. At the position $z$ and at the time $t$, the modulus squared of a component of the spin along $\pm z$ can be inferred as,
\begin{eqnarray}
P(\pm|z)=\frac{|\alpha_{\pm}|^2\exp\Big(-\frac{2\Delta^2}{1+4\Delta^4\hbar^2t^2/m^2}(z\mp \frac{\lambda\Delta \hbar t}{m})^2\Big)}{\sum_{\pm}|\alpha_{\pm}|^2\exp\Big(-\frac{2\Delta^2}{1+4\Delta^4\hbar^2t^2/m^2}(z\mp \frac{\lambda\Delta \hbar t}{m})^2\Big)}.
\end{eqnarray}
Because the particle spins are never truly observed, but rather are inferred from position detections, $P(\pm|z)$ is more or less the probability that a particle would continue to translate in space along $\pm z$ given another SG device was set up and the particle passed through $z$, i.e. the probability of the spin being up or down. If the goal is to infer spin values $\pm_x$ or $\pm_y$, one uses a different orientations of the Stern-Gerlach device and a different unitary evolution. 

%

\end{document}